\documentclass[astrosymb,times, preprint]{aastex701}

\usepackage{amsmath}

\newcommand{\lasp}{Laboratory for Atmospheric and Space Physics, University of Colorado, 600 UCB, Boulder, CO 80309, USA}
\newcommand{\casa}{Center for Astrophysics and Space Astronomy, University of Colorado, 593 UCB, Boulder, CO 80309}
\newcommand{\cuboulder}{Department of Astrophysical and Planetary Sciences, University of Colorado, Boulder, CO 80309, USA}
\newcommand{\vanderbilt}{Department of Physics and Astronomy, Vanderbilt University, Nashville, TN 37235, USA}

\newcommand{\Lya}{Ly$\alpha$}
\newcommand{\halp}{H$\alpha$}
\newcommand{\hbet}{H$\beta$}
\newcommand{\hgam}{H$\gamma$}
\newcommand{\hdel}{H$\delta$}

\newcommand{\SiIII}{\ion{Si}{3}}

\received{June 25, 2025}
\revised{August 8, 2025}
\accepted{August 25, 2025}
\submitjournal{AJ}

\begin{document}

\title{FUMES IV: Optical and Far-ultraviolet Spectra of a Flare on the M Dwarf GJ 4334}

\author[0000-0002-7119-2543]{Girish M. Duvvuri}
\affiliation{\vanderbilt}
\email{girish.duvvuri@vanderbilt.edu}

\author[0000-0002-4489-0135]{J. Sebastian Pineda}
\affiliation{\lasp}
\email{sebastian.pineda@lasp.colorado.edu}

\author[0000-0001-9828-3229]{Aylin Garc{\'i}a Soto}
\affiliation{Department of Physics and Astronomy, Dartmouth College, Hanover NH 03755, USA}
\email{Aylin.Garcia.Soto.GR@dartmouth.edu}

\author[0000-0002-3321-4924]{Zachory K. Berta-Thompson}
\affiliation{\cuboulder}
\affiliation{\casa}
\email{Zachory.BertaThompson@colorado.edu}

\author[0000-0002-1176-3391]{Allison Youngblood}
\affiliation{Exoplanets and Stellar Astrophysics Laboratory, NASA Goddard Space Flight Center, Greenbelt, MD 20771}
\email{allison.a.youngblood@nasa.gov}

\author[0000-0002-1002-3674]{Kevin France}
\affiliation{\cuboulder}
\affiliation{\lasp}
\affiliation{\casa}
\email{Kevin.France@lasp.colorado.edu}

\author[0000-0003-4150-841X]{Elisabeth R. Newton}
\affiliation{Department of Physics and Astronomy, Dartmouth College, Hanover NH 03755, USA}
\email{Elisabeth.R.Newton@dartmouth.edu}

\author[0000-0002-3481-9052]{Keivan G. Stassun}
\affiliation{\vanderbilt}
\email{keivan.stassun@vanderbilt.edu}

\correspondingauthor{Girish M. Duvvuri}
\email{girish.duvvuri@vanderbilt.edu}

\begin{abstract}

On 2017-09-20 we observed GJ 4334, an M5V dwarf rotating with a period of 23.5 days, simultaneously with both the Space Telescope Imaging Spectrograph aboard \emph{Hubble} (1160 -- 1710 \r{A}) and the Dual Imaging Spectrograph mounted on the 3.5m telescope at Apache Point Observatory (3750 -- 5050; 5800 -- 6950 \r{A}) as part of a larger survey of intermediately active M dwarfs. GJ 4334 flared during the observation, starting with a rise in the flux of optical chromospheric emission lines, followed by the rapid rise and decay of multiple far\added{-}ultraviolet emission lines formed in the transition region, followed by the slow decay of the optical lines. We find significant broadening and asymmetries in the optical emission lines that are potentially from bulk plasma motion, a post-flare elevated flux in both the optical and far\added{-}ultraviolet, and trends in the rise and decay timescales of the Balmer series such that higher-order lines rise earlier and decay faster than lower-order lines. The equivalent durations of the flare in individual lines range from 800 -- $3 \times 10^4$ seconds, mapping to flare energies of $1 \times 10^{28}$ -- $3 \times 10^{29}\,$erg for each line. To contextualize GJ 4334's flare behavior we measure and compare its optical flare frequency distribution with \emph{TESS} to EV Lacertae, a similar mass but faster rotating M dwarf, and find that GJ 4334 has an excess of large flares relative to the power-law established by the majority of its smaller flares. This dataset is a rare opportunity to characterize flares near a critical transition in stellar magnetic activity.

\end{abstract}


\section{Introduction} \label{sec:intro}

Stellar flares are the sudden brightening of a star following a release of magnetic energy in the star's upper atmosphere, usually associated with a magnetic reconnection event \citep{Benz:2010}. Electrons are suddenly accelerated to a non-thermal velocity distribution and are beamed down towards the chromosphere, where they collide with ambient plasma and heat two flows in opposite directions \citep{Fisher:1989}: a ``chromospheric condensation'' layer that continues to move deeper into the stellar atmosphere so long as the non-thermal electron flux can provide the layer with enough energy, and an upward and outward expansion of plasma resulting in an observable ``chromospheric evaporation'' that potentially launches prominences of plasma above the surface with enough velocity to become coronal mass ejections (CMEs, \citealp{Munro:1979}). This dynamic sequence of plasma motion is accompanied by a similarly dynamic evolution of the star's radiation, with the contributions of different atmospheric layers to the stellar spectrum changing depending on the local density, temperature, and non-thermal electron flux. During quiescence, many individual emission features can be associated with specific layers of the stellar upper atmosphere; coronal X-ray emission, far\added{-}ultraviolet transition region lines like \SiIII\ 1206 \r{A}, and chromospheric Balmer H$\alpha$ 6563 \r{A}. The flare's disruption of these layers complicates these emission features' formation and localization, but tracking their changes over time offers some insight into the plasma's physical properties during the flare.

Multiwavelength observations of flares on stars and the Sun have shown that while many elements of this canonical flare model apply to both the Sun and other stars\added{,} and some aspects of flares may scale similarly, others do not. For example, \added{\citet{Aarnio:2011,Aarnio:2012}} found a relationship between X-ray flux and coronal mass ejections for solar flares that appears to also apply to \added{pre-main sequence stellar} flares five orders of magnitude stronger than the strongest solar flare used to constrain the relationship, but the existence of \added{similarly strong} ``superflares'' \citep{Favata:1999, Shibayama:2013} on active GKM dwarfs has no solar analog. A difference between solar and \added{M dwarf} flares is that \Lya\ varies dramatically during solar flares but is much less responsive in M dwarf flares, possibly because the quiescent \Lya\ emission of M dwarfs is optically thicker, \citep{Gershberg:2024}. Understanding the details of these similarities and differences requires spectral snapshots during the evolution of a flare, which is observationally challenging for stellar flares. Flare spectra are typically obtained from dedicated monitoring programs targeting individual stars already known to flare frequently, or serendipitously from observations intended for other purposes such as from radial velocity spectrographs studying exoplanets \citep{Fuhrmeister:2011}, programs characterizing the spectra of a group of stars (again, typically for exoplanet purposes, such as \citealp{Loyd:2018a, Loyd:2018b, Froning:2019, Diamond-Lowe:2024}), and in the ultraviolet some notable instances where flares confound attempts to detect planetary atmospheric escape \citep{Rockcliffe:2023,Rockcliffe:2025}.


This last source of flare spectra is not simply accidental, but rather a consequence of the scientific interests linking flares and planetary atmospheres, particularly for exoplanets orbiting young stars that flare frequently and/or strongly. The net energy irradiating and heating these exoplanets' atmospheres includes contributions from flares\added{,} and affects planetary atmospheric structure and composition. Flare excess high-energy radiation has a variety of effects based on how deeply photons of a particular energy can penetrate the planetary atmosphere.  Extreme ultraviolet (EUV, 100 -- 912 \r{A}) photons are absorbed in the outermost layers of the planet's exosphere, predominantly because of these photons' interactions with hydrogen atoms, depositing their energy by photoionization and producing free electrons that heat the atmosphere and contribute to the escape of lighter species \citep{Dong:2017}. In the most extreme cases, the collective ``XUV'' (X-ray + EUV, 1 -- 912 \r{A}) radiation can drive a hydrodynamic outflow of material, essentially pouring a planet's air out into space \citep{Koskinen:2022}. Both far\added{-}ultraviolet (FUV, 912 -- 1700 \r{A}) and near\added{-}ultraviolet (NUV, 1700 -- 3200 \r{A}) \added{radiation} can drive photochemical reactions that may mask biosignature detection by producing hazes \citep{Horst:2017}\added{,} or masquerade as a biosignature by producing an excess of O\textsubscript{3} \citep{Hu:2012,Tian:2014,Harman:2015,Schaefer:2016}. Characterizing the spectra of individual flares is necessary to model an atmosphere's instantaneous response to such events, while understanding the frequency and diversity of flares at different energies is necessary to model the cumulative impact of flares on an atmosphere's long-term evolution. \citet{France:2020} used contemporaneous X-ray and FUV measurements of a flare on Barnard's Star to produce a panchromatic flare spectrum, modeled the response of a hypothetical planet to repeated flares of varying strength, and found that because of the heating of an atmosphere post-flare, it was more susceptible to erosion by a subsequent flare, making the flare frequency a key determinant of whether the planet's atmosphere could persist over Gyr timescales.

Young stars of all masses flare more frequently than their older counterparts, starting their magnetic behavior at a high activity ``saturation'' level that eventually crosses some transition age and/or rotation period which triggers a gradual ``decay'' phase characterized by decreasing activity as the star gets older and/or slows down its rotation \citep{Skumanich:1972,Johnstone:2021,Pass:2024}. The Far Ultraviolet M\added{-}dwarf Evolution Survey \citep[hereafter referred to as FUMES I]{Pineda:2021a} was designed to characterize elements of this saturation-decay evolution by filling in an observational gap: survey programs characterizing exoplanet hosts were biased towards older, slowly rotating M dwarfs \citep{France:2016} and high activity M dwarfs were covered by flare monitoring programs \citep{Hawley:2003, Kowalski:2013}, leaving the exact transition age/rotation period underconstrained with a lack of ultraviolet data for M dwarfs at intermediate ages and/or rotation periods \citep{Shkolnik:2014}. \citet[FUMES II]{Youngblood:2021} characterized the transition region \Lya\ emission of this sample, and \citet[FUMES III]{Duvvuri:2023} used a combination of the \emph{Hubble} FUV data and optical spectra to study the stochastic variability of magnetic activity indicators for most of the FUMES sample.


One star, GJ 4334, was excluded from the analysis of FUMES III because we were fortunate enough to spectroscopically observe a flare simultaneously in the far\added{-}ultraviolet with Space Telescope Imaging Spectrograph aboard \emph{Hubble} (1160 -- 1710 \r{A}) and in the optical with the Dual Imaging Spectrograph mounted on the Apache Point Observatory 3.5m telescope (3750 -- 5050; 5800 -- 6950 \r{A}). Three years later, GJ 4334 flared again during a follow-up \emph{XMM}-\emph{Newton} observation. Since the beginning of the \emph{TESS} mission, GJ 4334 has flared $> 75$ times during the net $\approx 86$~days of observing time over four \emph{TESS} sectors (a fifth sector of data is \added{now} available but was not analyzed in this work). \added{GJ 4334 was first reported to be a flare star in \citet{Hoffmeister:1967}, which observed multiple flares in an unspecified filter; at least one event flared greater than one magnitude}. This star, which was selected only for being a nearby M dwarf with an intermediate rotation period, could be an older version of the well-known flare star EV Lac; not flaring quite as frequently or dramatically, but often enough to reliably provide flare spectra in limited observation time.

There are few spectroscopic analogs to this dataset with similar wavelength coverage, all of which were much more massive co-ordinated campaigns for famous stars like the young rapid rotators AD Leo \citep{Hawley:1991, Hawley:2003}, EV Lac \citep{Osten:2005}, GJ 1243 \citep{Kowalski:2019}, and AU Mic \citep{Tristan:2023}. The only similar campaign for an old star \added{(age $>$ 1 Gyr)} is for Proxima Cen \citep{MacGregor:2021, Howard:2022}, which rotates much more slowly than GJ 4334 (89 days for Proxima Cen \citealp{Klein:2021} rather than GJ 4334's 23.5 days \citealp{Newton:2016}). Dedicating a similar effort to GJ 4334 would yield new insights into flare behavior near the critical transition of magnetic activity from saturation to decay, but the spectra in hand are already powerful constraints on flare models because of the range of formation temperatures and physical conditions probed by the observed emission features. This paper presents the observational setups of both wavelength regimes and describes the preparation of spectroscopic timeseries for different emission features in Section~\S\ref{sec:rainbows}, compares and analyzes spectra at specific times and lightcurves of the integrated flux of emission features in Section~\S\ref{sec:analysis}, discusses how these observations relate to previous flare observations and models in Section~\S\ref{sec:apo_hst:literature}, characterizes the \emph{TESS} flare frequency distribution of GJ 4334 in Section~\S\ref{sec:tess}, situates GJ 4334 in stellar parameter space in Section~\S\ref{sec:stellar}, and concludes by summarizing our findings and discussing their implications in Section~\S\ref{sec:conclusion}.

\section{Flare Observations, Data Reduction, and Spectroscopic Timeseries}\label{sec:rainbows}

\subsection{Apache Point Observatory/ Dual Imaging Spectrograph}
\begin{figure*}
    \centering
    \plotone{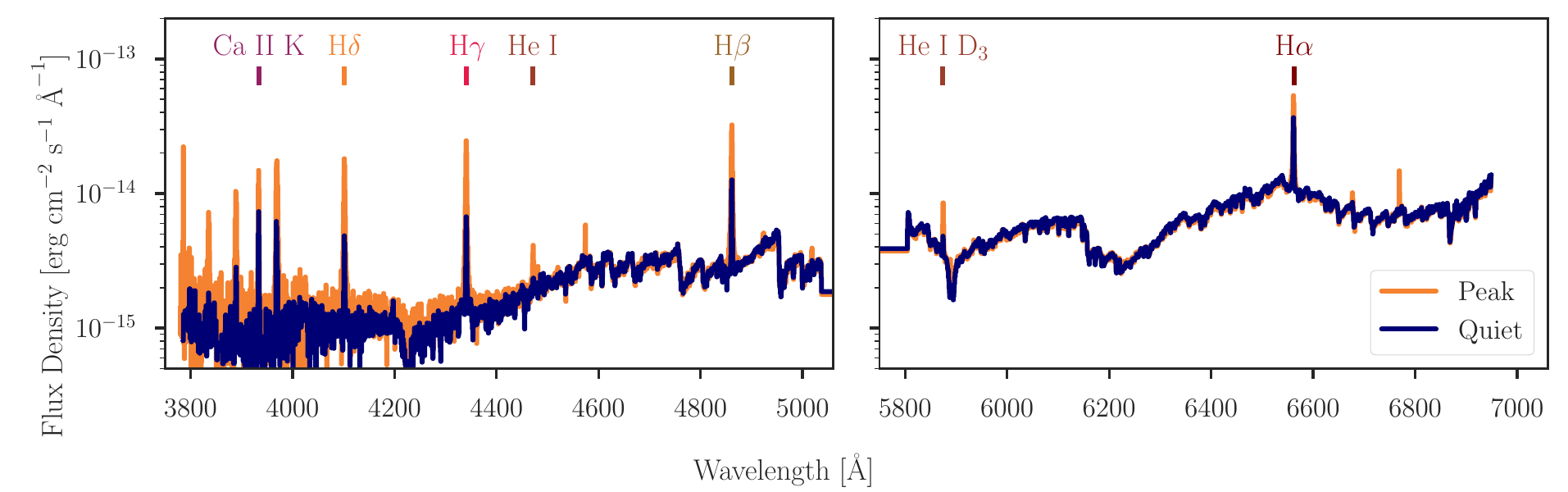}
    \caption{The quiescent APO optical spectrum is plotted in dark blue while the spectrum at flare peak is plotted in orange, with separate panels for the two arms of DIS. The flare only seems to brighten lines relative to the continuum level, leaving the majority of the continuum unchanged. The exception is the continuum below 4500 \r{A} which appears slightly enhanced. A subset of the optical lines analyzed in this work have been highlighted with ion labels above the line peak.}
    \label{fig:apo_full_spec}
\end{figure*}

The observational setup and reduction procedure for both the APO and \emph{HST} data of GJ 4334 have been previously described in Sections \S2.1 and \S3.1 of \citet{Duvvuri:2023} respectively. The subset of GJ 4334 APO observations from Table 2 of \citet{Duvvuri:2023} has been reproduced here in Table \ref{tab:observations_apo}. We observed GJ 4334 using the Dual Imaging Spectrograph (DIS), a dual-channel spectrograph mounted on the 3.5m telescope at APO. We centered the blue and red arms of the spectrograph on 4400 and 6400 $\textrm{\AA}$ with the B1200 and R1200 gratings respectively, resulting in a wavelength resolving power $7000 < R = \frac{\lambda}{\Delta\lambda} < 12000$ increasing from blue to red wavelengths. We reduced the DIS data using a modified version of \texttt{pyDIS} \citep{Davenport:2016}, a long-slit spectroscopy reduction package developed for DIS, and the modified version we used is available in the Zenodo repository \citep{duvvuri_2022_6909473} associated with \citet{Duvvuri:2023}. Each wavelength arm was treated independently for the bias-subtraction, flat-fielding, wavelength-tracing (using a cubic spline), and wavelength calibration (using a HeNeAr lamp). We extracted the spectra using a simple aperture boxcar extraction. The flux calibration observation was taken before the GJ 4334 data and the airmass correction does not account for other systematic changes during the night (such as weather, although no major events were noted during the observation period) so we used continuum regions from the averaged spectrum of the pre-flare quiescent exposures to normalize subsequent exposures for further analysis. Figure \ref{fig:apo_full_spec} compares the averaged quiescent spectrum and the original flare peak spectrum (no continuum normalization) and highlights the spectral emission lines analyzed in this section. The very beginning of the first DIS exposure, 2017-09-20T09:21:14.261, is the reference zero-point time for both the APO and \emph{HST} data analysis in this work.

\begin{deluxetable}{ccccc}
\tablecaption{APO/ARC 3.5m DIS/1.5" \label{tab:observations_apo}}
\tablehead{\colhead{UT Date} & \colhead{Star} & \colhead{Initial Airmass} & \colhead{Exposure Duration [s]} & \colhead{$N_{\mathrm{exposures}}$}}
\startdata
2017-09-20 & G 191B2B\tablenotemark{*} & 1.26 & 360 & 1 \\
2017-09-20 & GJ 4334 & 1.28 & 360 & 3 \\
2017-09-20 & GJ 4334 & 1.3 & 420 & 13 
\enddata
\tablenotetext{*}{This star was used as a flux standard.}
\end{deluxetable}

\subsection{Hubble/Space Telescope Imaging Spectrograph}
\begin{figure*}
    \centering
    \plotone{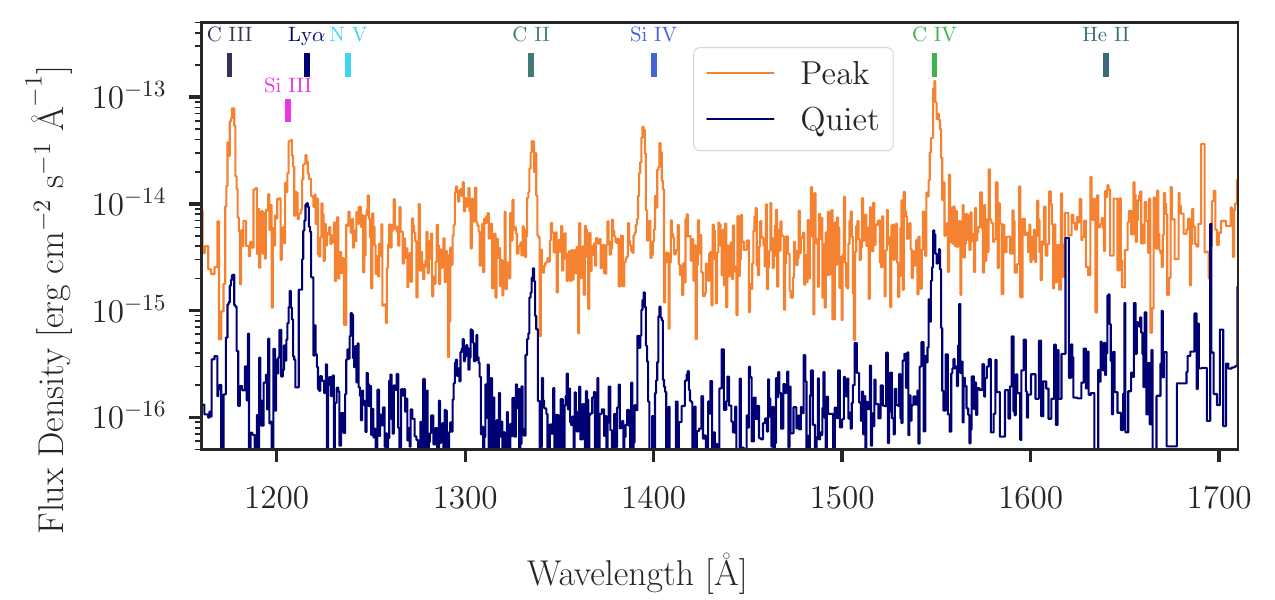}
    \caption{The quiescent \emph{HST} FUV spectrum is plotted in dark blue while the spectrum at flare peak is plotted in orange. The flare spectrum has not been staggered by multiplying a constant factor, but the offset is a physical consequence of the flare brightening the FUV spectrum across the observation bandpass. A subset of the FUV lines analyzed in this work have been highlighted with ion labels above the line peak.}
    \label{fig:hst_full_spec}
\end{figure*}

The \emph{HST} data for GJ 4334 were taken using the Space Telescope Imaging Spectrograph (STIS) using the G140L grating in TIME-TAG mode as part of GO program 14640\footnote{\url{https://archive.stsci.edu/proposal\_search.php?id=14640\&mission=hst}} and the observational program as a whole is discussed in \citet{Pineda:2021a}\added{; the specific observations are available via \dataset[doi:10.17909/gkv0-6x65]{https://doi.org/10.17909/gkv0-6x65}}. Just as in \citet{Duvvuri:2023} we used the \texttt{spectralPhoton} \citep{Loyd:2018a, Loyd:2018b} package to divide individual photon events recorded in TIME-TAG mode into pseudo-exposures\added{,} and to extract the spectrum using the trace of the pipeline-produced \texttt{x1d} spectrum with a constant wavelength bin width of 0.6 $\textrm{\AA}$ between 1140 -- 1715 $\textrm{\AA}$. For analyzing the GJ 4334 data, we produced three sets of pseudo-exposures: two divided into bins of equal time duration of 1 minute and 3 minutes, and a third divided into bins with an equal number of counts across the spectrum. The division of the ``count-sliced" spectroscopic time series was chosen to maximize the signal-to-noise of the first pseudo-exposure while ensuring that it ended strictly during the pre-flare quiescent period, and the remainder of the pseudo-exposures were divided based on having the same number of counts as this quiescent pseudo-exposure. Figure \ref{fig:hst_full_spec} compares the quiescent and flare peak \emph{HST} spectra and highlights a number of spectral emission features.

\section{Flare Analysis}\label{sec:analysis}

\subsection{Spectroscopic Time Series}\label{sec:timeseries}

We identified a number of spectral emission features in the APO and \emph{HST} spectra and created spectroscopic timeseries objects using \texttt{chromatic}\footnote{\url{https://github.com/zkbt/chromatic}}, listed in Table~\ref{tab:slr_defs}. Each spectroscopic time series associated with an emission feature is defined by a total wavelength range (the ``Spectrum Wavelength Boundaries'' in Table~\ref{tab:slr_defs}), a subset of times demarcated as quiescent, and at least one emission line defined by a central wavelength and a range of pixels over which to numerically integrate its flux (the ``Integration Wavelength Boundaries'' in Table~\ref{tab:slr_defs}). \added{These boundaries were determined manually by looking for the width necessary to encompass the line profile at its broadest during the flare}. Within each spectroscopic timeseries, any wavelength pixels that are not associated with a line are defined as the continuum. To account for variations in the absolute flux calibration over the course of the night we normalize each spectrum by a factor that aligns the continuum regions surrounding H$\beta$ and H$\alpha$ for the blue and red arms respectively to the median value of the pre-flare quiescent exposures. These lines were chosen because their continua seemed unchanged between the quiescent period and the flare peak without any normalization factors (see Figure~\ref{fig:apo_full_spec}).  \added{As a quantitative measure of the consistency of the continuum shape: the mean variation of the slope across the \halp\ continuum is 14\% while the mean variation of the slope across the \hbet\ continuum is 10\%.}

\begin{figure}
    \epsscale{0.7}
    \plotone{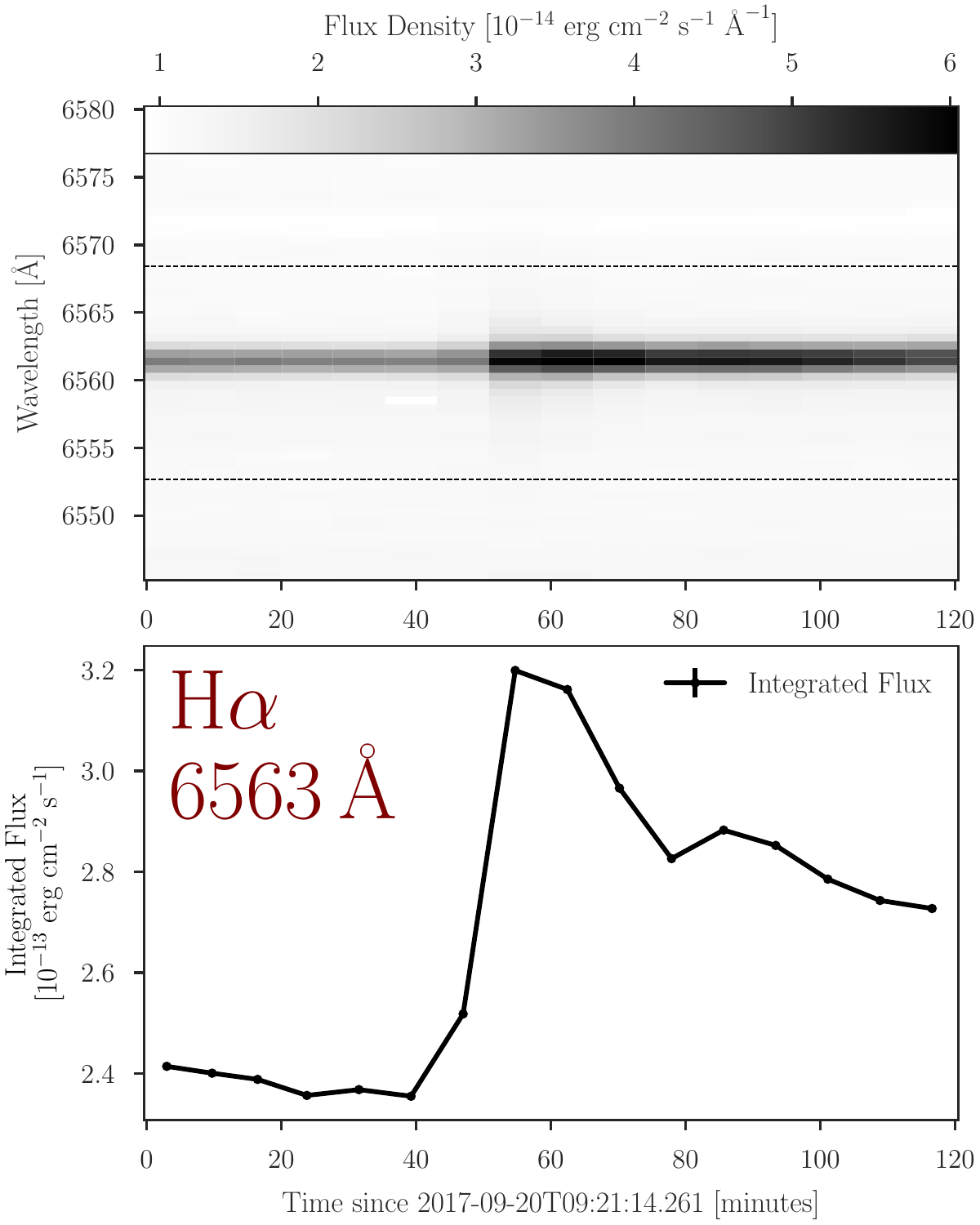}
    \caption{The top panel is a 2D representation of the spectroscopic timeseries for \halp, with darker values indicating higher flux density values according to the colorbar scale at the top of the image. The $y$-axis is wavelength, with two dashed lines indicating the integration boundaries for calculating the numerically integrated flux at each timestep, while the $x$-axis is the time since the start of the first APO exposure. The bottom panel plots the lightcurve of the numerically integrated flux along the same $x$-axis. The complete figureset (21 images for all lines listed in Table~\ref{tab:slr_defs}) is available in the online journal.}
    \label{fig:collapse_int}
\end{figure}

\begin{figure}
    \epsscale{0.7}
    \plotone{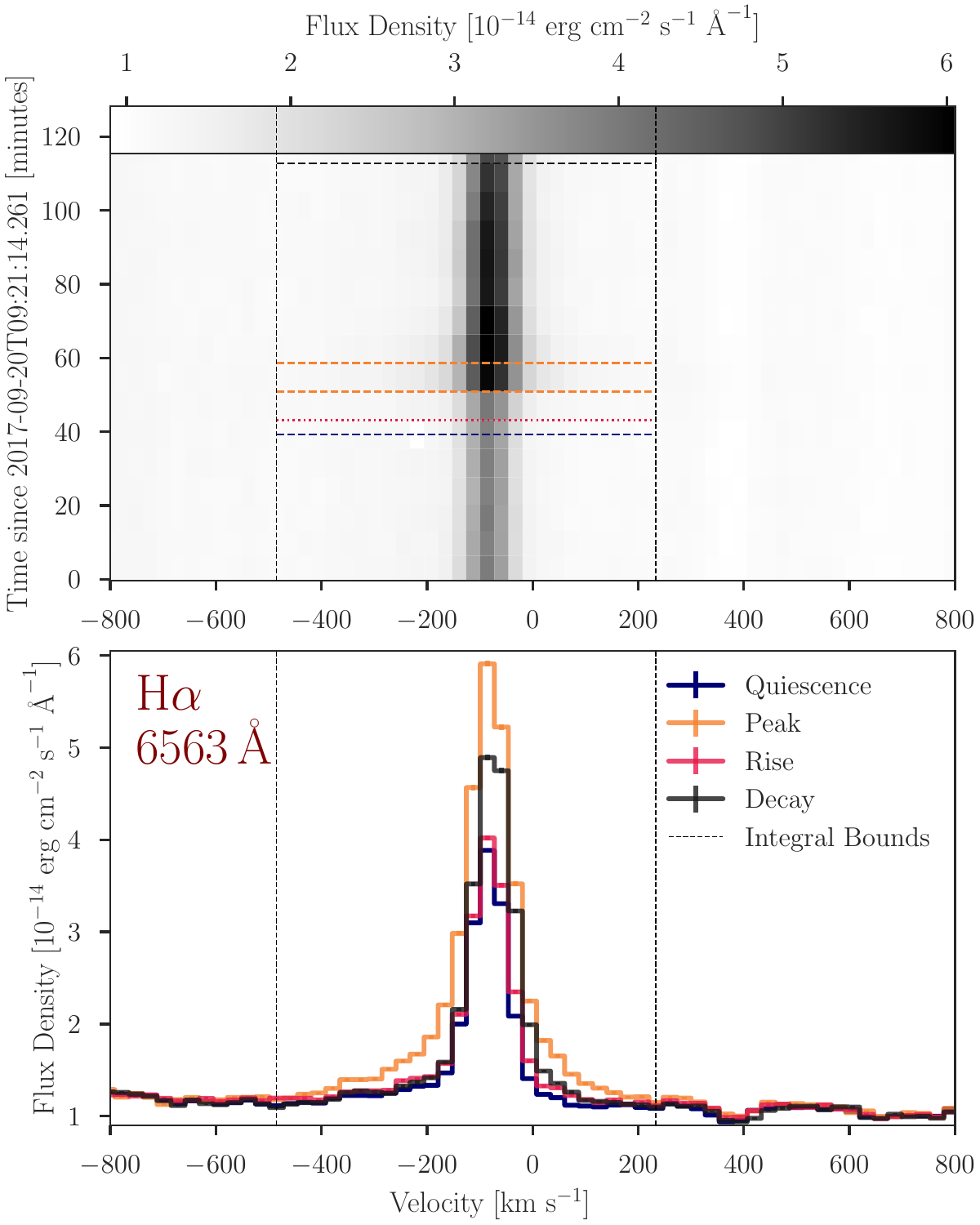}
    \caption{The top panel is a 2D representation of the spectroscopic timeseries for \halp, with darker values indicating higher flux density values according to the colorbar scale at the top of the image. The $x$-axis is velocity while the $y$-axis is the time since the start of the first APO exposure. The vertical black dashed lines demarcate the numerical integration boundaries as in Figure~\ref{fig:collapse_int}, and the horizontal dashed line demarcate specific range of time. The bottom panel plots the spectra at particular snapshots in time along the same $x$-axis: blue for the quiescent spectrum (averaged over all times below the blue dashed line in the top panel), orange for the peak of the flare (the spectrum during the timestamp between the two dashed orange lines in the top panel), red for the spectrum during the rise phase (the timestamp between the red and orange dashed lines in the top panel), and \added{dark grey} for the final timestamp during the decay phase (the timestamp above the \added{dark grey} dashed line in the top panel). The complete figureset (21 images for all lines listed in Table~\ref{tab:slr_defs}) is available in the online journal.}
    \label{fig:collapse_spec}
\end{figure}

Figures~\ref{fig:collapse_int} and \ref{fig:collapse_spec} present two different ways to visualize the spectroscopic time series for H$\alpha$. Both types of 2D spectroscopic time series figure have complete figuresets for each emission feature listed in Table \ref{tab:slr_defs} (21 figures in each figureset). In both figures the heatmap represents the spectroscopic time series as a whole, with wavelength or velocity on one directional axis, time on the other, and the pixels colored darker with increasing flux density. In Figure \ref{fig:collapse_int} time increases along the $x$-axis and wavelength along the $y$-axis of the heatmap in the top panel. The black dashed lines in the heatmap mark the wavelength/velocity boundaries for numerically integrating the flux, and the lightcurve of the integrated flux is plotted in solid black in the bottom panel. The numerically integrated flux in these figures does not subtract the quiescent flux or the continuum underneath.

Figure \ref{fig:collapse_spec} rotates the heatmap so that velocity runs along the $x$-axis and time along the $y$, and the bottom panel is the comparison of spectra instead of the integrated flux lightcurve. The black dashed lines (now running vertically) are shown in both panels and the heatmap now has additional horizontal dashed lines representing three different regions of time: all times below the navy blue line are during quiescence, the row of pixels between the orange dashed lines are the time of peak integrated flux, the row of pixels between the bottom orange and the red lines is during the rise phase of the flare, and the row of pixels above the \added{dark grey} line is the final exposure taken during the decay phase of the flare (for \emph{HST} spectra this is the final pseudo-exposure of the first orbit). The spectra from the peak (orange), rise (red), and decay (\added{dark grey}) timestamps are individual exposures plotted in the bottom spectrum comparison figure while the average of all quiescent exposures is plotted in blue. 

\begin{longrotatetable}
\begin{deluxetable}{cccc}
\tablecaption{Emission features analyzed in this work \label{tab:slr_defs}. \added{The ``Integration Wavelength Boundaries'' were manually defined to contain the line profile.}}
\tablehead{\colhead{Line Name} & \colhead{Velocity Central Wavelength} & \colhead{Spectrum Wavelength Boundaries} & \colhead{Integration Wavelength Boundaries}\\
\colhead{--} & \colhead{[\r{A}]} & \colhead{[\r{A}]} & \colhead{[\r{A}]}}
\startdata
\ion{H}{1} H$\alpha$ & 6563.29 & 6488.5 -- 6634.1 & 6552.7 -- 6568.4\\
\ion{Na}{1} D\textsubscript{1} & 5895.92 & 5814.1 -- 5958.4 & 5892.7 -- 5896.8\\
\ion{Na}{1} D\textsubscript{2} & 5889.95 & 5814.1 -- 5958.4 & 5887.0 -- 5891.0\\
\ion{He}{1} D\textsubscript{3} & 5875.62 & 5814.1 -- 5958.4 & 5870.7 -- 5877.1\\
\ion{He}{1} 4471 & 4472.94 & 4393.5 -- 4548.3 & 4468.8 -- 4474.4\\
\ion{H}{1} H$\beta$ & 4862.69 & 4784.6 -- 4937.3 & 4853.5 -- 4867.6\\
\ion{H}{1} H$\gamma$ & 4340.47 & 4262.8 -- 4417.8 & 4333.8 -- 4346.9\\
\ion{H}{1} H$\delta$ & 4101.73 & 4064.3 -- 4138.3 & 4094.1 -- 4108.4\\
\ion{Ca}{2} H +  H$\epsilon$ & 3968.9 & 3934.94 -- 4002.3 & 3962.7 -- 3974.5\\
\ion{Ca}{2} K & 3933.1 & 3899.2 -- 3966.4 & 3930.6 -- 3936.2\\
\ion{C}{1} 1658 & 1657.91 & 1615.2 -- 1667.4 & 1654.2 -- 1662.6\\
\ion{He}{2} 1640 & 1640.38 & 1615.2 -- 1667.4 & 1639.2 -- 1644.6\\
\ion{C}{4} 1548/50 & 1548.2 & 1522.8 -- 1575.0 & 1543.8 -- 1555.2\\
\ion{Si}{4} 1403 + \ion{O}{4} 1401 & 1402.71 & 1371.6 -- 1423.8 & 1399.2 -- 1407.0\\
\ion{Si}{4} 1394 & 1393.76 & 1371.6 -- 1423.8 & 1389.0 -- 1399.2\\
\ion{C}{2} 1334/5 & 1334.52 & 1310.4 -- 1362.6 & 1331.4 -- 1340.4\\
\ion{Si}{1}  -- \ion{Si}{3} + \ion{O}{1} & 1304.37 & 1276.8 -- 1329.0 & 1301.4 -- 1308.0\\
\ion{N}{5} 1238/42 & 1238.8 & 1221.6 -- 1273.8 & 1236.0 -- 1246.2\\
\ion{H}{1} Lyman-$\alpha$ & 1215.67 & 1181.4 -- 1233.6 & 1210.2 -- 1221.6\\
\ion{Si}{3} 1206 & 1206.51 & 1181.4 -- 1233.6 & 1203.6 -- 1210.2\\
\ion{C}{3} 1175 & 1175.987 & 1150.8 -- 1203.0 & 1171.2 -- 1181.4\\
\enddata
\tablenotetext{*}{Emission features in consecutive rows that share the same spectrum boundaries are features measured separately within the same spectroscopic timeseries.}
\end{deluxetable}
\end{longrotatetable}

\subsubsection{Notable Features in the Spectra}\label{sec:spectra}
\begin{figure*}
    \centering
    \includegraphics[width=\linewidth]{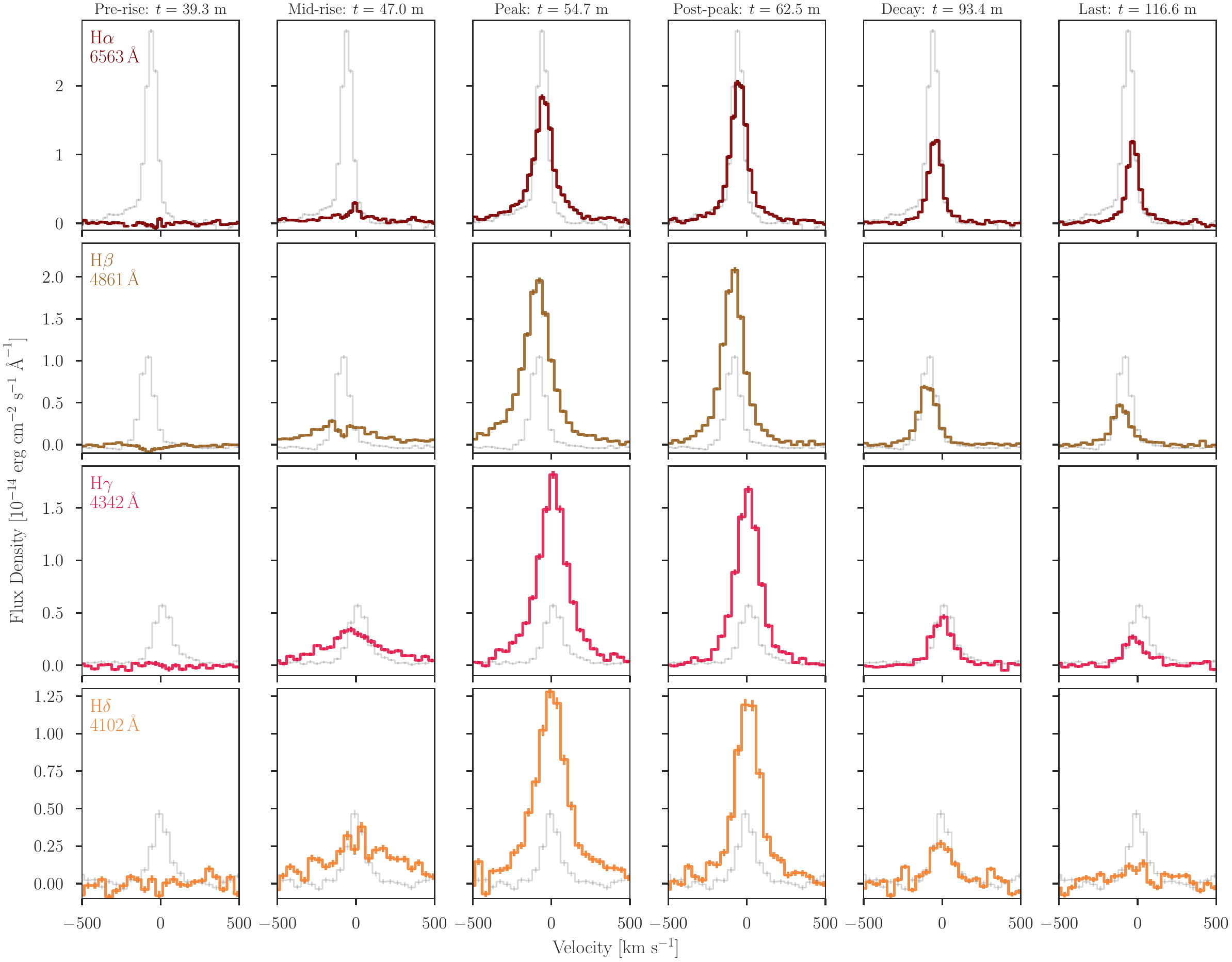}
    \caption{Snapshots of spectrum at different points relative to the peak of the H$\alpha$ lightcurve. Grey background points are the quiescent spectrum and the solid colored foreground points are flare-excess, taking the observed emission and subtracting the quiescent spectrum to highlight only emission from the flare itself. All velocities are calculated relative to the vacuum wavelength of the transition labeled in each row. From top to bottom, the rows represent different Balmer lines and each column corresponds to different points in time in each column. Each time column is titled with a position relative to the H$\alpha$ lightcurve and the time after the start of the observation period in minutes.}
    \label{fig:balmer_multitime}
\end{figure*}

\begin{figure*}
    \centering
    \includegraphics[width=\linewidth]{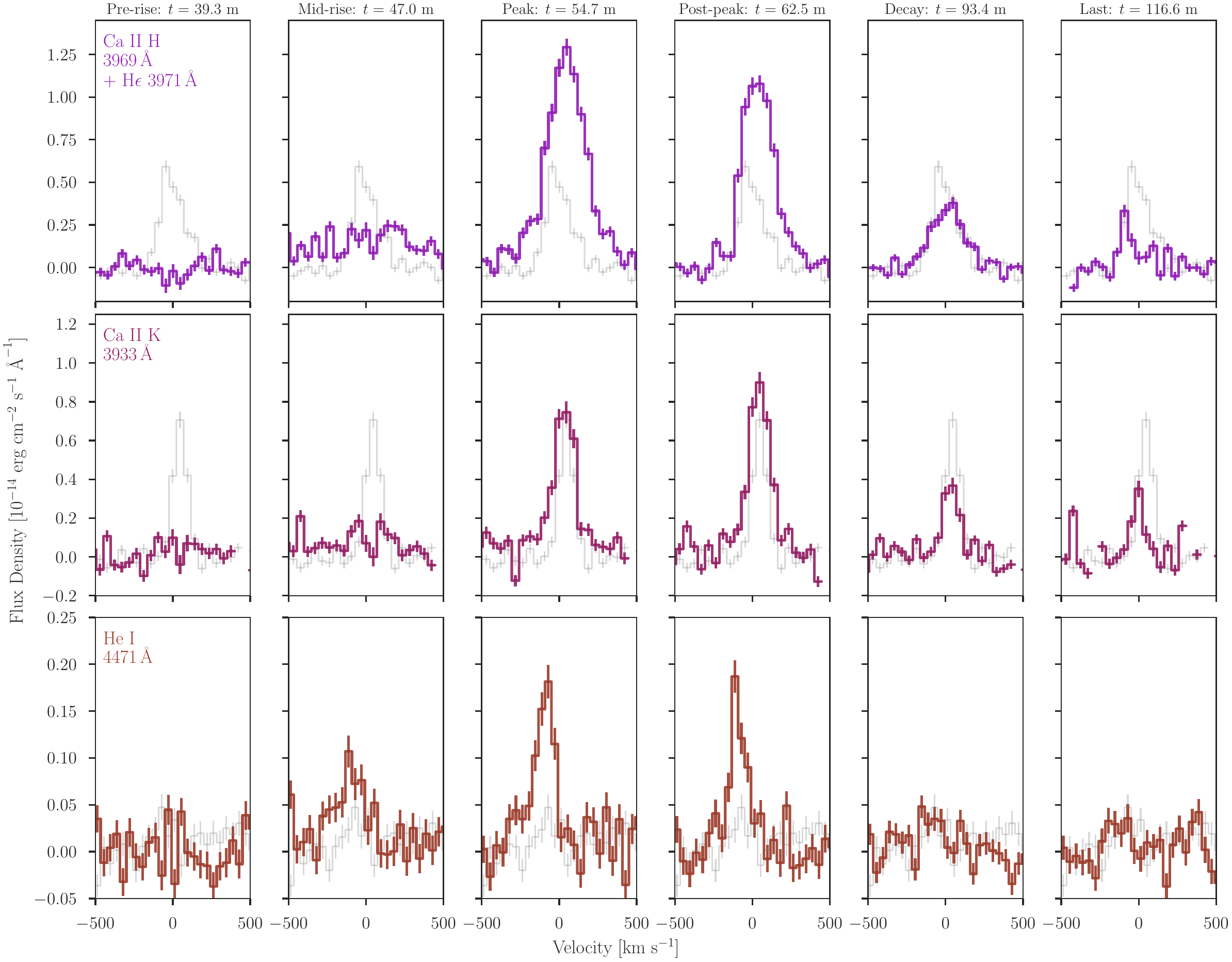}
    \caption{Similar to Figure \ref{fig:balmer_multitime} but for \ion{Ca}{2} H blended with H$\epsilon$, \ion{Ca}{2} K, and \ion{He}{1} 4472 $\textrm{\AA}$ instead. The grey background points are the quiescent spectrum while the foreground colored points represent the quiescent-subtracted flare-excess emission for the aforementioned lines in each row and at different points in time in each column. Each time column is titled with a position relative to the H$\alpha$ lightcurve and the time after the start of the observation period in minutes.}
    \label{fig:coll_multitime}
\end{figure*}

\begin{figure*}
    \plotone{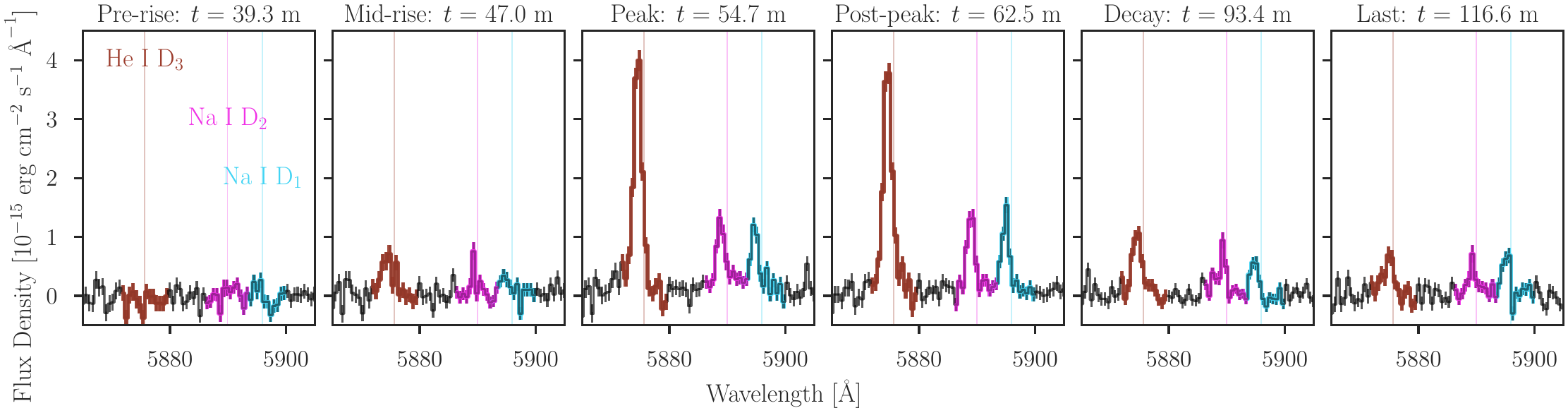}
    \caption{Similar to Figures \ref{fig:balmer_multitime} and \ref{fig:coll_multitime}, but for the Fraunhofer D lines \ion{Na}{1} D\textsubscript{1} 5895, \ion{Na}{1} D\textsubscript{2} 5890, and \ion{He}{1} 5876 $\textrm{\AA}$ instead. Because the quiescent emission is in fact a large absorption feature from the \ion{Na}{1} doublet, and the three lines are closely spaced, this figure plots the quiescent-subtracted flare-excess flux density against wavelength. Each panel plots vertical lines corresponding to the vacuum wavelengths of each D line, colored cyan, magenta, and maroon for lines D\textsubscript{1}, D\textsubscript{2}, and D\textsubscript{3} respectively. The points in the spectrum within 200 km s\textsuperscript{-1} of each D line are colored according to the same pattern.}
    \label{fig:d123_multitime}
\end{figure*}

\begin{figure*}
    \centering
    \includegraphics[width=\linewidth]{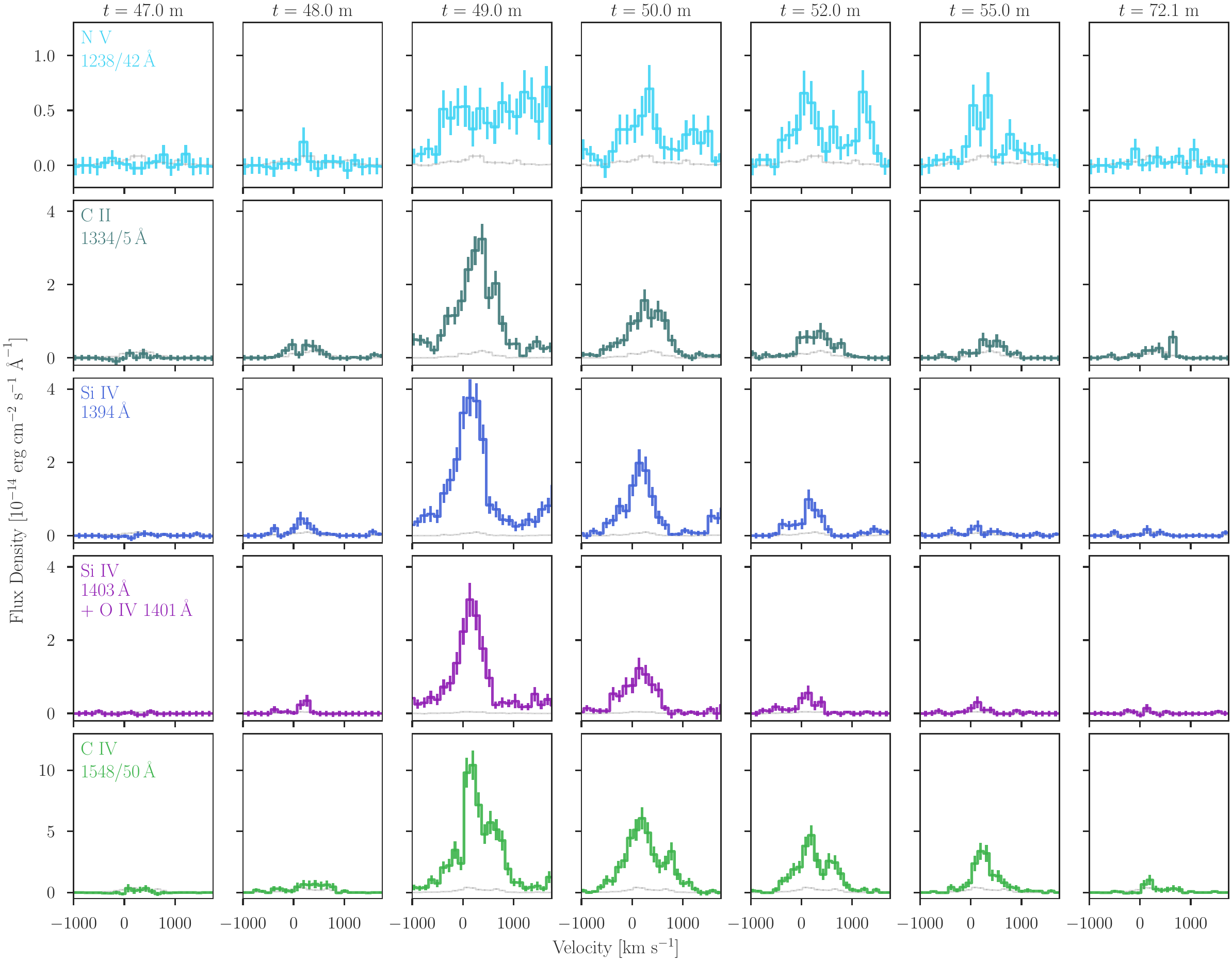}
    \caption{Similar to Figures~\ref{fig:balmer_multitime} and \ref{fig:coll_multitime} for a collection of doublets: \ion{N}{5} 1238/42 \r{A} (light blue, topmost row), \ion{C}{2} 1134/5 \r{A} (teal, second row), \ion{Si}{4}  1394 \r{A} (dark blue, third row) and \ion{Si}{4} 1403 \r{A} + \ion{O}{4} 1401 \r{A} (purple, fourth row) plotted separately, and \ion{C}{4} 1548/50 \r{A} (green, bottom). The times plotted in each column are single-minute pseudo-exposures for minutes 47, 48, 49, 50, 52, 55, and 72 from left to right. The faint gray spectra in the background are the quiescent emission while the solid step plots in the foreground are the flare excess spectra after subtracting off the quiescent emission. The flare evolves much more rapidly in the FUV than the optical, with a rise phase that lasts less than 2 minutes from quiescence to peak and the decay is slightly slower but still faster than the optical. The spectra at minute 72 still show a slight flare excess comparable to the original quiescent emission.}
    \label{fig:spec_hst_doublets_multitime}
\end{figure*}

\begin{figure*}
    \centering
    \includegraphics[width=\linewidth]{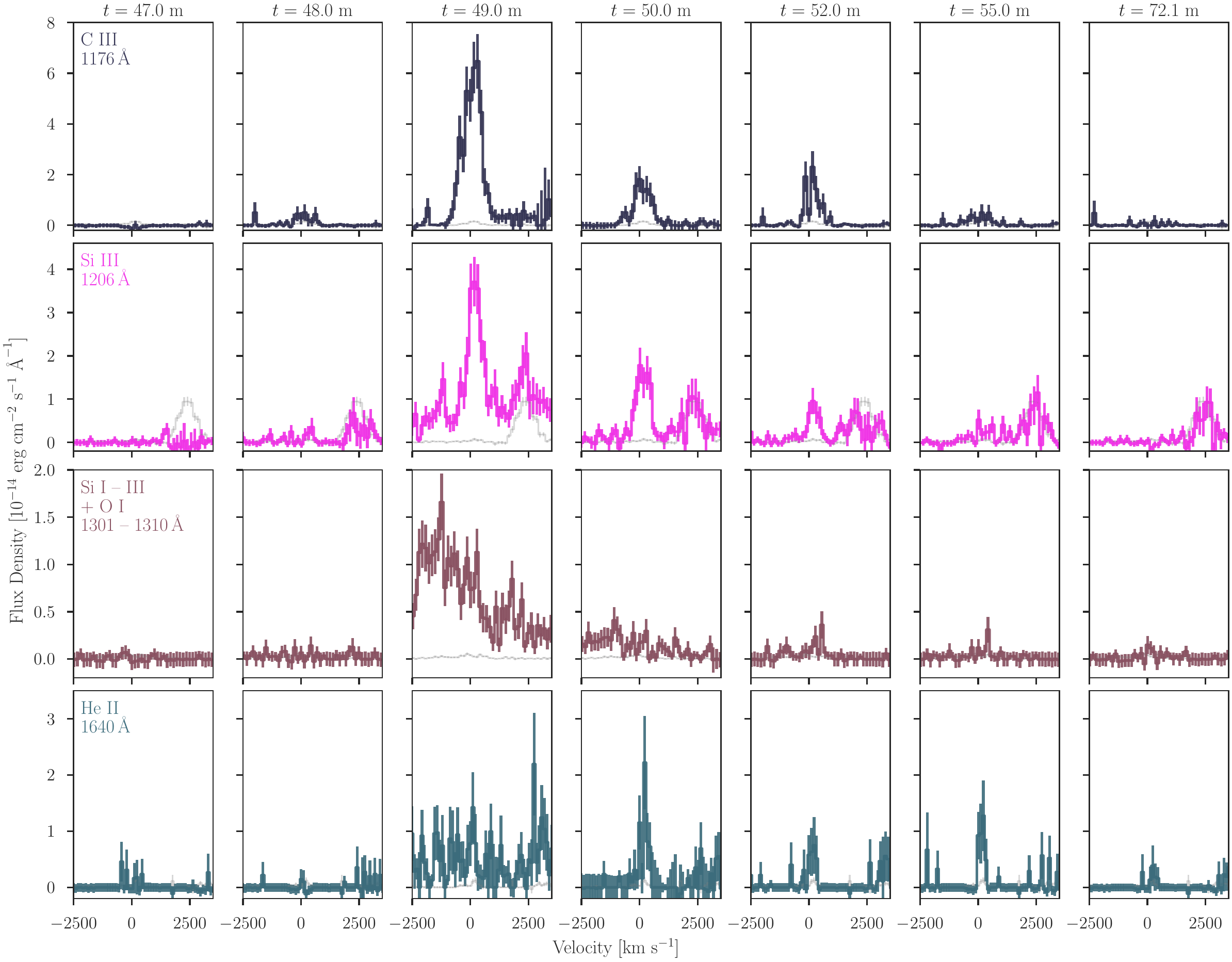}
    \caption{Similar to Figure~\ref{fig:spec_hst_doublets_multitime} for \ion{C}{3} 1175 \r{A} (bluish grey, top row), \ion{Si}{3} 1206 \r{A} and Lyman-$\alpha$ (pink, second row), a blend of \ion{Si}{1} -- \ion{Si}{3} + \ion{O}{1} from 1301 -- 1310 \r{A} (brown, third row), and \ion{He}{2} 1640 \r{A} (teal, bottom row). The evolution of these FUV lines is similarly rapid, but the bottom two panels also highlight a number of flare-brightened lines only visible during flare peak.}
    \label{fig:spec_hst_other_multitime}
\end{figure*}

\begin{figure*}
\plotone{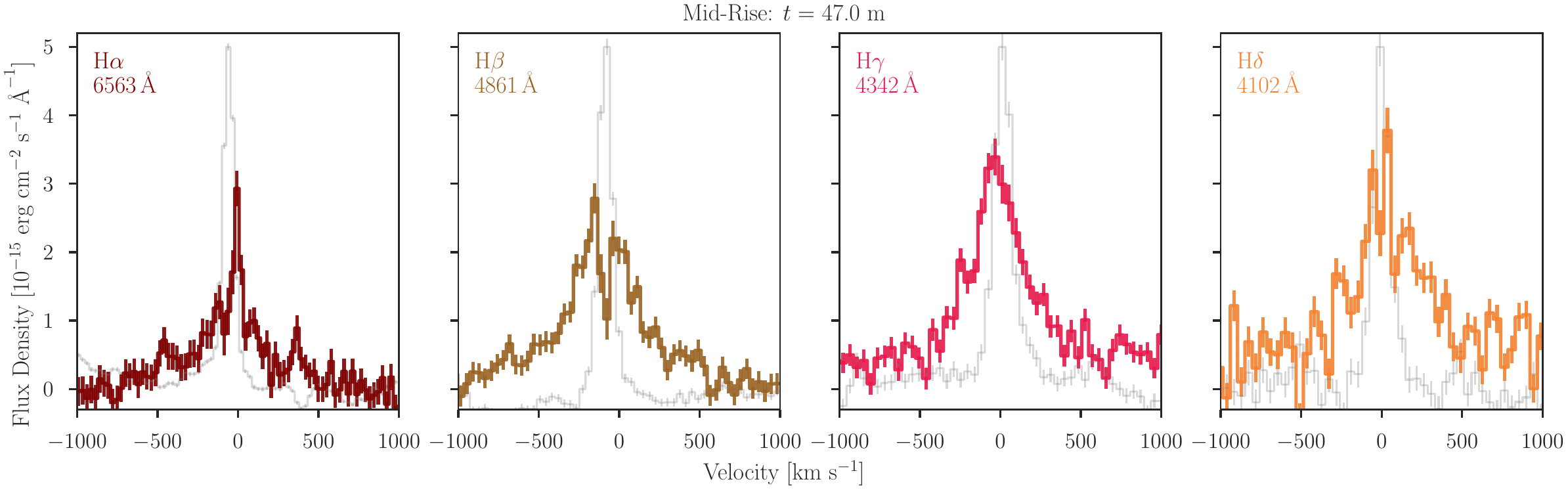}
    \caption{A closer look at the mid-rise timestamp for H$\alpha$ (left), H$\beta$ (middle), and H$\gamma$ (right) similar to the second column of Figure \ref{fig:balmer_multitime}. The grey line in the background of each panel is the quiescent emission line normalized to more easily compare the velocity profiles between the lines. The solid step plots with errorbars are the quiescent-subtracted emission from each panel's corresponding emission line during the rise phase of the flare. The excess H$\alpha$ emission shows a well-defined narrow component redward of the quiescent emission with a more amorphous and smaller amplitude blue wing. The excess H$\beta$ emission shows both blue and red wing components of comparable magnitude, although the blue wing is slightly more prominent. Finally, the H$\gamma$ excess has both blue and red wing components but the blue wing is clearly larger. Another point to note is that the fluxes of the excess H$\beta$ and H$\gamma$ are comparable to the excess H$\alpha$ and to each other, unlike the standard pattern of fluxes decreasing with higher Balmer order present during quiescence.}
    \label{fig:balmer_rise}
\end{figure*}

\begin{figure*}
\plotone{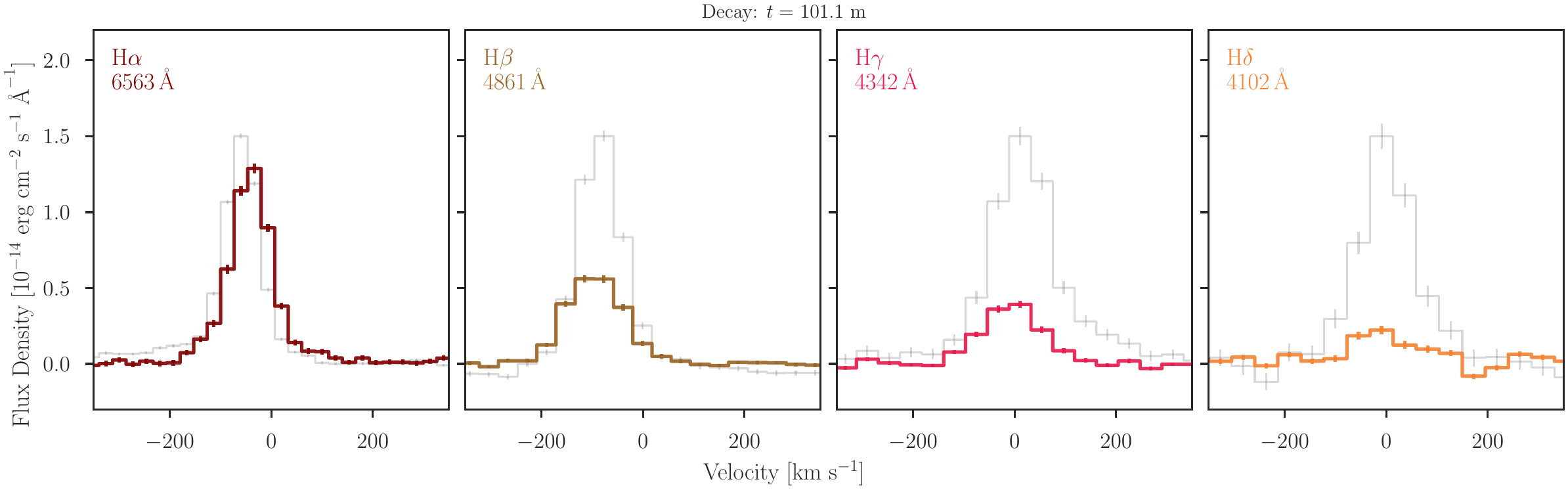}
    \caption{Similar to Figure~\ref{fig:balmer_rise} and the rightmost column of \ref{fig:balmer_multitime}, this figure is a closer look at a timestamp during the decay phase for H$\alpha$ (left), H$\beta$ (middle-left), H$\gamma$ (middle-right), and H$\delta$ (right). The grey line in the background of each panel is the (normalized for legibility) quiescent emission line and the solid step plots with errorbars are the quiescent-subtracted emission from each panel's corresponding emission line during the decay phase of the flare. The H$\alpha$ flare excess emission is slightly redshifted relative to the quiescent line while \hbet\ and \hgam\ are slightly blueshifted relative to quiescence (pixel widths of $27$, $38$, $43$, and $46$ km$\,$s$^{-1}$ for \halp -- \hdel\ respectively). The line profile is dominated by the line spread function in the line core, but extended wings are still present for at least \halp and \hbet}
    \label{fig:balmer_decay}
\end{figure*}

Figures \ref{fig:balmer_multitime}, \ref{fig:coll_multitime}, \ref{fig:d123_multitime}, \ref{fig:spec_hst_doublets_multitime}, and \ref{fig:spec_hst_other_multitime} present a comparison between the spectra of sets of emission features at different times. Each row corresponds to an emission feature, labeled in the first column of each row, and each column corresponds to an individual exposure, labeled on top of each column. Each panel plots the quiescent emission feature in faint grey and the quiescent-subtracted emission for the exposure in a solid color (corresponding to the color of the text label). The sets of lines within each figure are: the Balmer series H$\alpha$ through H$\delta$ in Figure~\ref{fig:balmer_multitime}; \ion{Ca}{2} H 3969 $\textrm{\AA}$ + H$\epsilon$ 3971 $\textrm{\AA}$, \ion{Ca}{2} K 3933 $\textrm{\AA}$, and \ion{He}{1} 4471 $\textrm{\AA}$ in Figure~\ref{fig:coll_multitime}; the Fraunhofer D lines (\ion{He}{1} D\textsubscript{3} 5876 $\textrm{\AA}$, \ion{Na}{1} D\textsubscript{2} 5890 $\textrm{\AA}$, and \ion{Na}{1} D\textsubscript{1} 5896 \textrm{\AA}) in Figure~\ref{fig:d123_multitime}; FUV transition region resonance line doublets (\ion{N}{5} 1238/42 $\textrm{\AA}$, \ion{C}{2} 1334/5 $\textrm{\AA}$, \ion{Si}{4} 1398/1403 $\textrm{\AA}$, and \ion{C}{4} 1548/50 $\textrm{\AA}$) in Figure~\ref{fig:spec_hst_doublets_multitime}; and finally FUV regions with multiplets and/or complicated blends (\ion{C}{3} 1175 $\textrm{\AA}$, \ion{Si}{3} 1206 $\textrm{\AA}$ + Lyman-$\alpha$ 1216 $\textrm{\AA}$, a span from 1301 -- 1310 $\textrm{\AA}$ containing lines from \ion{Si}{1} -- \ion{Si}{3} and \ion{O}{1}, and \ion{He}{2} 1640 $\textrm{\AA}$) in Figure~\ref{fig:spec_hst_other_multitime}. Note that the chosen timestamps for the optical lines are different from those of the FUV; the optical lines' timestamps are chosen to highlight different stages of the flare (pre-flare, mid-rise, peak, post-peak, decay, last exposure) spanning minutes 39 -- 117 while the FUV lines' timestamps are chosen to highlight the rapid evolution of the flare from minutes 47 -- 72. The FUV flare-excess flux is zero at minute 47 during the rise phase of the flare in the optical lines, shows only a slight enhancement a minute later, and peaks dramatically at minute 49. The FUV lines then decay well before the optical lines peak and the FUV lines end the first orbit with only a slight flare excess comparable to the quiescent emission.

During that minute 49 pseudo-exposure, which also aligns with the peak pseudo-exposure plotted in Figure~\ref{fig:hst_full_spec}, there is a significant FUV enhancement across the bandpass that includes contributions from a combination of lines and recombination edges \citep{Phillips:1992}, some of which can be seen in the panels for \ion{N}{5} in the top row of  Figure~\ref{fig:spec_hst_doublets_multitime}, \ion{Si}{1} -- \ion{Si}{3} in the 1301 -- 1310 \r{A} interval in the third row of Figure~\ref{fig:spec_hst_other_multitime}, and \ion{He}{2} 1640 \r{A} in the bottom row of Figure~\ref{fig:spec_hst_other_multitime}. While the zoomed out view in Figure~\ref{fig:hst_full_spec} might suggest that this is a flat continuum potentially in the tail of a blackbody, there is no consistent slope in the bottom envelope beneath observed emission lines to attempt fitting a blackbody temperature. The increased flux at the blue end of the optical spectrum is too subtle to identify a slope and fit a blackbody to the spectrum with the lines masked, but may include contributions from many \ion{Fe}{2} lines.

Something qualitatively similar to the swamping of \ion{N}{5} by other lines is the discrepancy in how \ion{Si}{3} 1206 \r{A} and other lines near Lyman-$\alpha$ are dramatically enhanced relative to Lyman-$\alpha$. While there is a clear Lyman-$\alpha$ excess that persists throughout the flare, it seems to be a nearly constant enhancement much less than that experienced by these other lines (echoing similar findings by \citealt{Loyd:2018a}). It is important to note that the observed Lyman-$\alpha$ emission is only from the wings of the intrinsic stellar profile and the core is extinguished by the interstellar medium. The shape of the observed Lyman-$\alpha$ wing emission is smeared out by the low resolution of the G140L grating's line-spread function and any measured flux is a lower limit and unknown fraction of the total Lyman-$\alpha$ emission. Thus we cannot say that the Lyman-$\alpha$ line as a whole is weaker than \ion{Si}{3} during the flare peak, but merely the wings.

If the Lyman-$\alpha$ profile is anything like the Balmer H$\alpha$ however, we can expect that the during the rise phase, the profile of the flare excess emission emphasizes the wings more than the quiescent line profile. Figure~\ref{fig:balmer_rise} is a zoomed in view of the second column of Figure~\ref{fig:balmer_multitime}, showing the Balmer series H$\alpha$ -- H$\delta$ during the middle of the rise phase of the flare. Each panel plots the flare excess emission, so the apparent ``self-absorption'' near the line core should be understood as the core not being enhanced as strongly as the wings (this ``self-absorption'' is positive so there is still some flare excess contribution to the core in this spectral snapshot). This enhancement to the wings is asymmetrical for each of the Balmer lines and different between each one. The red wing of the H$\alpha$ excess emission is much narrower and stronger than the blue wing, while H$\beta$ shows roughly equal enhancements on both sides with a slight edge favoring the blue wing, the blue wing of the H$\gamma$ excess is slightly narrower and stronger than its red wing, and H$\delta$ is the closest to being symmetrical overall (although there are nearby flare-enhanced lines confounding careful analysis). Similar excesses are seen in the other non-hydrogen optical lines during the rise phase (see Figures~\ref{fig:coll_multitime} and \ref{fig:d123_multitime}) although they are more symmetrical. The \ion{Ca}{2} H and K lines show a persistent blue excess relative to quiescence during the decay phase while the sodium doublet lines maintain an apparently constant blueshift relative to the vacuum wavelength which cannot be definitively attributed to the flare since they lack a quiescent emission line to compare to.

These excesses indicate some combination of gradients in the velocity distribution of hydrogen atoms in the line forming regions and bulk plasma motion. The red wing Balmer excess could be from the chromospheric condensation layer of plasma excited by the electron beam moving downwards deeper into the stellar chromosphere, while the blue wings are the hotter and optically thinner chromospheric evaporation flow. This narrative is supported by the persistence of these asymmetries in the decay phase of the Balmer lines, plotted in Figure~\ref{fig:balmer_decay}. Even at this low resolution, the flare excess H$\alpha$ is clearly redshifted relative to the quiescent emission and H$\beta$ -- H$\delta$ are blueshifted relative to quiescence. 
 
The differences in which wings are more or less prominent are a function of the different optical depths of each line and the influence of Stark broadening on each transition. These are qualitative observations based on the general physical model of flares \citep{Benz:2010}, but a more quantitative approach to determine properties of the flare electron beam is possible. \citet{Kowalski:2017} used RADYN \citep{Allred:2015} to model the Balmer line broadening and Balmer decrements of a megaflare on YZ CMi while \citet{GarciaSoto:2025} used a grid of RADYN models from \citet{Kowalski:2024} to model the \halp\ and \hbet\ profiles of a smaller flare on TIC 415508270. This optical + FUV dataset is an opportunity for similar experiments tested against a more expansive set of lines probing more layers of the atmosphere and incorporating more physical processes; the goal would be to construct a model capable of simultaneously explaining the evolving behavior of \Lya\, the Balmer series, the FUV inter-line spectrum (potentially a continuum like that in \citealt{Froning:2019}), transition region resonance doublets, and the \ion{He}{2} 1640 \r{A} triplet. Developing this model is beyond the scope of this paper's empirical characterization of the flare and will be left to future work.

We fit Voigt or Gaussian line profiles to each spectroscopic snapshot but found no trends in line profile properties as a function of wavelength, line formation temperature, or oscillator strength. The wings of the optical emission lines had large Lorentzian width parameters that decayed rapidly, tracking the evolution of the line flux. The centroids shifted according to the asymmetries already discussed, but within the widths of the instrumental line spread functions. The FUV data were too low resolution to fit separate narrow and broad Gaussian components as done in \citet{Froning:2019,France:2020}. The bulk of our analysis is therefore limited to the integrated fluxes of each line, but observing a similar flare at higher resolution would resolve many degeneracies and distinguish between causal explanations of the features we observe.

\subsection{Integrated Flux Lightcurves}\label{sec:lightcurves}
\begin{figure*}
    \centering
    \includegraphics[width=0.75\linewidth]{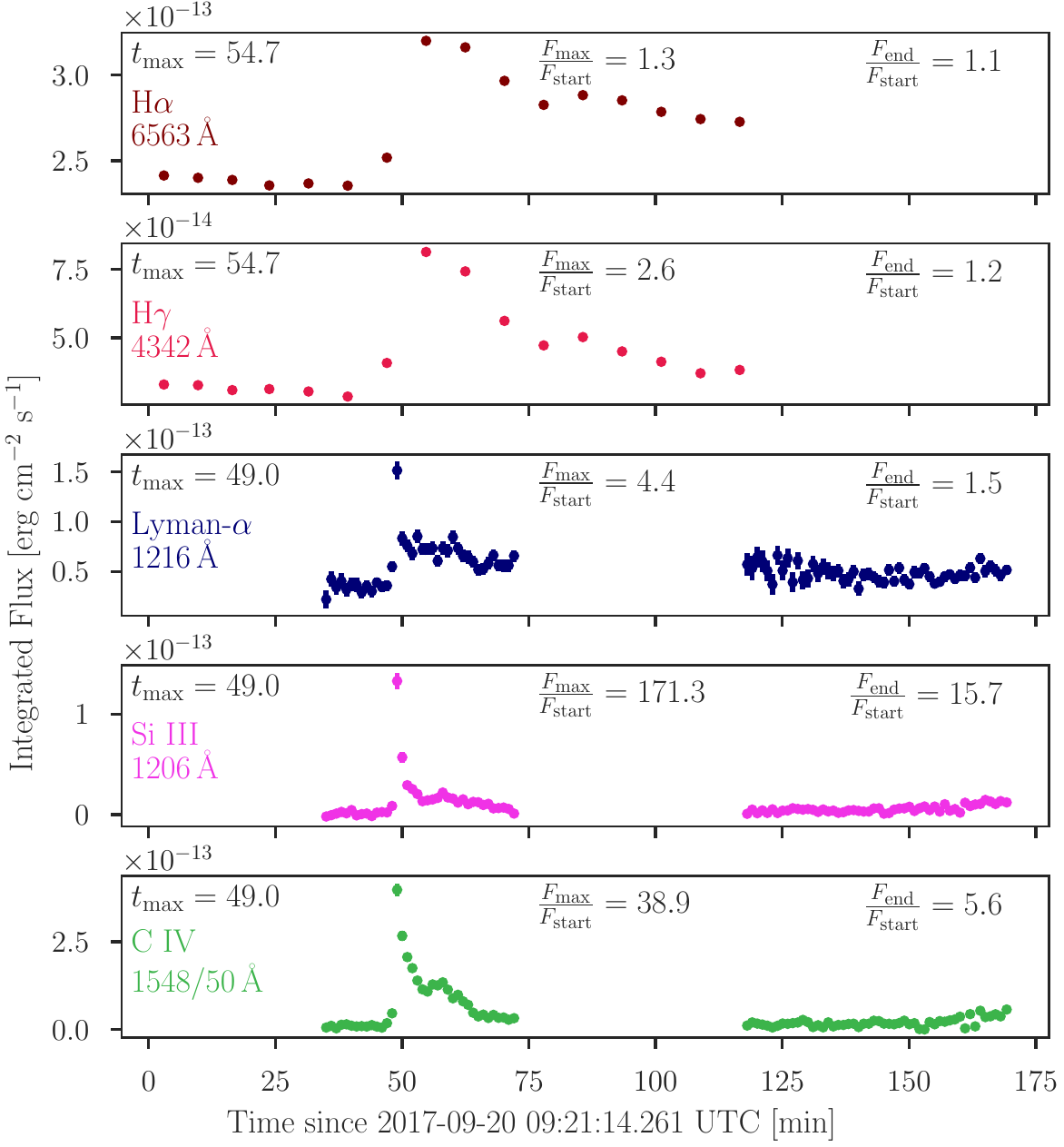}
    \caption{The top two panels show the lightcurves of integrated flux for H$\alpha$ (brown, top row) and H$\gamma$ (red, second row) from the APO spectra while the bottom three panels plot Lyman-$\alpha$ (dark blue, third row), \ion{Si}{3} 1206 \r{A} (pink, fourth row), and \ion{C}{4} 1548/50 \r{A} (green, bottom row) from the \emph{HST} spectra. The top of each panel includes the peak time on the left, at 54.7 minutes for the optical and 49 minutes for the FUV, the ratio of peak flux measurement to the initial (varying from 1.3 -- 171.3 depending on the line) in the middle, and the ratio of the final flux measurement to the initial on the right (varying from 1.1 -- 15.7). Each panel has an independent $y$-scale because of the dynamic range of peak fluxes depending on the line. Note that\added{:} the FUV lines peak and decay almost entirely within the rise phase of the optical lines\added{,} both optical and FUV lines show both a primary and secondary bump\added{,} the secondary bump of the optical takes place while \emph{Hubble} is unable to observe\added{,} and the secondary bumps of the FUV happen during the optical peaks.}
    \label{fig:hst_apo_long_lightcurve}
\end{figure*}

By numerically integrating within the \added{manually defined} ``Integration Wavelength Boundaries'' of each spectroscopic timeseries specified in Table~\ref{tab:slr_defs}, we can compare the relative timing and structure of lightcurves for each emission feature. Figure~\ref{fig:hst_apo_long_lightcurve} compares five such lightcurves to each other, from top to bottom: H$\alpha$, H$\gamma$, (observed) Lyman-$\alpha$, \ion{Si}{3} 1206 \r{A}, and the \ion{C}{4} 1548/50 \r{A} doublet. Each panel has text describing the time that the lightcurve peaks ($t_\mathrm{max}$), the ratio of the peak value to the initial ($\frac{F_\mathrm{peak}}{F_\mathrm{start}}$), and the ratio of the final value to the initial ($\frac{F_\mathrm{end}}{F_\mathrm{start}}$). These measures require no assumption of a functional form and already show interesting differences between the lines' lightcurves.

As noted before, the FUV lightcurves evolve rapidly, rising, peaking, and falling all within the rise phase and peak of the optical. Also noted before is the fact that H$\alpha$ has a much more muted response than H$\gamma$, with a $\frac{F_\mathrm{peak}}{F_\mathrm{start}} = 1.3$ as opposed to H$\gamma$'s 2.6, which is still less than Lyman-$\alpha$'s 4.4 and much less than the other FUV lines' values of 171.3 and 38.9 for \ion{Si}{3} and \ion{C}{4} respectively. This arrangement holds for the final decay phase values of $\frac{F_\mathrm{end}}{F_\mathrm{start}}$ as well. Structurally, all the lightcurves have a primary peak and a secondary bump, but the secondary bump of the FUV lines takes place during the peak of the optical and the secondary bump of the optical takes place while \emph{Hubble} is behind the Earth and unable to observe GJ 4334. If there was a corresponding enhancement to the FUV lines during the secondary optical bump, it may contribute to the elevated post-flare level seen in the FUV during the second \emph{Hubble} orbit. It is important to note that if we had only observed GJ 4334 during that second orbit, we would measure a spectrum that differed significantly from the first orbit's measured pre-flare quiescence, not only in the overall flux level but also in the ratios between lines. This elevated ``pseudo-quiescent'' activity level has also been seen in the FUV after a flare on GJ 486 \citep{Diamond-Lowe:2024} and implies that other M dwarfs observed on single occasions in the FUV may have mischaracterized FUV spectra if they flare frequently.

\subsubsection{Balmer Series}\label{sec:apo_hst:lightcurves:balmer}
\begin{figure*}
    \plottwo{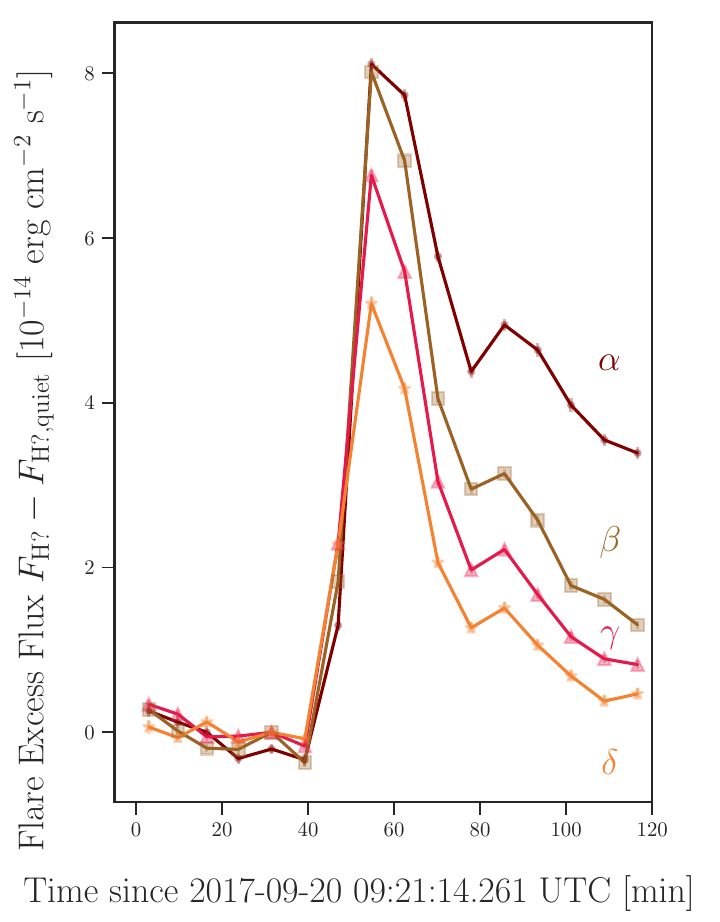}{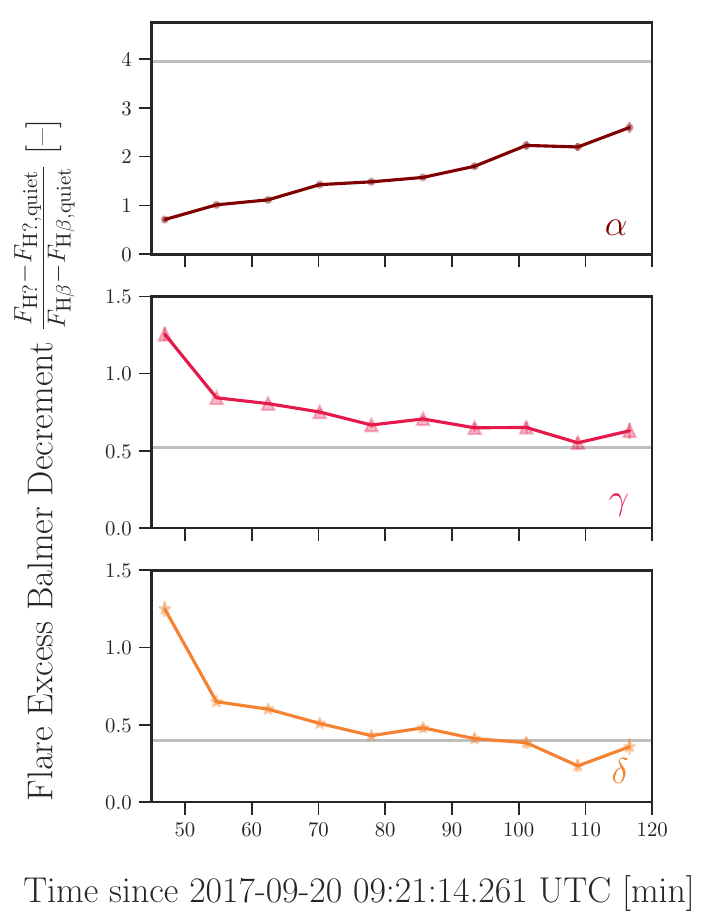}
    \caption{Lightcurves of Balmer line flare excess fluxes on the left, and the flare excess Balmer decrement (ratio relative to H$\beta$) on the right. After subtracting off the quiescent emission, the flare excess fluxes of all the Balmer series peak at similar values, a significant departure from the ratios of the quiescent emission lines, but the decay of each Balmer line is steeper from H$\alpha$ to H$\delta$. The Balmer decrements of the quiescent lines are plotted as horizontal grey lines in the right panels and the flare peak decrement is much lower (almost 1 instead of the quiescence's 4) for H$\alpha$ and higher for H$\gamma$ and H$\delta$. The H$\alpha$ flare excess decrement appears to increase linearly towards the quiescent value while the higher order decrements show a more exponential decay towards their quiescent values. There are errorbars on these plots but they are miniscule\added{; the values of the largest errorbars are $8.5 \times 10^{-16}$ erg cm$^{-2}$ s$^{-1}$ for the flux panel on the left and 0.12 for the decrement panel on the right.}}
    \label{fig:flux_dec_panels}
\end{figure*}
The Balmer line \halp\ is the transition between the $n=3 \rightarrow n=2$ states with the higher order lines \hbet, \hgam\ starting from $n=4, 5, ...$ and so on. The ratio of the Balmer lines to \hbet, the \hbet\ Balmer decrement, is a diagnostic of how these levels of hydrogen are populated relative to each other across the visible portions of the stellar atmosphere. Figure~\ref{fig:flux_dec_panels} plots the lightcurves of the flare excess flux of \halp\ -- \hdel\ on the left and the flare excess Balmer decrement on the right. At flare peak, all of these lines have fairly similar integrated fluxes and the excess \hbet\ flux roughly equals that of \halp, so the flare excess \halp\ decrement starts out very close to 1. The remainder of the \halp\ decrement lightcurve is a steady, nearly linear, rise to the quiescent decrement level (plotted as a gray horizontal line), suggesting a similarly steady repopulation of the $n=3$ state. The \hgam\ and \hdel\ decrements on the other hand start out above 1 and show a more exponential decay towards the quiescent level, indicating that the $n=5$ and $n=6$ states have a sudden excess relative to either $n=3$ or $n=4$. This could be due to the sudden recombination of electrons from the flare electron beam into preferentially higher states followed by a relaxation towards the quiescent default as the excited hydrogen atoms thermalize by collisions.

\subsubsection{Parameterization and Fitting}\label{sec:apo_hst:lightcurves:fitting}
\begin{figure}
    \plotone{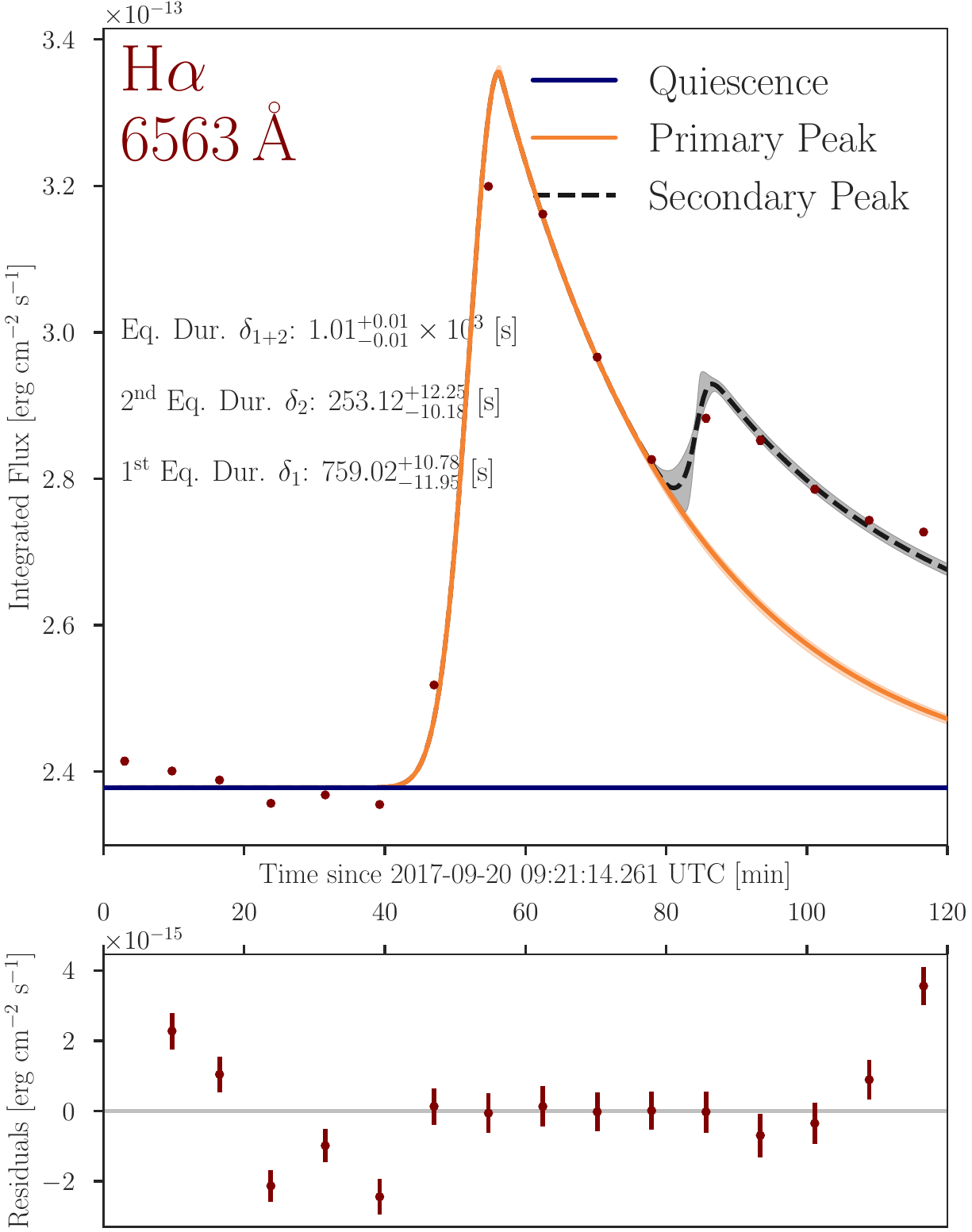}
    \caption{The flare model fit for H$\alpha$ with residuals in the bottom panel. The equivalent  durations for each component and their total is written on the top panel, with lines for the quiescent level (blue), the contribution from the primary peak (orange), and the contribution of the secondary bump (\added{dark grey}). The complete figureset (21 images for each emission feature listed in Table~\ref{tab:slr_defs}) is available in the online journal.}
    \label{fig:flare_residuals}
\end{figure}

To more carefully compare the relative timing, rise/decay timescales, and total energy of the lightcurves, we fit them with a sum of two components, with each component resembling the expression used in Equation 3 of \citet{Feinstein:2022}. A single component is a piecewise function combining a Gaussian rise and exponential decay, such that 

\begin{equation}
     \mathcal{F}_i(t) =
    \begin{cases} 
      A_i \exp{\left(-\frac{(t - t_i)^2}{2r_i^2}\right)} & t \leq t_i \\
      A_i \exp{\left(-\frac{t -t_i}{d_i}\right)} & t> t_i
      \end{cases}
\end{equation}

and the lightcurve for a given line is
\begin{equation}
    \mathcal{F_\mathrm{line}}(t) = q_\mathrm{line} + \mathcal{F_\mathrm{line,\, primary}}(t) + \mathcal{F}_\mathrm{line,\, secondary}(t)
\end{equation}
 where $q_\mathrm{line}$ is the quiescent integrated flux of the line, $A_i$ is the amplitude of a component $i$, $r_i$ is the Gaussian rise timescale, $d_i$ is the exponential decay timescale, and $t_i$ is the peak time of the component. Defining the flare relative to the peak time this way means that there is no strict ``start'' time of the flare, but we define flare onset as $t_\mathrm{primary} - 3r_\mathrm{primary}$ as the moment when there is a small but noticeable contribution to the flare in the analytic expression. The ``equivalent duration'' of the flare for a given line is
 \begin{equation}
     \delta_\mathrm{line} = \int_t \left(\frac{\mathcal{F_\mathrm{line}}(t)}{q_\mathrm{line}} - 1\right) \ dt
 \end{equation}

\noindent and describes the flare in terms of the time required for the quiescent emission to contribute as much flux as the flare did. The total energy of the flare in a given line is $q_\mathrm{line} \times \delta_\mathrm{line} \times 4\pi d^2$ where $d$ is the distance to the star. While fitting the 
 FUV lines we held the quiescent flux level fixed to the median of the pre-flare datapoints to prevent the equivalent duration calculation from growing extremely large when the quiescent fluxes were close to the noise floor.
 Figure~\ref{fig:flare_residuals} plots the model fit and residuals for H$\alpha$ as an example from the full figureset available in the online version of the paper to examine each individual line lightcurve's fit. The flare model fit parameters and derived quantities are listed in Appendix~\ref{sec:flare_params}.

\subsubsection{Notable Features in Integrated Flux Lightcurves}



\begin{figure*}[ht]
  \begin{minipage}[b]{0.5\linewidth}
    \centering
    \includegraphics[width=.7\linewidth]{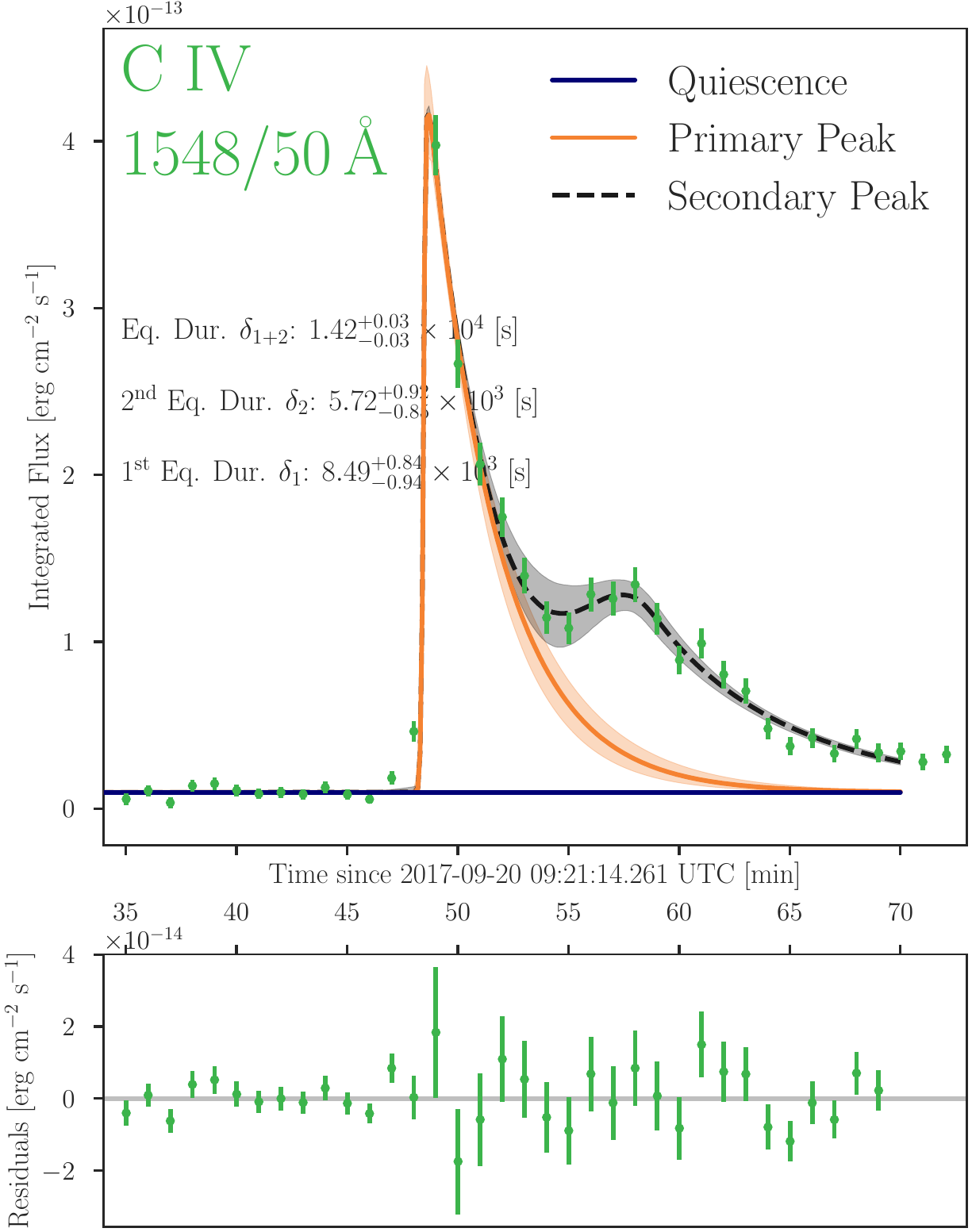} 
    \vspace{4ex}
  \end{minipage}
  \begin{minipage}[b]{0.5\linewidth}
    \centering
    \includegraphics[width=.7\linewidth]{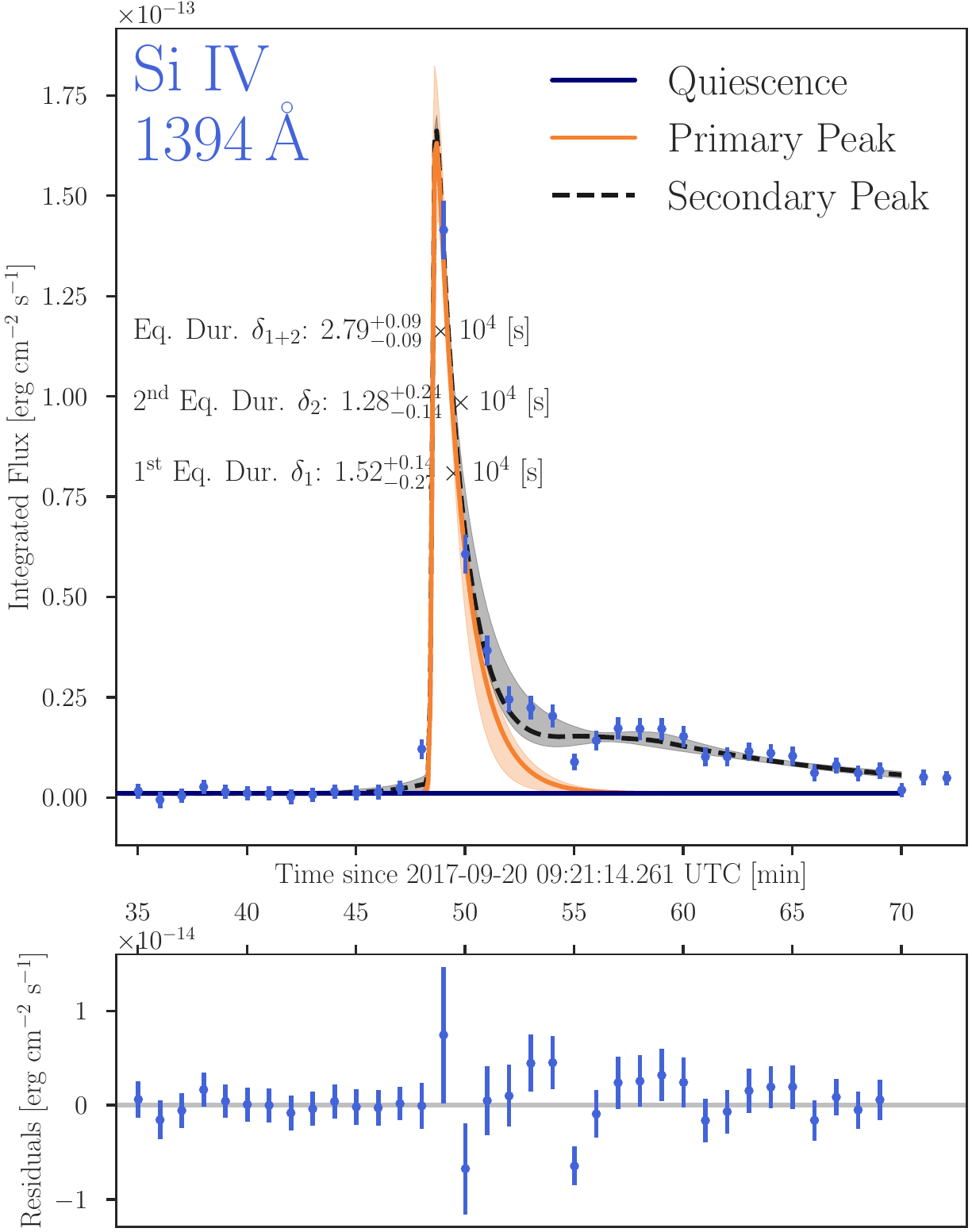} 
    \vspace{4ex}
  \end{minipage} 
  \begin{minipage}[b]{0.5\linewidth}
    \centering
    \includegraphics[width=.7\linewidth]{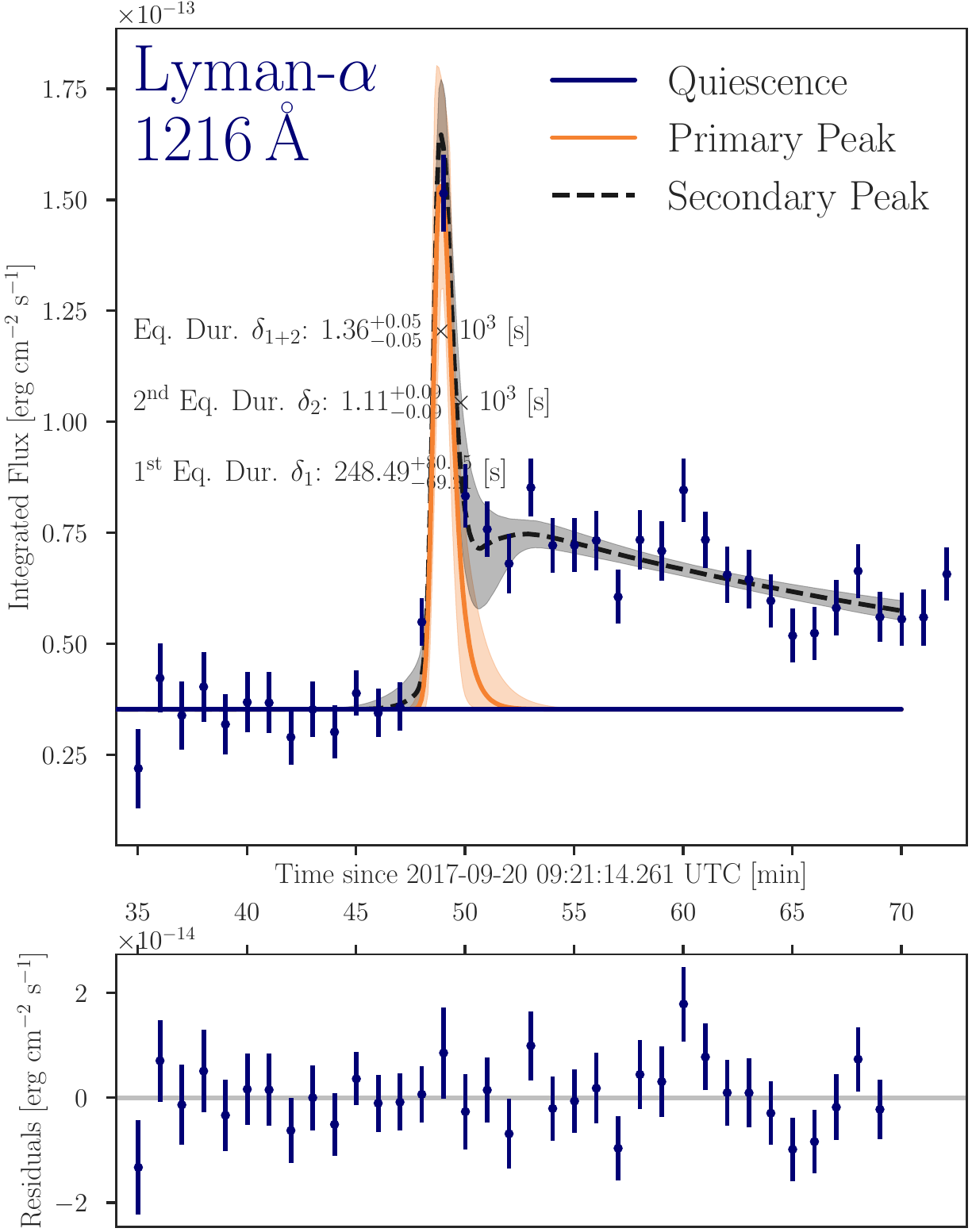} 
    \vspace{4ex}
  \end{minipage}
  \begin{minipage}[b]{0.5\linewidth}
    \centering
    \includegraphics[width=.7\linewidth]{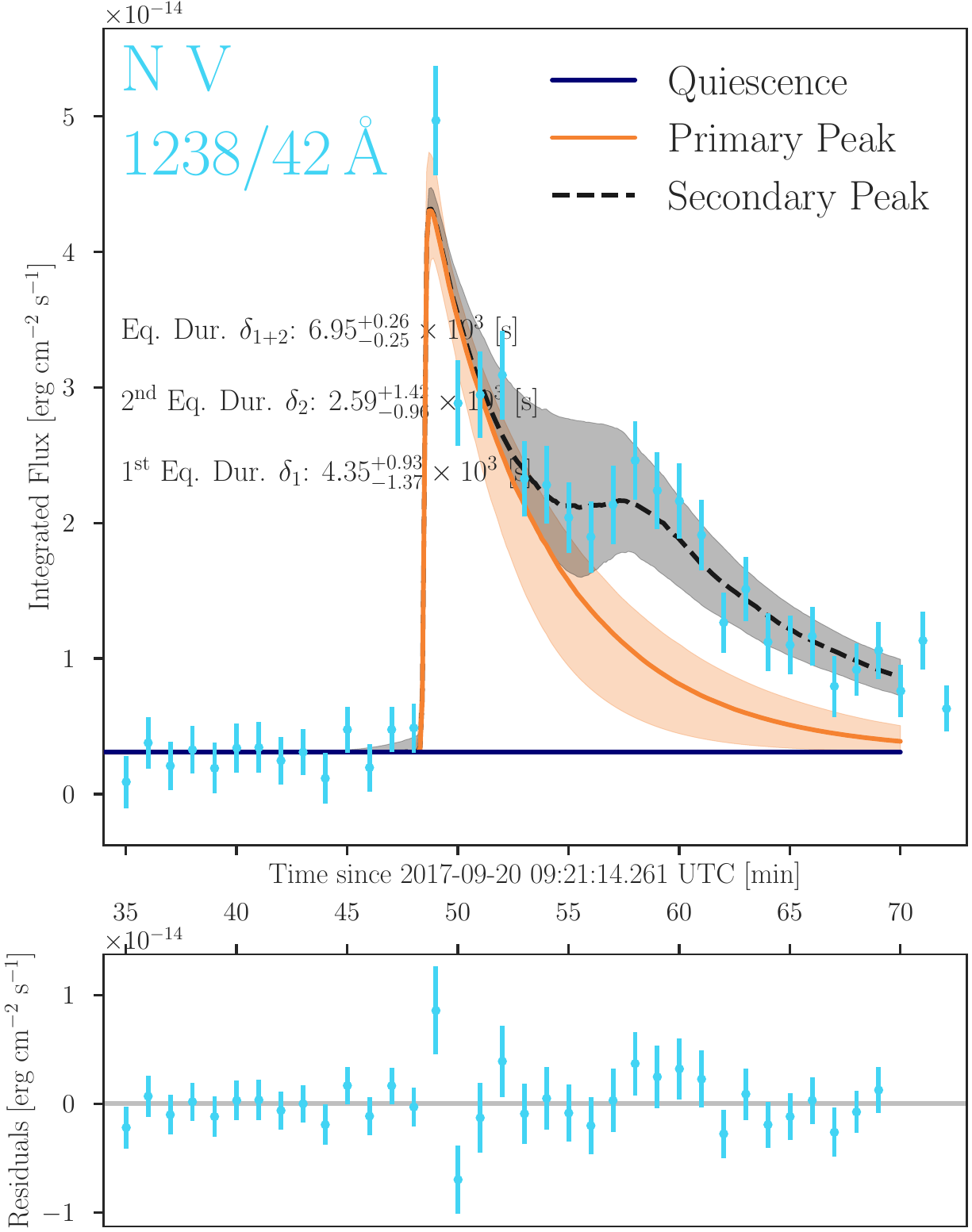} 
    \vspace{4ex}
  \end{minipage}
  \caption{A collection of flare profiles for FUV lines showing the contributions from both the primary peak (orange) and secondary bump (\added{dark grey}) with residuals in the bottom panel of each subfigure. The residuals show a hint of quasi-periodic oscillations, but we could not identify a statistically significant peak in a periodogram. The shapes and size of the secondary bump relative to the primary peak differ significantly among these four lines as well as the other FUV lines \added{available} in the figureset \label{fig:fuv_flares}}
\end{figure*}

\begin{figure}
    \plotone{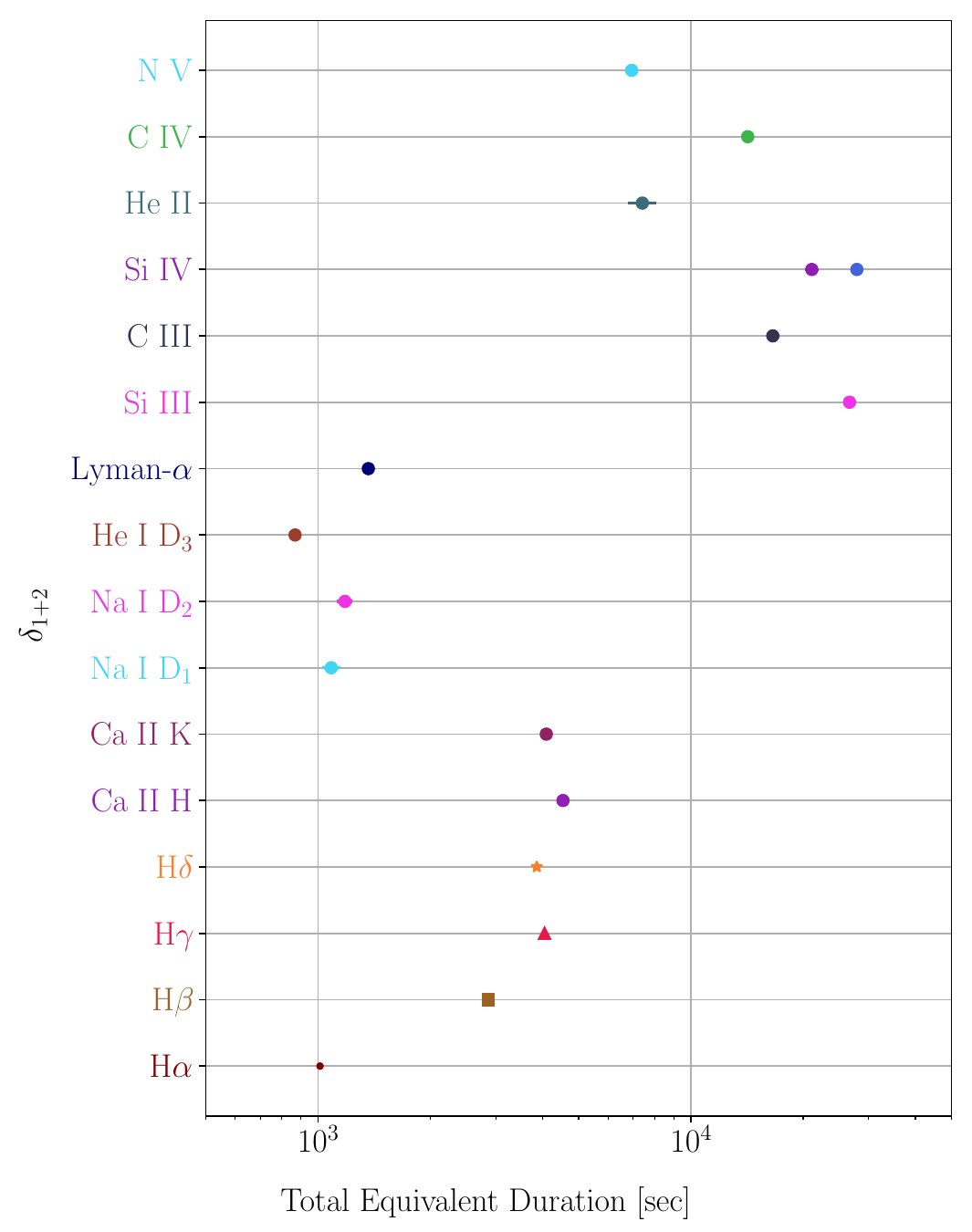}
    \caption{A comparison of total equivalent durations across most of the lines, ranging from nearly 1000 seconds -- 30,000 seconds.}
    \label{fig:equivalent_durations}
\end{figure}

Figure~\ref{fig:fuv_flares} plots the lightcurve models of multiple FUV lines to highlight some notable features. \ion{C}{4} and \ion{Si}{4} form in overlapping temperature ranges with the contribution function of \ion{C}{4} peaking at a slightly higher temperature (atomic data from \texttt{CHIANTI v10} \citealp{Dere:1997,DelZanna:2021}), but their lightcurves are very different. The ratio of the secondary/primary amplitudes of the \ion{Si}{4} lightcurve is much smaller than that of \ion{C}{4} but the decay of \ion{Si}{4}'s secondary component is so much slower that the total equivalent durations end up being similar (see Figure~\ref{fig:equivalent_durations} which compares many lines' equivalent durations). The observed Lyman-$\alpha$'s secondary component looks different from the rest of the FUV features, with a very slow decay leading to an extended triangular shape, almost like an intermediary between the rest of the FUV lines and the optical. All the FUV lines have their primary peak and rise at the same times within uncertainties, so the variation is largely in the primary component's decay and all aspects of the secondary component. This suggests a common heating is responsible for all of the FUV lines' primary components but some more complex \added{set of} phenomena is responsible for the secondary. Both double-peak and peak-bump structures have been seen for FUV flares before \citep{Loyd:2018a,Loyd:2018b} but it is not clear why the secondary should vary so much depending on which FUV line one looks at.

\begin{figure*}
    \plotone{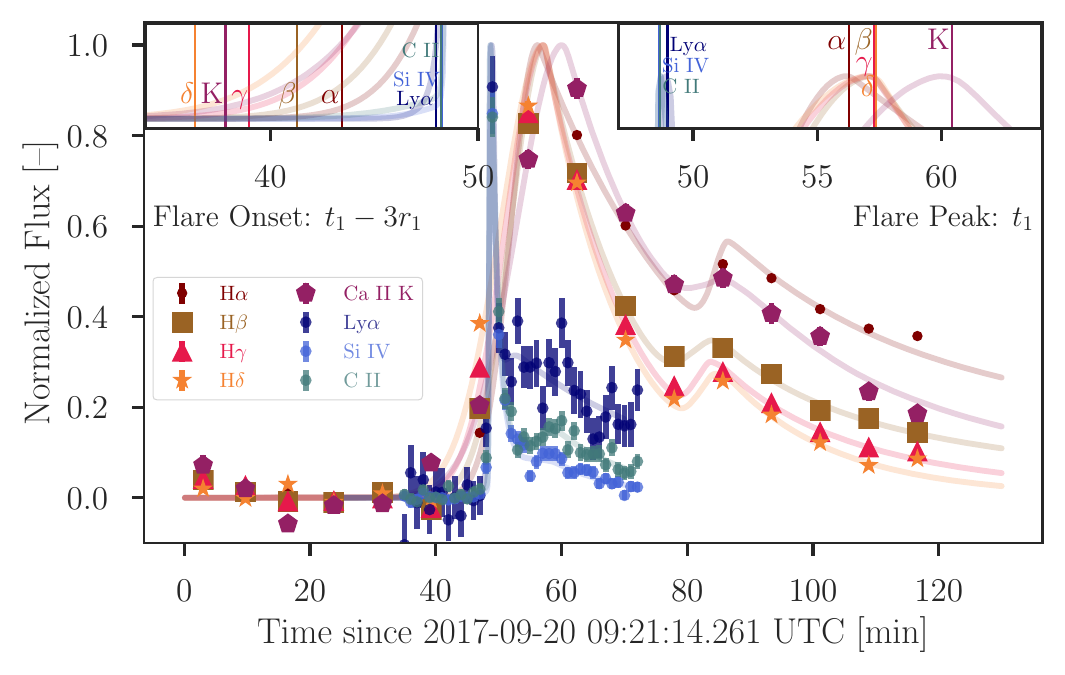}
    \caption{Lightcurves of the data and model fits for Balmer series, \ion{Ca}{2} K, Lyman-$\alpha$, \ion{Si}{4} 1394 \r{A}, and \ion{C}{2} 1334/5 \r{A}. The inset panels on the top zoom into the flare onset (left) and flare peak (right) times to show the relative timing of these lines' flare profiles. There is a trend in the flare onset of the Balmer series, with higher order lines starting first and a delay in the peak of \ion{Ca}{2} K relative to the other optical lines. The FUV lines all rise and peak at the same time within uncertainties, but the onset and peak of the FUV lines are both within the onset -- peak interval of the optical lines.}
    \label{fig:flare_timing}
\end{figure*}

Figure~\ref{fig:flare_timing} plots the model fits for the Balmer series, the \ion{Ca}{2} K line, Lyman-$\alpha$, \ion{Si}{4} 1398 \r{A}, and the \ion{C}{2} 1334/5 \r{A} doublet to show trends in the relative timing of these lines. \halp\ peaks earlier than \hbet\ -- \hdel\ which all peak at almost exactly the same time but the flare onset occurs in order from \hdel\ -- \halp\ at fairly even intervals. \ion{Ca}{2} K starts between \hdel\ and \hgam\ but rises more slowly and peaks well after all the Balmer lines, an instance slightly resembling the Neupert effect \citep{Neupert:1968}. The Neupert effect was first identified as the lightcurve of soft X-ray emission being slightly delayed relative to the microwave radio emission during a solar flare, then later refined to note that the time-integral of the hard X-ray emission resembled the lightcurve of the soft X-rays, implying that different energy bands were emitted from related heating/cooling processes. Similar Neupert-like effects have been identified in correlations between \textit{U}-band photometry and the EUV flux for AD Leo \citep{Hawley:2003}, the soft X-ray flux and NUV continuum for AU Mic \citep{Tristan:2023}, and for the inactive M dwarf L-98-59 in \citet{Behr:2023}. The existence of a time lag between the Balmer series and \ion{Ca}{2} K has been seen before \citep{Kowalski:2013}, but in this dataset there is no clear relationship between the \ion{Ca}{2} K and Balmer line lightcurve derivatives/integrals to definitively declare that we have seen a Neupert-effect linking these lines' emission. Section 5.3 of \citet{Kowalski:2013} discusses the potential Neupert-like behavior of \ion{Ca}{2} K and posits that the delay in \ion{Ca}{2} K may be due to it being associated with chromospheric evaporation, and the blue excess of \ion{Ca}{2} K during the rise and decay phases (top row, columns 3 and 6 of Figure~\ref{fig:coll_multitime}) support this hypothesis. At higher optical cadence it would be possible to more thoroughly explore Neupert-effect relations between the FUV lines and the optical to determine how the FUV lines might relate to the chromospheric heating.

\section{Flare Literature Comparison}\label{sec:apo_hst:literature}
We can compare the flare we observed to three groups of observations in the literature: optical spectroscopic monitoring programs \citep{Fuhrmeister:2018,Maehara:2021,Kowalski:2013}, serendipitous FUV flares \citep{Loyd:2018a,Loyd:2018b, Loyd:2014,Feinstein:2022,Froning:2019,Diamond-Lowe:2024}, and 
 a couple of rare multiwavelength observations targeting AD Leo \citep{Hawley:1991,Hawley:2003} and EV Lac \citep{Osten:2005}.

The Great Flare of AD Leo in 1985 \citep{Hawley:1991} has a primary peak with a secondary bump 11 minutes in the optical photometry UBVR filters, and narrow-band photometry of the Balmer series H$\alpha$ -- H$\gamma$ showing a much more flat-topped lightcurve with a subtle secondary bump, but both features occurring much later than the optical photometry. The UV behavior is very different, with a strong factor of 20 response in Lyman-$\alpha$ and comparable factors of 10 -- 50 in the \ion{N}{5}, \ion{Si}{4}, and \ion{C}{4} doublets. This suggests that the flare we observe is not simply weaker in degree, but different in character, possibly because the electron beam deposited its energy higher up in the stellar atmosphere, well above the regions of high density in the chromosphere required to brighten the optical continuum. However, this higher elevation or lower density also seems to have enabled this flare to heat chromospheric plasma to transition region temperatures, resulting in the dramatic brightening of the optically thin FUV transition region lines.
    
\citet{Hawley:2003} reported observations of multiple flares on AD Leo during March 2000. The FUV lines showed multiple morphologies across the flares observed during this campaign, including double-peaked (flare 3), an approximately symmetric rise and decay (flares 4 and 7), a peak + bump (flare 6 and 8). Flare 8 seems most similar to the GJ 4334 flare, with the FUV emission lines rising, peaking, and decaying all during the rise phase of the optical emission lines. \citet{Hawley:2003} had faster cadence observations of the Balmer series and show a Neupert-effect relation such that the FUV emission lightcurve is similar to the derivative of the optical emission. This would suggest that the FUV emission lines trace the flare heating rate while the optical emission lines trace the resulting change in temperature. This raises the question of whether the extended optical and FUV emission line fluxes indicate a continuous heating process instead of an elevation of material into flare loops. \citet{Hawley:2003} also observe that the FUV lines show different decay phase behaviors between sets of lines, possibly correlated with formation temperature, where the cooler \ion{C}{2} 1334, \ion{Si}{2} 1265, \ion{Si}{3} 1206, and \ion{O}{1} 1305 decay much more rapidly than the hotter \ion{C}{4} 1548/50 and \ion{N}{5} 1238/42. The secondary bump in the GJ 4334 is more prominent than any of the flares, but the decay timescales do follow the same grouping with slower decays for \ion{C}{4} and \ion{N}{5} than most of the cool lines. Neither flare 3 or 8 show a significant post-flare elevated pseudo-quiescence in the FUV lines, but the optical emission lines remain elevated towards the end of the observation like the GJ 4334 flare.

The optical spectra of the EV Lac multiwavelength flare \citet{Osten:2005} were affected by clouds and so a detailed comparison is not possible, while the bulk of the ultraviolet analysis focused on the velocity distributions of the lines which we are less sensitive to due to the resolution of the G140L grating. \citet{Osten:2005} do not describe the morphology of the flare in the ultraviolet or mention any post-flare elevated pseudoquiescent level.

\cite{Loyd:2014} displays a number of FUV flare morphologies where the only shown peak + bump lightcurve is one of the AD Leo flares in \citet{Hawley:2003}. The fiducial FUV flare constructed by \citet{Loyd:2018a} and GJ 674 flare observed by \citet{Froning:2019} are also double-peaked rather than a peak + bump. Flare B from \citet{Feinstein:2022} shows a peak + bump morphology in \ion{N}{5} and the orbit following the flare remains at an elevated level, although this is interrupted by two more flares and is therefore not really pseudo-quiescent. \citet{Diamond-Lowe:2024} observed two FUV flares on GJ 486 without a complete return to pre-flare quiescence, and later measured FUV line fluxes higher than expected based on line correlations and prior observations. They suggest that M dwarfs can go through periods of elevated activity with heightened FUV flux that may last days, which would call into question the accuracy of any M dwarf FUV spectrum based on a single observation.

\citet{Fuhrmeister:2018} show multiple asymmetric profiles of some of the optical emission lines discussed in this work and discuss multiple explanations in the literature for red and blue asymmetries. For red asymmetries, their list of potential explanations is: coronal rain, cool plasma from above the stellar surface falling down along magnetic field lines; a rising upflow of material absorbing blue wing flux; and chromospheric condensation, a downward flow of cool plasma within the chromosphere. It is possible that all of these mechanisms contribute, but at different points in time during the flare, for example the coronal rain explanation would only be present during the decay phase of the flare. They associate blue asymmetries with rising material, such as filaments, during the early stages of the flare but do not address potential causes of blue asymmetry during the decay phase of a flare.

\citet{Namekata:2022} studied the evolution of H$\alpha$ during a solar flare with both spatially resolved spectra and disk-integrated ``Sun-as-a-star'' observations. They find that the Sun-as-a-star spectral profile most closely matches the profile of the large halo around the flare heated region, and the core of the flare-heated region with the most H$\alpha$ intensity does not match the disk-integrated spectral profile. They also use observations near 1600 \r{A} to show that the time-evolution of this transition region emission peaks impulsively during the flare impulsive rise phase like the flare shown here, and they show a relationship between this transition region emission and the time-evolution of the H$\alpha$ redshift velocity that suggests both are tracing the heating of the upper chromosphere and transition region.

\section{TESS Analysis}\label{sec:tess}
\emph{TESS} is often the best option to determine the flare rate and probability of observing a flare for a given star, but the flares \emph{TESS} is sensitive to are those strong enough to brighten the optical continuum, which is not true of the flare we observed and is likely not true for the majority of stellar flares in general \added{(\citealt{Paudel:2021} showed that less than a third of NUV continuum flares had an optical \emph{TESS} counterpart)}. Despite their samples being unrepresentative of the most common flares for any particular star, photometric flare surveys are still the best way to compare flare statistics between multiple stars. We wanted to determine how GJ 4334's flare behavior compared to faster, more well-studied, rotators and to quantify how unexpected the flare we observed was if we only had the most commonly available flare statistics to inform us, so we identified flares and fit a flare frequency distribution (FFD) to four sectors of GJ 4334 and five sectors of EV Lac \emph{TESS} data. EV Lac was chosen because it is also on the fully convective side of the partial/fully convective boundary and flares frequently enough that one can plan observing campaigns targeting it in multiple wavelengths with a reasonable expectation of observing flares \citep{Osten:2005}.

\subsection{Optical Flare Detection Using \texttt{stella} and \texttt{altaipony}}
We used two open-source codes, \texttt{stella} \citep{Feinstein:2020} and \texttt{altaiPony} \citep{Ilin:2021}, to identify flares. \texttt{altaiPony} uses an iterative sigma-clipping algorithm that identifies consecutive points above the typical noise of a detrended lightcurve \citep{Davenport:2016} to identify flares while \texttt{stella} uses a convolutional neural network trained on 2-minute cadence \emph{TESS} data to assign a robust probability of a given datapoint being part of a flare. We used a cutoff probability of 0.9 based on visual inspection of the lightcurves. Both codes identify the datapoints belonging to a particular flare and calculate the flare's equivalent duration. We found that \texttt{stella} was more sensitive to small flares than \texttt{altaiPony} and both failed to accurately characterize the equivalent durations of large complex flares because both codes treated the large flares as multiple smaller flares. We flagged these complex flares, redefined them to be single flares, and calculated their equivalent durations by numerically integrating the lightcurves over manually chosen intervals. We identified a total of 87 flares for GJ 4334 (1.02 flares per day) and 465 flares for EV Lac (3.90 flares per day).

\subsection{Flare Frequency Distributions}

To understand the rates of flares of different energies for both GJ~4334 and EV~Lac, we considered the flare frequency distribution, described as a single power-law function (also known as a Pareto distribution) above some minimum equivalent duration. There are various parameterizations in the literature \citep[e.g.,][]{Lacy:1976, Hilton:2011, Jackman:2021}, but we employed the following for the differential FFD as a function of the equivalent duration, $\delta$,

\begin{equation}
    \frac{dN}{d \delta} = k \, (\alpha - 1) \, \delta_{m}^{\alpha -1} \, \delta^{-\alpha} \; ,
\end{equation}

\noindent where $\alpha$ is the canonical power-law index ($\alpha > 1$), $\delta_{m}$ is the minimum threshold equivalent duration to which the power-law distribution applies, and $k$ is the normalization for the observed rate of flares. This form has the advantage of simplifying the normalization when considering the cumulative number of flares expected greater than some $\delta$,

\begin{equation}
N( \geq \delta) = k \left(  \frac{\delta}{\delta_{m}} \right)^{1 - \alpha} \; .
\label{eq:cumulativeFFD}
\end{equation}

\noindent In Equation~\ref{eq:cumulativeFFD}, $k$ explicitly describes the total number of flares with equivalent durations matching or exceeding the minimum threshold. $k$ can thus be treated as a Poisson variable for the stochastic flaring process, and can be divided by the total observing time to get the expected flare rate normalization ($N$ and $k$ have the same units).

To fit these FFDs to the TESS data for both EV~Lac and GJ~4334, we first had to determine the minimum threshold equivalent durations. At low flare energies the data can mask the presence of real flares, rendering the flare statistics incomplete. This is evident as a flattening of the cumulative flare frequency distribution for the smallest equivalent durations. We defined the minimum applicable threshold for the power-law FFDs as the smallest value such that the data could be well described by a single power-law distribution. We used the Anderson-Darling goodness-of-fit test comparing the data against the expected maximum-likelihood fit with a power-law distribution \citep[see also][]{Getman:2021}, at different choices of equivalent duration minimum, $\delta_{m}$. For $\delta_{m}$ values that are too small the incomplete flare statistics bend the distribution making a single power-law a poor fit to the data. We chose $\delta_{m}$ as the value of minimum equivalent duration where the Anderson-Darling test statistic first bottoms out to consistent values with increasing $\delta_{m}$. For EV~Lac and GJ~4334 these tests yielded minimum values of about 3~s and 5~s, respectively. Our subsequent fits thus utilized only those data for each star above the noted threshold. Alternative approaches dealing with the incomplete flare statistics for low flare energies are also evident in the literature \citep{Jackman:2021}; however, we focus here on the probability of large TESS events, rather than the weakest events, in order to compare flare probabilities for the multi-wavelength flares observed with this work ($\delta \gtrsim 10^{3}$).

We used a Bayesian formalism within PyMC \citep{PyMC:2023} for the regression analysis of the TESS equivalent duration data sets\footnote{Often in the literature, least-squares approaches for linear fits in log-log space are used to characterize stellar FFDs; however, even if the resulting power-law slopes are similar, the corresponding uncertainties do not correctly reflect the physical assumptions attributed to flaring statistics, and cannot reliably be used for statistical comparisons.}. Data are not binned, rather each data point contributes to the likelihood through its probability given the assumed power-law distribution function, and the normalization $k$ is treated as a Poisson variable constrained by the number of detected flares above the minimum threshold, $\delta_m$ \citep[see also, e.g., ][]{Loyd:2018a}. We further include the data uncertainty on the measured equivalent durations as part of this analysis by drawing the data points from their respective normal distributions using heteroskedastic errors. The main parameter of the model is the power-law index, $\alpha$, which is treated with a log-uniform prior.

\begin{figure}
    \centering
    \includegraphics[width=0.5\textwidth]{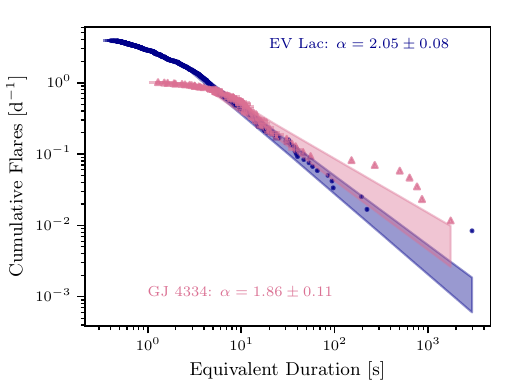}
    \caption{Flare frequency distributions in equivalent duration space (including uncertainties) for TESS data of EV~Lac (blue dots) and GJ~4334 (red triangles) show significant overlap. GJ~4334 appears to exhibit an excess of very large flares relative to EV~Lac despite their similar physical properties and slower rotation. The prevalence of large flares also flattens the apparent slope of the FFD for GJ~4334. The results of our regression analysis (see text) are shown as shaded swaths for the central 68\% confidence interval on the best fit cumulative FFD (Equation~\ref{eq:cumulativeFFD}). }
    \label{fig:ffd}
\end{figure}

We show the results of this analysis in Figure~\ref{fig:ffd}, showing the data sets for both EV~Lac (blue) and GJ~4334 (red), and the best cumulative FFDs that represent these data sets with shaded central 68\% confidence intervals. There is considerable overlap in the flare frequency distributions of both targets, which is not especially surprising as these are both active \added{M }dwarfs with similar properties. GJ~4334 does appear to have a slightly shallower FFD slope, a consequence of more frequent very strong flares ($\delta > 10^{2}$), although it is consistent with EV Lac to within 1$\sigma$. The small difference is driven by an excess of large flares that depart from the power-law established by the majority of the smaller flares. Using the median FFD for GJ~4334, we can further address the probability that we would have detected a large optical flare in our APO monitoring of this star. With 2 hours of monitoring the probability of detecting at least one $\delta = 10^{2}$~s event is $\sim$0.5\%. The observed multi-wavelength flare on GJ~4334 exhibited equivalent durations in excess of $10^{3}$~s. It is highly unlikely that we would have detected such an event with APO based on the TESS flaring statistics --- the APO/\emph{HST} event exemplifies a different kind of flare, not represented well in the broadband optical surveys. 

\begin{table}[h]
    \centering
    \caption{FFDs}
    \begin{tabular}{l c c c}
        \hline
         Star & $\delta_{m}$ [s] & $k^{*}$ [d$^{-1}$] & $\alpha$  \\
         \hline
         \hline
         EV~Lac & 3  & $1.48\pm 0.11 $  & $2.05 \pm 0.08$  \\ 
         GJ~4334 & 5 & $0.80\pm^{0.11}_{0.09} $ & $1.86 \pm 0.11$\\ 
         \hline
         \multicolumn{4}{p{0.4\linewidth}}{ $^{*}$Flare rate normalization corresponds to events at minimum equivalent duration, $\delta_{m}$.} \\
    \end{tabular}
    \label{tab:ffd_param}
\end{table}

\section{GJ 4334 in Stellar Parameter Space}\label{sec:stellar}

\citet{Pineda:2021b} self-consistently determined the physical parameters of the FUMES sample along with other M dwarfs with UV spectra and found that GJ 4334 has a mass $M_\star = 0.294 \pm 0.007$, radius $R_\star = 0.307 \pm 0.012$, effective temperature $T_\mathrm{eff} = 3260^{+69}_{-67} \,$K, and  Rossby number $Ro = 0.286$. This places GJ 4334 on the fully convective side of the convective boundary (0.35 $M_\odot$, \citealp{Gossage:2024}), like the more active stars YZ CMi ($M_\star = 0.316$, $R_\star = 0.328$, $T_\mathrm{eff} = 3293$)  and EV Lac ($M_\star = 0.320$, $R_\star = 0.331$, $T_\mathrm{eff} = 3370$) in the \citet{Pineda:2021b} sample. However, to ensure a consistent methodology across all the stars considered, the properties from \citet{Pineda:2021b} are determined from photometric relations. It would be worth doing a thorough study of GJ 4334 with additional infrared spectra to model both its atmosphere and interior across a grid of masses and metallicities to characterize its convective behavior.

GJ 4334's rotation period is 23.5 days, identified in the MEarth survey as an intermediate rotator by \citet{Newton:2016}. In that work, GJ 4334 was referred to as LHS 543a and a lightcurve showing dramatic flares and rapid spot evolution was  plotted in Figure~9. A followup MEarth paper, \citet{Mondrik:2019}, measured flare rates for a number of M dwarfs including GJ 4334 and found a tentative enhanced flare rate for intermediate rotators. If the flare behavior of these stars is different because they are at a critical transition in their magnetic topology, maybe that results in an excess of large flares leading to the tail seen in Figure~\ref{fig:ffd}. \citet{Medina:2022} identifies two main populations within the parameter space of flare rate as a function of either rotation period or Rossby number, where stars rotating faster than 10 days have high flare rates and those rotating more slowly than 100 days have very low flare rates. That interval between 10 -- 100 days, where GJ 4334 sits with its rotation period of 23 days, is very sparsely populated but the few examples available still have high flare rates comparable to the fast rotator population. \citet{Medina:2022} does not compare these stars' flare rates at specific bins of energy or equivalent duration, so it is possible that those intermediate rotators identified in that work may also have an excess at high energies like GJ 4334. \citet{Gossage:2024} identifies a spike in the empirical convective turnover time of M dwarfs at masses near 0.3 $M_\odot$ (which GJ 4334 is close to) that may be associated with a critical transition in magnetic field topology which could result in a similar irregularity in flare behavior relative to stars on either side of the spike.

\section{Summary and Conclusions}\label{sec:conclusion}
We observed a flare on the intermediate rotator GJ 4334 simultaneously with optical and far\added{-}ultraviolet spectroscopy and identified a number of features in emission lines' spectral profiles and integrated flux lightcurves:
\begin{itemize}
    \item The optical emission lines display broadened and asymmetric profiles, with H$\alpha$ having a persistent red asymmetry in contrast to H$\beta$ and H$\gamma$ which have persistent blue asymmetries.
    \item The Balmer lines decay such that the lowest order H$\alpha$ decays slowest, and the higher order lines H$\beta$ -- H$\delta$ decay increasingly rapidly.
    \item \ion{Ca}{2} K peaks slightly later than the remainder of the optical emission lines. The Balmer lines also rise in order from H$\delta$ -- H$\alpha$.
    \item The FUV lines rise, peak, and decay during the rise phase of the optical lines.
    \item Both the optical and FUV lines show a primary peak and a secondary bump, but the secondary bump of the FUV lines takes place during the primary peak of the optical lines and shows very different behaviors between lines.
    \item The FUV lines all rise and peak at similar times but decay very differently, with a potential trend in formation temperature where \ion{C}{4} 1548/50 and \ion{N}{5} 1238/42 decay more slowly than most of the other lines. The lightcurve of the Lyman-$\alpha$ wing flux has a distinct shape from all the rest, with a rapid primary decay but a very slow secondary bump decay that keeps Lyman-$\alpha$ slightly elevated relative to quiescence for the duration of the observation.
    \item Some of the optical and all of the FUV emission lines remain elevated at a pseudo-quiescent level beyond the duration of the observation. 
    \item \added{This raises the possibility that the ``quiescence'' we observe is itself a pseudo-quiescence or decay phase from a previous flare, particularly given the slope of the datapoints in the \halp\ integrated flux lightcurve seen in the bottom panel of Figure~\ref{fig:collapse_int}. This would make all our measured equivalent durations underestimates because the pseudo-quiescent denominator would be greater than the unobserved ``true quiescence'' (if such a thing can even be well-defined between the many small flares occuring frequently across the stellar disk).}
\end{itemize}

The rapid rise of the FUV lines during the optical rise phase strongly suggests that it is related to the impulsive heating of the chromosphere, but the data we have do not definitively identify whether or not it is a Neupert-effect behavior where the FUV rise is dominated by the non-thermal electron flux and the optical is the thermal radiation consequence of that heated plasma interacting with its surroundings. The complication with a Neupert-effect explanation is that the optical rise precedes the FUV, so while the two lightcurves may be related to common heating/energy transport processes, the explanation cannot be identical to that of the Neupert X-ray/radio relation. Another puzzling element is why this secondary bump in the FUV should have such different behaviors depending on the line(s) involved and whether it can be tied to a particular physical property like formation temperature. One mechanism we considered was that optically thick lines like Lyman-$\alpha$ have a slower decay and stronger secondary bump, and that the carbon lines \ion{C}{2} 1334/5 \r{A}, \ion{C}{3} 1175 \r{A}, and \ion{C}{4} 1548/50 \r{A} become optically thicker: all have stronger secondary bumps because the lines are broadened enough for photons to diffuse between lines of the multiplet/doublet, delaying the decay overall and that the secondary bump is related to a turbulent heating process causing this broadening. This would fail to explain why \ion{N}{5} also has a slow decay and significant secondary bump.

An alternative explanation is that the secondary bump is an abundance-driven behavior related to the ``first ionization potential'' (FIP) effect and/or its inverse \citep{Laming:2021} where elements whose first ionization potential is above/below 10 eV experience different enhancements/depletions during flares and in active regions (carbon and nitrogen are above the boundary while silicon is below) due to the pondermotive forces of Alfv$\acute{\text{e}}$n waves. Without additional lightcurves for low-FIP elements, we cannot definitively state that this is a FIP-effect. If it is abundance driven, then some additional explanation is required to understand the differences between ionization stages of the same element and to understand how Alfv$\acute{\text{e}}$n wave heating might affect each species differently.

\citet{Skumanich:1986} posits that the total flare rate stays fairly consistent across activity evolution with only the flare energies decaying over time. \added{On the other hand, \citet{Davenport:2019} found a consistent power-law slope and decreasing intercept as age increased, which would indicate the ratio of different energy flares is consistent but the overall rate is decreasing.} \citet{Skumanich:1986}, citing a correlation from \citet{Doyle:1985} between flare rate and quiescent X-ray emission, also posits that the quiet X-ray flux is powered by microflares, so \added{either the weakening of flare energies or decrease in flare rate over time} would lead to decaying X-ray flux.  If small flares like this one are elevating material from the chromosphere to the transition region and corona (suggested by the elevated post-flare FUV emission), and these flares are extremely frequent, then most of the material in the corona is likely associated with a recent microflare. This could explain why less active stars have relatively large amplitude activity cycles \citep{Ayres:2025}: the most active stars flare often enough that their coronae are always ``full'' because of flares lifting material into flare loops\added{,} while less active stars' coronae are ``empty'' during activity cycle troughs with few or no flares but fill up as the flare rate increases towards the peak of the activity cycle. 

Analysis of how high-energy radiation from flares affects planet atmospheres should take into account the fact that \emph{TESS} flare rates likely underpredict the delivery of X-ray and ultraviolet light, not only because of the fallacious assumption that the flare spectrum can be described by a blackbody spectrum missing line emission \citep{Jackman:2023}, but also because there is an entire category of FUV-only flares that \emph{TESS} is completely insensitive to \citep{Jackman:2024a}, possibly because these flares only significantly brighten line emission. According to the \emph{TESS} flare-frequency distribution calculated in Section~\S\ref{sec:tess}, the probability of observing a flare with an equivalent duration of 100 seconds during the APO observation would be $0.5\%$, let alone flares with equivalent durations of $10^3$ -- $10^4$ seconds as we found in different optical and FUV emission lines. Moreover, the wide spread in equivalent durations between lines shows that one cannot assume that the flare spectrum is well-described by a single spectrum scaled by an amplitude as a function of time, but that depending on the timescale of the phenomena being modeled, one either needs the flare spectrum averaged over the entirety of the flare or a more dynamic series of flare spectra snapshots during the flare's evolution. Higher cadence observations of optical chromospheric emission lines should also accompany any high-energy stellar observations to understand the relationship between these wavelength regimes and determine whether flare rates/properties of optical emission lines can predict the rates/properties of flares at higher energies.

The \emph{TESS} analysis did however show that GJ 4334 displays an excess of large flares (equivalent durations $> 100$ seconds) compared to the power-law relationship established by its smaller flares, and this excess leads to a marginally flatter slope (still consistent within 1$\sigma$) than the one found for the younger and more active EV Lac, which lacks this excess of large flares. The large flares likely correlate with an enhanced rate of coronal mass ejections, which implies that GJ 4334 is subject to an enhanced loss of angular momentum. \added{The angular momentum loss rate would partly depend on the likelihood of the M dwarf's magnetic field confining the plasma launched by these flares \citep{Alvarado-Gomez:2018,Alvarado-Gomez:2019,Sun:2022}}. If this burst of angular momentum loss occurs for all intermediate rotators, it would explain their apparent rarity \citep{Newton:2016}, and provide a new phenomenon for magnetic dynamo models to reproduce and explain.

GJ 4334 is clearly an active flare star, but follow-up observations at X-ray and UV wavelengths will help determine whether this star flares frequently enough that it can be considered a slower version of a flare star like EV Lac. If GJ 4334 can be reliably counted on to flare with high energies and at high-energy wavelengths, then it should be targeted to understand flare behavior in a region of stellar parameter space that remains unexplored and to provide examples of M dwarf flare spectra at intermediate ages that can be used for planet atmosphere models.  We also need to find the few intermediate rotator M dwarfs bright enough to characterize their flare rates and then determine whether GJ 4334 is truly anomalous or is a nearby and relatively bright representative of a larger population of intermediate rotators that flare a little extra dramatically before they transition to the decay phase of activity evolution.

\begin{acknowledgments}
G.M.D. thanks those he discussed this flare with, including but not limited to discussions with Alex Brown, P. Christian Schneider, David Wilson, Adina Feinstein, Christopher S. Moore, Crisel Suarez Bustamante, Ilija Medan, and Lyra Cao. \added{We thank the referee for their patient and constructive review of a long manuscript. Their comments significantly improved the final work.} This work used atomic data from CHIANTI, a collaborative project involving George Mason University, the University of Michigan (USA), University of Cambridge (UK) and NASA Goddard Space Flight Center (USA). Support for Program numbers HST-GO 14633, 14640, 16197 were provided by NASA through a grant from the Space Telescope Science Institute, which is operated by the Association of Universities for Research in Astronomy, Incorporated, under NASA contract NAS5-26555. This work is partially based upon effort supported by NASA under award 80NSSC22K0076 and based on observations obtained with the Apache Point Observatory 3.5 m telescope, which is owned and operated by the Astrophysical Research Consortium. This research has made use of the Astrophysics Data System, funded by NASA under Cooperative Agreement 80NSSC21M00561.

\end{acknowledgments}

\begin{contribution}
G.M.D. was responsible for writing and submitting the manuscript. J.S.P., K. F., and A.Y.\ designed the Far Ultraviolet M\added{-}dwarf Evolution Survey that obtained the \emph{Hubble} data analyzed in this work. J.S.P took the optical data at APO, and ran the FFD analysis, contributing to that section. J.S.P.\ and Z.K.B-T.\ advised G.M.D.\ on the project from its conception through to execution. A. G. S. analyzed the \emph{TESS} data to identify flares, characterized their equivalent durations, and contributed to the \emph{TESS} section. Z.K.B-T.\ is the primary developer of {\tt{chromatic}}, a software package for the analysis of spectroscopic timeseries data that was central to the project. K. G. S. contributed to the discussion of coronal mass ejections. E. R. N. contributed to the section on stellar parameters. All co-authors edited the manuscript and contributed to the introduction and discussion sections.


\end{contribution}

%

\facilities{\emph{HST}(STIS), APO:3.5m (DIS), \emph{TESS}}

\software{
Astropy \citep{AstropyCollaboration:2013, AstropyCollaboration:2018, AstropyCollaboration:2022},
bibmanager \citep{bibmanager:2020},
CHIANTI \citep{Dere:1997, DelZanna:2021},
emcee \citep{Foreman-Mackey:2013},
matplotlib \citep{matplotlib:2007},
numpy \citep{numpy:2020},
pandas \citep{pandas:2010, pandas:2020},
seaborn \citep{seaborn:2021},
chromatic (\url{https://github.com/zkbt/chromatic}),
spectralPhoton \citep{Loyd:2018a, Loyd:2018b},
pyDIS \citep{Davenport:2016},
Zenodo \citep{Zenodo:2013}
}

\appendix

\section{Flare Profile Fitting Parameter Tables}\label{sec:flare_params}

\begin{deluxetable}{ccccc}[h]
\tablecaption{Primary Peak Time and Timescale Flare Profile Parameters}
\tabletypesize{\footnotesize}
\tablehead{\colhead{Line Name} & \colhead{Primary Peak $t_1$} & \colhead{Primary Rise $r_1$} & \colhead{Primary Decay $d_1$} & \colhead{Primary Onset $t_1 - 3r_1$} \\
\colhead{[--]} & \colhead{[min]} & \colhead{[min]} & \colhead{[min]} & \colhead{[min]}}
\startdata
\ion{Si}{1}  -- \ion{Si}{3} + \ion{O}{1} & $49.09^{+0.28}_{-0.27}$ & $0.32^{+0.13}_{-0.13}$ & $0.34^{+0.18}_{-0.17}$ & $48.13^{+0.16}_{-0.17}$ \\
\ion{Si}{3} 1206 & $48.69^{+0.21}_{-0.13}$ & $0.13^{+0.11}_{-0.08}$ & $1.15^{+0.18}_{-0.33}$ & $48.31^{+0.11}_{-0.13}$ \\
\ion{Si}{4} 1394 & $48.64^{+0.20}_{-0.11}$ & $0.12^{+0.11}_{-0.08}$ & $1.25^{+0.20}_{-0.39}$ & $48.28^{+0.12}_{-0.15}$ \\
\ion{Si}{4} 1403 + \ion{O}{4} 1401 & $48.67^{+0.17}_{-0.12}$ & $0.12^{+0.10}_{-0.08}$ & $1.24^{+0.17}_{-0.21}$ & $48.31^{+0.11}_{-0.13}$ \\
\ion{N}{5} 1238/42 & $48.63^{+0.17}_{-0.10}$ & $0.10^{+0.12}_{-0.07}$ & $5.39^{+1.69}_{-2.16}$ & $48.32^{+0.13}_{-0.20}$ \\
\ion{H}{1} Lyman-$\alpha$ & $48.96^{+0.35}_{-0.30}$ & $0.33^{+0.20}_{-0.18}$ & $0.52^{+0.45}_{-0.32}$ & $47.97^{+0.29}_{-0.33}$ \\
\ion{C}{1} 1658 & $49.01^{+0.35}_{-0.32}$ & $0.32^{+0.23}_{-0.19}$ & $0.61^{+0.44}_{-0.37}$ & $48.05^{+0.33}_{-0.42}$ \\
\ion{C}{2} 1334/5 & $48.66^{+0.18}_{-0.13}$ & $0.15^{+0.10}_{-0.09}$ & $1.30^{+0.14}_{-0.14}$ & $48.22^{+0.14}_{-0.14}$ \\
\ion{C}{3} 1175 & $49.11^{+0.25}_{-0.27}$ & $0.36^{+0.12}_{-0.13}$ & $0.34^{+0.25}_{-0.18}$ & $48.04^{+0.15}_{-0.14}$ \\
\ion{C}{4} 1548/50 & $48.59^{+0.15}_{-0.11}$ & $0.12^{+0.10}_{-0.09}$ & $3.06^{+0.49}_{-0.45}$ & $48.22^{+0.15}_{-0.15}$ \\
\ion{He}{2} 1640 & $49.04^{+0.42}_{-0.33}$ & $0.23^{+0.27}_{-0.16}$ & $5.59^{+1.74}_{-2.45}$ & $48.29^{+0.31}_{-0.44}$ \\
\ion{H}{1} H$\alpha$ & $56.26^{+0.20}_{-0.21}$ & $4.28^{+0.13}_{-0.13}$ & $27.16^{+0.84}_{-0.98}$ & $43.43^{+0.23}_{-0.24}$ \\
\ion{H}{1} H$\beta$ & $57.26^{+0.15}_{-0.16}$ & $5.33^{+0.10}_{-0.10}$ & $14.86^{+0.43}_{-0.45}$ & $41.28^{+0.17}_{-0.17}$ \\
\ion{H}{1} H$\gamma$ & $57.26^{+0.20}_{-0.21}$ & $6.10^{+0.14}_{-0.14}$ & $13.49^{+0.47}_{-0.42}$ & $38.96^{+0.23}_{-0.23}$ \\
\ion{H}{1} H$\delta$ & $57.37^{+0.34}_{-0.36}$ & $6.99^{+0.24}_{-0.24}$ & $12.33^{+0.68}_{-0.62}$ & $36.39^{+0.43}_{-0.42}$ \\
\ion{Ca}{2} H +  H$\epsilon$ & $55.51^{+0.51}_{-0.36}$ & $5.11^{+0.32}_{-0.24}$ & $16.17^{+1.33}_{-1.73}$ & $40.17^{+0.44}_{-0.52}$ \\
\ion{Ca}{2} K & $60.45^{+0.62}_{-0.58}$ & $7.53^{+0.55}_{-0.50}$ & $19.99^{+2.91}_{-3.85}$ & $37.86^{+1.07}_{-1.13}$ \\
\ion{Na}{1} D\textsubscript{1} & $59.84^{+1.23}_{-1.22}$ & $7.78^{+1.11}_{-1.11}$ & $39.99^{+5.77}_{-11.89}$ & $36.50^{+2.35}_{-2.29}$ \\
\ion{Na}{1} D\textsubscript{2} & $57.57^{+1.16}_{-1.17}$ & $5.79^{+0.98}_{-0.96}$ & $41.89^{+5.23}_{-8.49}$ & $40.24^{+1.94}_{-2.03}$ \\
\ion{He}{1} D\textsubscript{3} & $58.16^{+0.41}_{-0.43}$ & $5.56^{+0.34}_{-0.35}$ & $19.02^{+1.47}_{-1.32}$ & $41.47^{+0.76}_{-0.70}$ \\
\ion{He}{1} 4471 & $56.05^{+1.17}_{-0.76}$ & $7.81^{+0.97}_{-0.63}$ & $20.59^{+2.13}_{-2.33}$ & $32.54^{+1.54}_{-1.81}$
\enddata
\end{deluxetable}

\begin{deluxetable}{cccccc}
\tablecaption{Secondary Bump Time and Timescale Flare Profile Parameters}
\tabletypesize{\scriptsize}
\tablehead{\colhead{Line Name} & \colhead{Secondary Bump $t_2$} & \colhead{Secondary Rise $r_2$} & \colhead{Secondary Decay $d_2$} & \colhead{Secondary Onset $t_2 - 3r_2$} & \colhead{Primary -- Secondary Lag $t_2 - t_1$} \\
\colhead{[--]} & \colhead{[min]} & \colhead{[min]} & \colhead{[min]} & \colhead{[min]} & \colhead{[min]}}
\startdata
\ion{Si}{1}  -- \ion{Si}{3} + \ion{O}{1} & $51.33^{+1.84}_{-1.31}$ & $2.05^{+1.91}_{-1.30}$ & $56.38^{+26.96}_{-22.34}$ & $45.00^{+3.23}_{-3.96}$ & $2.23^{+1.94}_{-1.38}$ \\
\ion{Si}{3} 1206 & $57.78^{+1.22}_{-5.73}$ & $4.66^{+1.13}_{-2.77}$ & $12.47^{+5.43}_{-2.97}$ & $43.80^{+3.06}_{-2.60}$ & $9.10^{+1.25}_{-6.01}$ \\
\ion{Si}{4} 1394 & $57.73^{+1.45}_{-6.87}$ & $4.34^{+1.61}_{-3.23}$ & $10.85^{+3.71}_{-2.87}$ & $44.45^{+3.13}_{-3.51}$ & $9.08^{+1.51}_{-7.06}$ \\
\ion{Si}{4} 1403 + \ion{O}{4} 1401 & $58.40^{+1.67}_{-5.24}$ & $4.92^{+1.77}_{-2.56}$ & $9.41^{+5.51}_{-3.15}$ & $43.26^{+3.91}_{-3.94}$ & $9.74^{+1.68}_{-5.36}$ \\
\ion{N}{5} 1238/42 & $58.54^{+1.23}_{-1.56}$ & $2.94^{+1.90}_{-1.88}$ & $12.27^{+10.13}_{-3.42}$ & $48.90^{+5.13}_{-4.11}$ & $9.90^{+1.23}_{-1.52}$ \\
\ion{H}{1} Lyman-$\alpha$ & $52.23^{+1.41}_{-2.13}$ & $1.79^{+1.10}_{-0.95}$ & $28.52^{+9.74}_{-5.76}$ & $46.45^{+2.06}_{-2.05}$ & $3.19^{+1.62}_{-2.22}$ \\
\ion{C}{1} 1658 & $59.49^{+3.55}_{-3.70}$ & $5.42^{+2.51}_{-2.86}$ & $50.61^{+32.80}_{-28.53}$ & $42.95^{+5.56}_{-4.94}$ & $10.48^{+3.58}_{-3.73}$ \\
\ion{C}{2} 1334/5 & $59.26^{+0.74}_{-0.84}$ & $5.75^{+0.80}_{-0.87}$ & $12.82^{+2.92}_{-2.24}$ & $41.93^{+2.23}_{-1.96}$ & $10.58^{+0.76}_{-0.88}$ \\
\ion{C}{3} 1175 & $50.67^{+0.80}_{-1.08}$ & $0.70^{+0.53}_{-0.42}$ & $11.07^{+1.33}_{-1.14}$ & $48.18^{+1.58}_{-1.04}$ & $1.48^{+0.99}_{-1.08}$ \\
\ion{C}{4} 1548/50 & $58.27^{+0.45}_{-0.45}$ & $3.04^{+0.98}_{-0.89}$ & $6.89^{+0.77}_{-0.62}$ & $49.17^{+2.37}_{-2.72}$ & $9.65^{+0.47}_{-0.45}$ \\
\ion{He}{2} 1640 & $58.03^{+3.86}_{-3.14}$ & $3.44^{+2.89}_{-2.25}$ & $31.54^{+41.87}_{-18.73}$ & $47.46^{+4.69}_{-5.64}$ & $8.89^{+3.88}_{-3.14}$ \\
\ion{H}{1} H$\alpha$ & $86.96^{+1.27}_{-1.85}$ & $2.56^{+1.26}_{-1.67}$ & $219.57^{+21.56}_{-34.94}$ & $79.33^{+3.11}_{-2.55}$ & $30.73^{+1.17}_{-1.83}$ \\
\ion{H}{1} H$\beta$ & $85.53^{+1.51}_{-2.70}$ & $4.60^{+1.16}_{-1.93}$ & $68.42^{+18.11}_{-12.33}$ & $71.81^{+3.07}_{-2.18}$ & $28.27^{+1.47}_{-2.67}$ \\
\ion{H}{1} H$\gamma$ & $84.71^{+1.65}_{-2.28}$ & $2.95^{+1.14}_{-1.54}$ & $36.60^{+6.57}_{-5.42}$ & $75.93^{+2.46}_{-2.02}$ & $27.49^{+1.61}_{-2.31}$ \\
\ion{H}{1} H$\delta$ & $85.63^{+1.86}_{-2.42}$ & $3.18^{+1.51}_{-1.77}$ & $21.43^{+4.75}_{-3.74}$ & $75.95^{+3.09}_{-2.81}$ & $28.31^{+1.79}_{-2.44}$ \\
\ion{Ca}{2} H +  H$\epsilon$ & $90.61^{+2.37}_{-3.92}$ & $10.58^{+3.84}_{-3.40}$ & $58.61^{+84.99}_{-23.93}$ & $59.20^{+7.44}_{-10.36}$ & $35.01^{+2.41}_{-3.85}$ \\
\ion{Ca}{2} K & $88.74^{+4.86}_{-4.11}$ & $6.02^{+6.77}_{-3.68}$ & $65.93^{+87.33}_{-32.07}$ & $70.78^{+7.08}_{-15.36}$ & $28.21^{+4.93}_{-4.01}$ \\
\ion{Na}{1} D\textsubscript{1} & $99.43^{+8.86}_{-17.13}$ & $8.42^{+8.97}_{-7.02}$ & $133.00^{+80.75}_{-83.98}$ & $68.39^{+33.62}_{-32.52}$ & $39.32^{+8.87}_{-16.45}$ \\
\ion{Na}{1} D\textsubscript{2} & $90.08^{+5.97}_{-6.00}$ & $11.91^{+5.84}_{-8.73}$ & $67.06^{+115.24}_{-58.24}$ & $55.43^{+24.13}_{-18.13}$ & $32.51^{+6.07}_{-5.90}$ \\
\ion{He}{1} D\textsubscript{3} & $83.89^{+2.07}_{-1.62}$ & $1.53^{+1.62}_{-1.06}$ & $159.37^{+61.58}_{-69.27}$ & $79.14^{+2.10}_{-3.13}$ & $25.76^{+2.15}_{-1.75}$ \\
\ion{He}{1} 4471 & $92.60^{+10.40}_{-8.56}$ & $10.85^{+6.45}_{-7.89}$ & $86.49^{+112.40}_{-75.92}$ & $61.79^{+22.77}_{-22.29}$ & $36.26^{+10.61}_{-8.52}$
\enddata
\end{deluxetable}

\begin{deluxetable}{cccc}
\tabletypesize{\footnotesize}
\tablecaption{Equivalent Duration Flare Profile Parameters}
\tablehead{\colhead{Line Name} & \colhead{Primary Equivalent Duration $\delta_1$} & \colhead{Secondary Equivalent Duration $\delta_2$} & \colhead{Total Equivalent Duration $\delta_{1 + 2}$}\\
\colhead{[--]} & \colhead{[sec]}  & \colhead{[sec]}  & \colhead{[sec]}}
\startdata
\ion{Si}{1}  -- \ion{Si}{3} + \ion{O}{1} & $2.52^{+0.27}_{-0.27} \times 10^{3}$ & $4.41^{+0.40}_{-0.40} \times 10^{3}$ & $6.92^{+0.37}_{-0.35} \times 10^{3}$ \\
\ion{Si}{3} 1206 & $1.30^{+0.12}_{-0.17} \times 10^{4}$ & $1.37^{+0.15}_{-0.12} \times 10^{4}$ & $2.66^{+0.09}_{-0.10} \times 10^{4}$ \\
\ion{Si}{4} 1394 & $1.52^{+0.14}_{-0.27} \times 10^{4}$ & $1.28^{+0.24}_{-0.14} \times 10^{4}$ & $2.79^{+0.09}_{-0.09} \times 10^{4}$ \\
\ion{Si}{4} 1403 + \ion{O}{4} 1401 & $1.30^{+0.11}_{-0.13} \times 10^{4}$ & $8.17^{+1.24}_{-1.10} \times 10^{3}$ & $2.11^{+0.09}_{-0.09} \times 10^{4}$ \\
\ion{N}{5} 1238/42 & $4.35^{+0.93}_{-1.37} \times 10^{3}$ & $2.59^{+1.42}_{-0.96} \times 10^{3}$ & $6.95^{+0.26}_{-0.25} \times 10^{3}$ \\
\ion{H}{1} Lyman-$\alpha$ & $248.49^{+80.15}_{-69.21}$ & $1.11^{+0.09}_{-0.09} \times 10^{3}$ & $1.36^{+0.05}_{-0.05} \times 10^{3}$ \\
\ion{C}{1} 1658 & $2.67^{+0.67}_{-0.65} \times 10^{3}$ & $5.81^{+0.85}_{-0.87} \times 10^{3}$ & $8.48^{+0.97}_{-0.99} \times 10^{3}$ \\
\ion{C}{2} 1334/5 & $3.61^{+0.25}_{-0.22} \times 10^{3}$ & $4.98^{+0.29}_{-0.32} \times 10^{3}$ & $8.59^{+0.24}_{-0.24} \times 10^{3}$ \\
\ion{C}{3} 1175 & $6.08^{+1.04}_{-1.00} \times 10^{3}$ & $1.05^{+0.11}_{-0.11} \times 10^{4}$ & $1.66^{+0.06}_{-0.06} \times 10^{4}$ \\
\ion{C}{4} 1548/50 & $8.49^{+0.84}_{-0.94} \times 10^{3}$ & $5.72^{+0.92}_{-0.85} \times 10^{3}$ & $1.42^{+0.03}_{-0.03} \times 10^{4}$ \\
\ion{He}{2} 1640 & $4.28^{+1.10}_{-1.48} \times 10^{3}$ & $3.13^{+1.45}_{-1.08} \times 10^{3}$ & $7.41^{+0.65}_{-0.60} \times 10^{3}$ \\
\ion{H}{1} H$\alpha$ & $759.02^{+10.78}_{-11.95}$ & $253.12^{+12.25}_{-10.18}$ & $1.01^{+0.01}_{-0.01} \times 10^{3}$ \\
\ion{H}{1} H$\beta$ & $2.12^{+0.03}_{-0.03} \times 10^{3}$ & $740.92^{+27.85}_{-26.59}$ & $2.87^{+0.03}_{-0.03} \times 10^{3}$ \\
\ion{H}{1} H$\gamma$ & $3.26^{+0.05}_{-0.05} \times 10^{3}$ & $790.61^{+43.83}_{-45.43}$ & $4.05^{+0.05}_{-0.05} \times 10^{3}$ \\
\ion{H}{1} H$\delta$ & $3.23^{+0.07}_{-0.07} \times 10^{3}$ & $632.25^{+63.52}_{-64.25}$ & $3.86^{+0.09}_{-0.09} \times 10^{3}$ \\
\ion{Ca}{2} H +  H$\epsilon$ & $3.52^{+0.17}_{-0.23} \times 10^{3}$ & $1.02^{+0.22}_{-0.16} \times 10^{3}$ & $4.54^{+0.13}_{-0.11} \times 10^{3}$ \\
\ion{Ca}{2} K & $3.13^{+0.22}_{-0.34} \times 10^{3}$ & $958.13^{+329.23}_{-203.92}$ & $4.09^{+0.15}_{-0.15} \times 10^{3}$ \\
\ion{Na}{1} D\textsubscript{1} & $996.96^{+72.50}_{-219.96}$ & $78.45^{+209.95}_{-51.19}$ & $1.08^{+0.06}_{-0.06} \times 10^{3}$ \\
\ion{Na}{1} D\textsubscript{2} & $1.07^{+0.08}_{-0.14} \times 10^{3}$ & $104.19^{+146.89}_{-60.11}$ & $1.18^{+0.06}_{-0.06} \times 10^{3}$ \\
\ion{He}{1} D\textsubscript{3} & $677.36^{+25.71}_{-24.44}$ & $189.47^{+24.30}_{-24.46}$ & $867.45^{+23.34}_{-23.44}$ \\
\ion{He}{1} 4471 & $751.68^{+50.76}_{-50.22}$ & $20.84^{+33.31}_{-15.39}$ & $781.09^{+46.64}_{-46.17}$
\enddata
\end{deluxetable}

\begin{deluxetable}{ccccc}
\tabletypesize{\footnotesize}
\tablecaption{Energy Flare Profile Parameters}
\tablehead{\colhead{Line Name} & \colhead{Primary Peak Amplitude $A_1$} & \colhead{Secondary Bump Amplitude $A_2$} & \colhead{Quiescent Flux $q_\mathrm{line}$} & \colhead{Line Flare Energy $E_\mathrm{line}$} \\
\colhead{[--]} & \colhead{[erg cm$^{-2}$ s$^{-1}$]} & \colhead{[erg cm$^{-2}$ s$^{-1}$]} & \colhead{[erg cm$^{-2}$ s$^{-1}$]} & \colhead{[erg]}}
\startdata
\ion{Si}{1}  -- \ion{Si}{3} + \ion{O}{1} & $7.07^{+0.31}_{-0.54} \times 10^{-14}$ & $4.94^{+0.62}_{-0.48} \times 10^{-15}$ & $1.24 \times 10^{-15}$ & $1.11^{+0.06}_{-0.06} \times 10^{28}$ \\
\ion{Si}{3} 1206 & $1.73^{+0.15}_{-0.16} \times 10^{-13}$ & $1.71^{+0.20}_{-0.15} \times 10^{-14}$ & $1.03 \times 10^{-15}$ & $3.52^{+0.12}_{-0.13} \times 10^{28}$ \\
\ion{Si}{4} 1394 & $1.81^{+0.16}_{-0.14} \times 10^{-13}$ & $1.60^{+0.46}_{-0.18} \times 10^{-14}$ & $9.84 \times 10^{-16}$ & $3.54^{+0.12}_{-0.11} \times 10^{28}$ \\
\ion{Si}{4} 1403 + \ion{O}{4} 1401 & $1.19^{+0.10}_{-0.11} \times 10^{-13}$ & $7.98^{+1.33}_{-1.00} \times 10^{-15}$ & $7.65 \times 10^{-16}$ & $2.09^{+0.09}_{-0.09} \times 10^{28}$ \\
\ion{N}{5} 1238/42 & $4.18^{+0.62}_{-0.35} \times 10^{-14}$ & $1.18^{+0.47}_{-0.40} \times 10^{-14}$ & $3.09 \times 10^{-15}$ & $2.77^{+0.10}_{-0.10} \times 10^{28}$ \\
\ion{H}{1} Lyman-$\alpha$ & $1.49^{+0.23}_{-0.21} \times 10^{-13}$ & $4.16^{+0.37}_{-0.34} \times 10^{-14}$ & $3.53 \times 10^{-14}$ & $6.21^{+0.24}_{-0.23} \times 10^{28}$ \\
\ion{C}{1} 1658 & $7.65^{+1.73}_{-2.26} \times 10^{-14}$ & $1.02^{+0.17}_{-0.14} \times 10^{-14}$ & $1.70 \times 10^{-15}$ & $1.85^{+0.21}_{-0.22} \times 10^{28}$ \\
\ion{C}{2} 1334/5 & $1.46^{+0.13}_{-0.10} \times 10^{-13}$ & $2.07^{+0.12}_{-0.12} \times 10^{-14}$ & $3.63 \times 10^{-15}$ & $4.02^{+0.11}_{-0.11} \times 10^{28}$ \\
\ion{C}{3} 1175 & $3.45^{+0.16}_{-0.25} \times 10^{-13}$ & $4.69^{+0.64}_{-0.53} \times 10^{-14}$ & $2.72 \times 10^{-15}$ & $5.82^{+0.21}_{-0.21} \times 10^{28}$ \\
\ion{C}{4} 1548/50 & $4.26^{+0.25}_{-0.24} \times 10^{-13}$ & $9.85^{+0.81}_{-0.85} \times 10^{-14}$ & $9.76 \times 10^{-15}$ & $1.79^{+0.04}_{-0.04} \times 10^{29}$ \\
\ion{He}{2} 1640 & $3.06^{+0.71}_{-0.62} \times 10^{-14}$ & $8.82^{+4.71}_{-2.90} \times 10^{-15}$ & $2.37 \times 10^{-15}$ & $2.27^{+0.20}_{-0.18} \times 10^{28}$ \\
\ion{H}{1} H$\alpha$ & $9.80^{+0.09}_{-0.08} \times 10^{-14}$ & $2.39^{+0.09}_{-0.08} \times 10^{-14}$ & $2.38^{+0.00}_{-0.00} \times 10^{-13}$ & $3.11^{+0.02}_{-0.02} \times 10^{29}$ \\
\ion{H}{1} H$\beta$ & $9.83^{+0.07}_{-0.07} \times 10^{-14}$ & $1.91^{+0.09}_{-0.09} \times 10^{-14}$ & $5.95^{+0.01}_{-0.01} \times 10^{-14}$ & $2.20^{+0.02}_{-0.02} \times 10^{29}$ \\
\ion{H}{1} H$\gamma$ & $7.96^{+0.07}_{-0.06} \times 10^{-14}$ & $1.37^{+0.07}_{-0.07} \times 10^{-14}$ & $3.09^{+0.01}_{-0.01} \times 10^{-14}$ & $1.61^{+0.01}_{-0.02} \times 10^{29}$ \\
\ion{H}{1} H$\delta$ & $6.10^{+0.07}_{-0.08} \times 10^{-14}$ & $1.11^{+0.09}_{-0.09} \times 10^{-14}$ & $2.38^{+0.02}_{-0.02} \times 10^{-14}$ & $1.19^{+0.02}_{-0.02} \times 10^{29}$ \\
\ion{Ca}{2} H +  H$\epsilon$ & $6.33^{+0.10}_{-0.10} \times 10^{-14}$ & $9.77^{+1.83}_{-1.50} \times 10^{-15}$ & $2.42^{+0.02}_{-0.03} \times 10^{-14}$ & $1.42^{+0.03}_{-0.03} \times 10^{29}$ \\
\ion{Ca}{2} K & $2.66^{+0.07}_{-0.07} \times 10^{-14}$ & $6.14^{+1.78}_{-1.17} \times 10^{-15}$ & $1.48^{+0.02}_{-0.02} \times 10^{-14}$ & $7.81^{+0.20}_{-0.22} \times 10^{28}$ \\
\ion{Na}{1} D\textsubscript{1} & $3.02^{+0.14}_{-0.18} \times 10^{-15}$ & $3.45^{+4.60}_{-2.36} \times 10^{-16}$ & $7.80^{+0.05}_{-0.05} \times 10^{-15}$ & $1.09^{+0.05}_{-0.05} \times 10^{28}$ \\
\ion{Na}{1} D\textsubscript{2} & $3.25^{+0.15}_{-0.17} \times 10^{-15}$ & $4.25^{+3.75}_{-2.52} \times 10^{-16}$ & $7.60^{+0.05}_{-0.06} \times 10^{-15}$ & $1.16^{+0.05}_{-0.05} \times 10^{28}$ \\
\ion{He}{1} D\textsubscript{3} & $1.13^{+0.03}_{-0.03} \times 10^{-14}$ & $1.97^{+0.24}_{-0.24} \times 10^{-15}$ & $2.56^{+0.01}_{-0.01} \times 10^{-14}$ & $2.86^{+0.07}_{-0.07} \times 10^{28}$ \\
\ion{He}{1} 4471 & $5.02^{+0.20}_{-0.20} \times 10^{-15}$ & $1.17^{+2.43}_{-0.85} \times 10^{-16}$ & $1.20^{+0.01}_{-0.01} \times 10^{-14}$ & $1.20^{+0.07}_{-0.07} \times 10^{28}$
\enddata
\end{deluxetable}


\bibliography{sample701}{}

\begin{thebibliography}{}
\expandafter\ifx\csname natexlab\endcsname\relax\def\natexlab#1{#1}\fi
\providecommand{\url}[1]{\href{#1}{#1}}
\providecommand{\dodoi}[1]{doi:~\href{http://doi.org/#1}{\nolinkurl{#1}}}
\providecommand{\doeprint}[1]{\href{http://ascl.net/#1}{\nolinkurl{http://ascl.net/#1}}}
\providecommand{\doarXiv}[1]{\href{https://arxiv.org/abs/#1}{\nolinkurl{https://arxiv.org/abs/#1}}}

\bibitem[{A.~N. {Aarnio} {et~al.}(2012){Aarnio}, {Matt}, \& {Stassun}}]{Aarnio:2012}
{Aarnio}, A.~N., {Matt}, S.~P., \& {Stassun}, K.~G. 2012, \bibinfo{title}{{Mass Loss in Pre-main-sequence Stars via Coronal Mass Ejections and Implications for Angular Momentum Loss},} \apj, 760, 9, \dodoi{10.1088/0004-637X/760/1/9}

\bibitem[{A.~N. {Aarnio} {et~al.}(2011){Aarnio}, {Stassun}, {Hughes}, \& {McGregor}}]{Aarnio:2011}
{Aarnio}, A.~N., {Stassun}, K.~G., {Hughes}, W.~J., \& {McGregor}, S.~L. 2011, \bibinfo{title}{{Solar Flares and Coronal Mass Ejections: A Statistically Determined Flare Flux - CME Mass Correlation},} \solphys, 268, 195, \dodoi{10.1007/s11207-010-9672-7}

\bibitem[{O. {Abril-Pla} {et~al.}(2023){Abril-Pla}, {Andreani}, {Carroll}, {Dong}, {Fonnesbeck}, {Kochurov}, {Kumar}, {Lao}, {Luhmann}, {Martin}, {Osthege}, {Vieira}, {Wiecki}, \& {Zinkov}}]{PyMC:2023}
{Abril-Pla}, O., {Andreani}, V., {Carroll}, C., {et~al.} 2023, \bibinfo{title}{{PyMC: a modern, and comprehensive probabilistic programming framework in Python},} PeerJ Computer Science, \dodoi{https://doi.org/10.7717/peerj-cs.1516}

\bibitem[{J.~C. {Allred} {et~al.}(2015){Allred}, {Kowalski}, \& {Carlsson}}]{Allred:2015}
{Allred}, J.~C., {Kowalski}, A.~F., \& {Carlsson}, M. 2015, \bibinfo{title}{{A Unified Computational Model for Solar and Stellar Flares},} \apj, 809, 104, \dodoi{10.1088/0004-637X/809/1/104}

\bibitem[{J.~D. {Alvarado-G{\'o}mez} {et~al.}(2018){Alvarado-G{\'o}mez}, {Drake}, {Cohen}, {Moschou}, \& {Garraffo}}]{Alvarado-Gomez:2018}
{Alvarado-G{\'o}mez}, J.~D., {Drake}, J.~J., {Cohen}, O., {Moschou}, S.~P., \& {Garraffo}, C. 2018, \bibinfo{title}{{Suppression of Coronal Mass Ejections in Active Stars by an Overlying Large-scale Magnetic Field: A Numerical Study},} \apj, 862, 93, \dodoi{10.3847/1538-4357/aacb7f}

\bibitem[{J.~D. {Alvarado-G{\'o}mez} {et~al.}(2019){Alvarado-G{\'o}mez}, {Drake}, {Moschou}, {Garraffo}, {Cohen}, {NASA LWS Focus Science Team: Solar-Stellar Connection}, {Yadav}, \& {Fraschetti}}]{Alvarado-Gomez:2019}
{Alvarado-G{\'o}mez}, J.~D., {Drake}, J.~J., {Moschou}, S.~P., {et~al.} 2019, \bibinfo{title}{{Coronal Response to Magnetically Suppressed CME Events in M-dwarf Stars},} \apjl, 884, L13, \dodoi{10.3847/2041-8213/ab44d0}

\bibitem[{ {Astropy Collaboration} {et~al.}(2013){Astropy Collaboration}, {Robitaille}, {Tollerud}, {Greenfield}, {Droettboom}, {Bray}, {Aldcroft}, {Davis}, {Ginsburg}, {Price-Whelan}, {Kerzendorf}, {Conley}, {Crighton}, {Barbary}, {Muna}, {Ferguson}, {Grollier}, {Parikh}, {Nair}, {Unther}, {Deil}, {Woillez}, {Conseil}, {Kramer}, {Turner}, {Singer}, {Fox}, {Weaver}, {Zabalza}, {Edwards}, {Azalee Bostroem}, {Burke}, {Casey}, {Crawford}, {Dencheva}, {Ely}, {Jenness}, {Labrie}, {Lim}, {Pierfederici}, {Pontzen}, {Ptak}, {Refsdal}, {Servillat}, \& {Streicher}}]{AstropyCollaboration:2013}
{Astropy Collaboration}, {Robitaille}, T.~P., {Tollerud}, E.~J., {et~al.} 2013, \bibinfo{title}{{Astropy: A community Python package for astronomy},} \aap, 558, A33, \dodoi{10.1051/0004-6361/201322068}

\bibitem[{ {Astropy Collaboration} {et~al.}(2018){Astropy Collaboration}, {Price-Whelan}, {Sip{\H{o}}cz}, {G{\"u}nther}, {Lim}, {Crawford}, {Conseil}, {Shupe}, {Craig}, {Dencheva}, {Ginsburg}, {VanderPlas}, {Bradley}, {P{\'e}rez-Su{\'a}rez}, {de Val-Borro}, {Aldcroft}, {Cruz}, {Robitaille}, {Tollerud}, {Ardelean}, {Babej}, {Bach}, {Bachetti}, {Bakanov}, {Bamford}, {Barentsen}, {Barmby}, {Baumbach}, {Berry}, {Biscani}, {Boquien}, {Bostroem}, {Bouma}, {Brammer}, {Bray}, {Breytenbach}, {Buddelmeijer}, {Burke}, {Calderone}, {Cano Rodr{\'\i}guez}, {Cara}, {Cardoso}, {Cheedella}, {Copin}, {Corrales}, {Crichton}, {D'Avella}, {Deil}, {Depagne}, {Dietrich}, {Donath}, {Droettboom}, {Earl}, {Erben}, {Fabbro}, {Ferreira}, {Finethy}, {Fox}, {Garrison}, {Gibbons}, {Goldstein}, {Gommers}, {Greco}, {Greenfield}, {Groener}, {Grollier}, {Hagen}, {Hirst}, {Homeier}, {Horton}, {Hosseinzadeh}, {Hu}, {Hunkeler}, {Ivezi{\'c}}, {Jain}, {Jenness}, {Kanarek}, {Kendrew}, {Kern}, {Kerzendorf}, {Khvalko}, {King}, {Kirkby}, {Kulkarni},
  {Kumar}, {Lee}, {Lenz}, {Littlefair}, {Ma}, {Macleod}, {Mastropietro}, {McCully}, {Montagnac}, {Morris}, {Mueller}, {Mumford}, {Muna}, {Murphy}, {Nelson}, {Nguyen}, {Ninan}, {N{\"o}the}, {Ogaz}, {Oh}, {Parejko}, {Parley}, {Pascual}, {Patil}, {Patil}, {Plunkett}, {Prochaska}, {Rastogi}, {Reddy Janga}, {Sabater}, {Sakurikar}, {Seifert}, {Sherbert}, {Sherwood-Taylor}, {Shih}, {Sick}, {Silbiger}, {Singanamalla}, {Singer}, {Sladen}, {Sooley}, {Sornarajah}, {Streicher}, {Teuben}, {Thomas}, {Tremblay}, {Turner}, {Terr{\'o}n}, {van Kerkwijk}, {de la Vega}, {Watkins}, {Weaver}, {Whitmore}, {Woillez}, {Zabalza}, \& {Astropy Contributors}}]{AstropyCollaboration:2018}
{Astropy Collaboration}, {Price-Whelan}, A.~M., {Sip{\H{o}}cz}, B.~M., {et~al.} 2018, \bibinfo{title}{{The Astropy Project: Building an Open-science Project and Status of the v2.0 Core Package},} \aj, 156, 123, \dodoi{10.3847/1538-3881/aabc4f}

\bibitem[{ {Astropy Collaboration} {et~al.}(2022){Astropy Collaboration}, {Price-Whelan}, {Lim}, {Earl}, {Starkman}, {Bradley}, {Shupe}, {Patil}, {Corrales}, {Brasseur}, {N{\"o}the}, {Donath}, {Tollerud}, {Morris}, {Ginsburg}, {Vaher}, {Weaver}, {Tocknell}, {Jamieson}, {van Kerkwijk}, {Robitaille}, {Merry}, {Bachetti}, {G{\"u}nther}, {Aldcroft}, {Alvarado-Montes}, {Archibald}, {B{\'o}di}, {Bapat}, {Barentsen}, {Baz{\'a}n}, {Biswas}, {Boquien}, {Burke}, {Cara}, {Cara}, {Conroy}, {Conseil}, {Craig}, {Cross}, {Cruz}, {D'Eugenio}, {Dencheva}, {Devillepoix}, {Dietrich}, {Eigenbrot}, {Erben}, {Ferreira}, {Foreman-Mackey}, {Fox}, {Freij}, {Garg}, {Geda}, {Glattly}, {Gondhalekar}, {Gordon}, {Grant}, {Greenfield}, {Groener}, {Guest}, {Gurovich}, {Handberg}, {Hart}, {Hatfield-Dodds}, {Homeier}, {Hosseinzadeh}, {Jenness}, {Jones}, {Joseph}, {Kalmbach}, {Karamehmetoglu}, {Ka{\l}uszy{\'n}ski}, {Kelley}, {Kern}, {Kerzendorf}, {Koch}, {Kulumani}, {Lee}, {Ly}, {Ma}, {MacBride}, {Maljaars}, {Muna}, {Murphy}, {Norman},
  {O'Steen}, {Oman}, {Pacifici}, {Pascual}, {Pascual-Granado}, {Patil}, {Perren}, {Pickering}, {Rastogi}, {Roulston}, {Ryan}, {Rykoff}, {Sabater}, {Sakurikar}, {Salgado}, {Sanghi}, {Saunders}, {Savchenko}, {Schwardt}, {Seifert-Eckert}, {Shih}, {Jain}, {Shukla}, {Sick}, {Simpson}, {Singanamalla}, {Singer}, {Singhal}, {Sinha}, {Sip{\H{o}}cz}, {Spitler}, {Stansby}, {Streicher}, {{\v{S}}umak}, {Swinbank}, {Taranu}, {Tewary}, {Tremblay}, {de Val-Borro}, {Van Kooten}, {Vasovi{\'c}}, {Verma}, {de Miranda Cardoso}, {Williams}, {Wilson}, {Winkel}, {Wood-Vasey}, {Xue}, {Yoachim}, {Zhang}, {Zonca}, \& {Astropy Project Contributors}}]{AstropyCollaboration:2022}
{Astropy Collaboration}, {Price-Whelan}, A.~M., {Lim}, P.~L., {et~al.} 2022, \bibinfo{title}{{The Astropy Project: Sustaining and Growing a Community-oriented Open-source Project and the Latest Major Release (v5.0) of the Core Package},} \apj, 935, 167, \dodoi{10.3847/1538-4357/ac7c74}

\bibitem[{T. {Ayres}(2025){Ayres}}]{Ayres:2025}
{Ayres}, T. 2025, \bibinfo{title}{{Landscape of Coronal X-Ray Variability and Cycles},} \aj, 169, 281, \dodoi{10.3847/1538-3881/adc570}

\bibitem[{P.~R. {Behr} {et~al.}(2023){Behr}, {France}, {Brown}, {Duvvuri}, {Bean}, {Berta-Thompson}, {Froning}, {Miguel}, {Pineda}, {Wilson}, \& {Youngblood}}]{Behr:2023}
{Behr}, P.~R., {France}, K., {Brown}, A., {et~al.} 2023, \bibinfo{title}{{The MUSCLES Extension for Atmospheric Transmission Spectroscopy: UV and X-Ray Host-star Observations for JWST ERS \& GTO Targets},} \aj, 166, 35, \dodoi{10.3847/1538-3881/acdb70}

\bibitem[{A.~O. {Benz} \& M. {G{\"u}del}(2010){Benz} \& {G{\"u}del}}]{Benz:2010}
{Benz}, A.~O., \& {G{\"u}del}, M. 2010, \bibinfo{title}{{Physical Processes in Magnetically Driven Flares on the Sun, Stars, and Young Stellar Objects},} \araa, 48, 241, \dodoi{10.1146/annurev-astro-082708-101757}

\bibitem[{P.~E. {Cubillos}(2020){Cubillos}}]{bibmanager:2020}
{Cubillos}, P.~E. 2020, {bibmanager: A BibTeX manager for LaTeX projects}, Zenodo, \dodoi{10.5281/zenodo.2547042}

\bibitem[{J. {Davenport} {et~al.}(2016){Davenport}, {De Val-Borro}, \& {Wilkinson}}]{Davenport:2016}
{Davenport}, J., {De Val-Borro}, M., \& {Wilkinson}, T.~D. 2016, {pydis: Possibly Useful}, v1.1 Zenodo, \dodoi{10.5281/zenodo.58753}

\bibitem[{J.~R.~A. {Davenport} {et~al.}(2019){Davenport}, {Covey}, {Clarke}, {Boeck}, {Cornet}, \& {Hawley}}]{Davenport:2019}
{Davenport}, J. R.~A., {Covey}, K.~R., {Clarke}, R.~W., {et~al.} 2019, \bibinfo{title}{{The Evolution of Flare Activity with Stellar Age},} \apj, 871, 241, \dodoi{10.3847/1538-4357/aafb76}

\bibitem[{G. {Del Zanna} {et~al.}(2021){Del Zanna}, {Dere}, {Young}, \& {Landi}}]{DelZanna:2021}
{Del Zanna}, G., {Dere}, K.~P., {Young}, P.~R., \& {Landi}, E. 2021, \bibinfo{title}{{CHIANTI{\textemdash}An Atomic Database for Emission Lines. XVI. Version 10, Further Extensions},} \apj, 909, 38, \dodoi{10.3847/1538-4357/abd8ce}

\bibitem[{K.~P. {Dere} {et~al.}(1997){Dere}, {Landi}, {Mason}, {Monsignori Fossi}, \& {Young}}]{Dere:1997}
{Dere}, K.~P., {Landi}, E., {Mason}, H.~E., {Monsignori Fossi}, B.~C., \& {Young}, P.~R. 1997, \bibinfo{title}{{CHIANTI - an atomic database for emission lines},} \aaps, 125, 149, \dodoi{10.1051/aas:1997368}

\bibitem[{H. {Diamond-Lowe} {et~al.}(2024){Diamond-Lowe}, {King}, {Youngblood}, {Brown}, {Howard}, {Winters}, {Wilson}, {France}, {Mendon{\c{c}}a}, {Buchhave}, {Corrales}, {Kreidberg}, {Medina}, {Bean}, {Berta-Thompson}, {Evans-Soma}, {Froning}, {Duvvuri}, {Kempton}, {Miguel}, {Pineda}, \& {Schneider}}]{Diamond-Lowe:2024}
{Diamond-Lowe}, H., {King}, G.~W., {Youngblood}, A., {et~al.} 2024, \bibinfo{title}{{High-energy spectra of LTT 1445A and GJ 486 reveal flares and activity},} \aap, 689, A48, \dodoi{10.1051/0004-6361/202450107}

\bibitem[{C. {Dong} {et~al.}(2017){Dong}, {Huang}, {Lingam}, {T{\'o}th}, {Gombosi}, \& {Bhattacharjee}}]{Dong:2017}
{Dong}, C., {Huang}, Z., {Lingam}, M., {et~al.} 2017, \bibinfo{title}{{The Dehydration of Water Worlds via Atmospheric Losses},} \apjl, 847, L4, \dodoi{10.3847/2041-8213/aa8a60}

\bibitem[{J.~G. {Doyle} \& C.~J. {Butler}(1985){Doyle} \& {Butler}}]{Doyle:1985}
{Doyle}, J.~G., \& {Butler}, C.~J. 1985, \bibinfo{title}{{Ultraviolet radiation from stellar flares and the coronal X-ray emission for dwarf-Me stars},} \nat, 313, 378, \dodoi{10.1038/313378a0}

\bibitem[{G.~M. Duvvuri {et~al.}(2022)Duvvuri, Pineda, Berta-Thompson, France, \& Youngblood}]{duvvuri_2022_6909473}
Duvvuri, G.~M., Pineda, J.~S., Berta-Thompson, Z.~K., France, K., \& Youngblood, A. 2022, FUMES III: Ultraviolet and Optical Variability of M Dwarf Chromospheres, Zenodo, \dodoi{10.5281/zenodo.6909473}

\bibitem[{G.~M. {Duvvuri} {et~al.}(2023){Duvvuri}, {Pineda}, {Berta-Thompson}, {France}, \& {Youngblood}}]{Duvvuri:2023}
{Duvvuri}, G.~M., {Pineda}, J.~S., {Berta-Thompson}, Z.~K., {France}, K., \& {Youngblood}, A. 2023, \bibinfo{title}{{FUMES. III. Ultraviolet and Optical Variability of M-dwarf Chromospheres},} \aj, 165, 12, \dodoi{10.3847/1538-3881/ac9b49}

\bibitem[{ {European Organization For Nuclear Research} \&  {OpenAIRE}(2013){European Organization For Nuclear Research} \& {OpenAIRE}}]{Zenodo:2013}
{European Organization For Nuclear Research}, \& {OpenAIRE}. 2013, Zenodo, CERN, \dodoi{10.25495/7GXK-RD71}

\bibitem[{F. {Favata} \& J.~H.~M.~M. {Schmitt}(1999){Favata} \& {Schmitt}}]{Favata:1999}
{Favata}, F., \& {Schmitt}, J.~H.~M.~M. 1999, \bibinfo{title}{{Spectroscopic analysis of a super-hot giant flare observed on Algol by BeppoSAX on 30 August 1997},} \aap, 350, 900, \dodoi{10.48550/arXiv.astro-ph/9909041}

\bibitem[{A. {Feinstein} {et~al.}(2020){Feinstein}, {Montet}, \& {Ansdell}}]{Feinstein:2020}
{Feinstein}, A., {Montet}, B., \& {Ansdell}, M. 2020, \bibinfo{title}{{stella: Convolutional Neural Networks for Flare Identification in TESS},} The Journal of Open Source Software, 5, 2347, \dodoi{10.21105/joss.02347}

\bibitem[{A.~D. {Feinstein} {et~al.}(2022){Feinstein}, {France}, {Youngblood}, {Duvvuri}, {Teal}, {Cauley}, {Seligman}, {Gaidos}, {Kempton}, {Bean}, {Diamond-Lowe}, {Newton}, {Ginzburg}, {Plavchan}, {Gao}, \& {Schlichting}}]{Feinstein:2022}
{Feinstein}, A.~D., {France}, K., {Youngblood}, A., {et~al.} 2022, \bibinfo{title}{{AU Microscopii in the Far-UV: Observations in Quiescence, during Flares, and Implications for AU Mic b and c},} \aj, 164, 110, \dodoi{10.3847/1538-3881/ac8107}

\bibitem[{G.~H. {Fisher}(1989){Fisher}}]{Fisher:1989}
{Fisher}, G.~H. 1989, \bibinfo{title}{{Dynamics of Flare-driven Chromospheric Condensations},} \apj, 346, 1019, \dodoi{10.1086/168084}

\bibitem[{D. {Foreman-Mackey} {et~al.}(2013){Foreman-Mackey}, {Hogg}, {Lang}, \& {Goodman}}]{Foreman-Mackey:2013}
{Foreman-Mackey}, D., {Hogg}, D.~W., {Lang}, D., \& {Goodman}, J. 2013, \bibinfo{title}{{emcee: The MCMC Hammer},} \pasp, 125, 306, \dodoi{10.1086/670067}

\bibitem[{K. {France} {et~al.}(2016){France}, {Loyd}, {Youngblood}, {Brown}, {Schneider}, {Hawley}, {Froning}, {Linsky}, {Roberge}, {Buccino}, {Davenport}, {Fontenla}, {Kaltenegger}, {Kowalski}, {Mauas}, {Miguel}, {Redfield}, {Rugheimer}, {Tian}, {Vieytes}, {Walkowicz}, \& {Weisenburger}}]{France:2016}
{France}, K., {Loyd}, R.~O.~P., {Youngblood}, A., {et~al.} 2016, \bibinfo{title}{{The MUSCLES Treasury Survey. I. Motivation and Overview},} \apj, 820, 89, \dodoi{10.3847/0004-637X/820/2/89}

\bibitem[{K. {France} {et~al.}(2020){France}, {Duvvuri}, {Egan}, {Koskinen}, {Wilson}, {Youngblood}, {Froning}, {Brown}, {Alvarado-G{\'o}mez}, {Berta-Thompson}, {Drake}, {Garraffo}, {Kaltenegger}, {Kowalski}, {Linsky}, {Loyd}, {Mauas}, {Miguel}, {Pineda}, {Rugheimer}, {Schneider}, {Tian}, \& {Vieytes}}]{France:2020}
{France}, K., {Duvvuri}, G., {Egan}, H., {et~al.} 2020, \bibinfo{title}{{The High-energy Radiation Environment around a 10 Gyr M Dwarf: Habitable at Last?},} \aj, 160, 237, \dodoi{10.3847/1538-3881/abb465}

\bibitem[{C.~S. {Froning} {et~al.}(2019){Froning}, {Kowalski}, {France}, {Loyd}, {Schneider}, {Youngblood}, {Wilson}, {Brown}, {Berta-Thompson}, {Pineda}, {Linsky}, {Rugheimer}, \& {Miguel}}]{Froning:2019}
{Froning}, C.~S., {Kowalski}, A., {France}, K., {et~al.} 2019, \bibinfo{title}{{A Hot Ultraviolet Flare on the M Dwarf Star GJ 674},} \apjl, 871, L26, \dodoi{10.3847/2041-8213/aaffcd}

\bibitem[{B. {Fuhrmeister} {et~al.}(2011){Fuhrmeister}, {Lalitha}, {Poppenhaeger}, {Rudolf}, {Liefke}, {Reiners}, {Schmitt}, \& {Ness}}]{Fuhrmeister:2011}
{Fuhrmeister}, B., {Lalitha}, S., {Poppenhaeger}, K., {et~al.} 2011, \bibinfo{title}{{Multi-wavelength observations of Proxima Centauri},} \aap, 534, A133, \dodoi{10.1051/0004-6361/201117447}

\bibitem[{B. {Fuhrmeister} {et~al.}(2018){Fuhrmeister}, {Czesla}, {Schmitt}, {Jeffers}, {Caballero}, {Zechmeister}, {Reiners}, {Ribas}, {Amado}, {Quirrenbach}, {B{\'e}jar}, {Galad{\'\i}-Enr{\'\i}quez}, {Guenther}, {K{\"u}rster}, {Montes}, \& {Seifert}}]{Fuhrmeister:2018}
{Fuhrmeister}, B., {Czesla}, S., {Schmitt}, J.~H.~M.~M., {et~al.} 2018, \bibinfo{title}{{The CARMENES search for exoplanets around M dwarfs. Wing asymmetries of H{\ensuremath{\alpha}}, Na I D, and He I lines},} \aap, 615, A14, \dodoi{10.1051/0004-6361/201732204}

\bibitem[{A. {Garc{\'\i}a Soto} {et~al.}(2025){Garc{\'\i}a Soto}, {Duvvuri}, {Newton}, {Howard}, {N{\'u}{\~n}ez}, \& {Douglas}}]{GarciaSoto:2025}
{Garc{\'\i}a Soto}, A., {Duvvuri}, G.~M., {Newton}, E.~R., {et~al.} 2025, \bibinfo{title}{{Short-term Balmer Line Emission Variability in M Dwarfs},} \apj, 982, 98, \dodoi{10.3847/1538-4357/adb615}

\bibitem[{R.~E. {Gershberg} {et~al.}(2024){Gershberg}, {Kleeorin}, {Pustilnik}, {Airapetian}, \& {Shlyapnikov}}]{Gershberg:2024}
{Gershberg}, R.~E., {Kleeorin}, N.~I., {Pustilnik}, L.~A., {Airapetian}, V.~S., \& {Shlyapnikov}, A.~A. 2024, \bibinfo{title}{{Physics of mid- and low-mass stars with solar-type activity and their impact on exoplanetary environments},} arXiv e-prints, arXiv:2411.11898, \dodoi{10.48550/arXiv.2411.11898}

\bibitem[{K.~V. {Getman} {et~al.}(2021){Getman}, {Feigelson}, \& {Garmire}}]{Getman:2021}
{Getman}, K.~V., {Feigelson}, E.~D., \& {Garmire}, G.~P. 2021, \bibinfo{title}{{X-Ray Superflares from Pre-main-sequence Stars: Flare Modeling},} \apj, 920, 154, \dodoi{10.3847/1538-4357/ac1746}

\bibitem[{S. {Gossage} {et~al.}(2024){Gossage}, {Kiman}, {Monsch}, {Medina}, {Drake}, {Garraffo}, {Yuxi}, {Lu}, {Wing}, \& {Wright}}]{Gossage:2024}
{Gossage}, S., {Kiman}, R., {Monsch}, K., {et~al.} 2024, \bibinfo{title}{{On Convective Turnover Times and Dynamos In Low-Mass Stars},} arXiv e-prints, arXiv:2410.20000, \dodoi{10.48550/arXiv.2410.20000}

\bibitem[{C.~E. {Harman} {et~al.}(2015){Harman}, {Schwieterman}, {Schottelkotte}, \& {Kasting}}]{Harman:2015}
{Harman}, C.~E., {Schwieterman}, E.~W., {Schottelkotte}, J.~C., \& {Kasting}, J.~F. 2015, \bibinfo{title}{{Abiotic O$_{2}$ Levels on Planets around F, G, K, and M Stars: Possible False Positives for Life?},} \apj, 812, 137, \dodoi{10.1088/0004-637X/812/2/137}

\bibitem[{C.~R. Harris {et~al.}(2020)Harris, Millman, van~der Walt, Gommers, Virtanen, Cournapeau, Wieser, Taylor, Berg, Smith, Kern, Picus, Hoyer, van Kerkwijk, Brett, Haldane, del R{\'{i}}o, Wiebe, Peterson, G{\'{e}}rard-Marchant, Sheppard, Reddy, Weckesser, Abbasi, Gohlke, \& Oliphant}]{numpy:2020}
Harris, C.~R., Millman, K.~J., van~der Walt, S.~J., {et~al.} 2020, \bibinfo{title}{Array programming with {NumPy},} Nature, 585, 357, \dodoi{10.1038/s41586-020-2649-2}

\bibitem[{S.~L. {Hawley} \& B.~R. {Pettersen}(1991){Hawley} \& {Pettersen}}]{Hawley:1991}
{Hawley}, S.~L., \& {Pettersen}, B.~R. 1991, \bibinfo{title}{{The Great Flare of 1985 April 12 on AD Leonis},} \apj, 378, 725, \dodoi{10.1086/170474}

\bibitem[{S.~L. {Hawley} {et~al.}(2003){Hawley}, {Allred}, {Johns-Krull}, {Fisher}, {Abbett}, {Alekseev}, {Avgoloupis}, {Deustua}, {Gunn}, {Seiradakis}, {Sirk}, \& {Valenti}}]{Hawley:2003}
{Hawley}, S.~L., {Allred}, J.~C., {Johns-Krull}, C.~M., {et~al.} 2003, \bibinfo{title}{{Multiwavelength Observations of Flares on AD Leonis},} \apj, 597, 535, \dodoi{10.1086/378351}

\bibitem[{E.~J. {Hilton}(2011){Hilton}}]{Hilton:2011}
{Hilton}, E.~J. 2011, \bibinfo{title}{{The Galactic M Dwarf Flare Rate},} PhD thesis, University of Washington, Seattle

\bibitem[{C. {Hoffmeister}(1967){Hoffmeister}}]{Hoffmeister:1967}
{Hoffmeister}, C. 1967, \bibinfo{title}{{New Flare Star S10113 And with Proper Motion},} Information Bulletin on Variable Stars, 203, 1

\bibitem[{S.~M. {H{\"o}rst}(2017){H{\"o}rst}}]{Horst:2017}
{H{\"o}rst}, S.~M. 2017, \bibinfo{title}{{Titan's atmosphere and climate},} Journal of Geophysical Research (Planets), 122, 432, \dodoi{10.1002/2016JE005240}

\bibitem[{W.~S. {Howard} {et~al.}(2022){Howard}, {MacGregor}, {Osten}, {Forbrich}, {Cranmer}, {Tristan}, {Weinberger}, {Youngblood}, {Barclay}, {Loyd}, {Shkolnik}, {Zic}, \& {Wilner}}]{Howard:2022}
{Howard}, W.~S., {MacGregor}, M.~A., {Osten}, R., {et~al.} 2022, \bibinfo{title}{{The Mouse That Squeaked: A Small Flare from Proxima Cen Observed in the Millimeter, Optical, and Soft X-Ray with Chandra and ALMA},} \apj, 938, 103, \dodoi{10.3847/1538-4357/ac9134}

\bibitem[{R. {Hu} {et~al.}(2012){Hu}, {Seager}, \& {Bains}}]{Hu:2012}
{Hu}, R., {Seager}, S., \& {Bains}, W. 2012, \bibinfo{title}{{Photochemistry in Terrestrial Exoplanet Atmospheres. I. Photochemistry Model and Benchmark Cases},} \apj, 761, 166, \dodoi{10.1088/0004-637X/761/2/166}

\bibitem[{J.~D. {Hunter}(2007){Hunter}}]{matplotlib:2007}
{Hunter}, J.~D. 2007, \bibinfo{title}{{Matplotlib: A 2D Graphics Environment},} Computing in Science and Engineering, 9, 90, \dodoi{10.1109/MCSE.2007.55}

\bibitem[{E. {Ilin}(2021){Ilin}}]{Ilin:2021}
{Ilin}, E. 2021, \bibinfo{title}{{AltaiPony - Flare science in Kepler, K2 and TESS light curves},} The Journal of Open Source Software, 6, 2845, \dodoi{10.21105/joss.02845}

\bibitem[{J.~A.~G. {Jackman} {et~al.}(2024){Jackman}, {Shkolnik}, {Loyd}, \& {Richey-Yowell}}]{Jackman:2024a}
{Jackman}, J. A.~G., {Shkolnik}, E.~L., {Loyd}, R.~O.~P., \& {Richey-Yowell}, T. 2024, \bibinfo{title}{{Optically quiet, but FUV loud: results from comparing the far-ultraviolet predictions of flare models with TESS and HST},} \mnras, 533, 1894, \dodoi{10.1093/mnras/stae1570}

\bibitem[{J.~A.~G. {Jackman} {et~al.}(2023){Jackman}, {Shkolnik}, {Million}, {Fleming}, {Richey-Yowell}, \& {Loyd}}]{Jackman:2023}
{Jackman}, J. A.~G., {Shkolnik}, E.~L., {Million}, C., {et~al.} 2023, \bibinfo{title}{{Extending optical flare models to the UV: results from comparing of TESS and GALEX flare observations for M Dwarfs},} \mnras, 519, 3564, \dodoi{10.1093/mnras/stac3135}

\bibitem[{J.~A.~G. {Jackman} {et~al.}(2021){Jackman}, {Wheatley}, {Acton}, {Anderson}, {Bayliss}, {Briegal}, {Burleigh}, {Casewell}, {G{\"a}nsicke}, {Gill}, {Gillen}, {Goad}, {G{\"u}nther}, {Henderson}, {Hodgkin}, {Jenkins}, {Pugh}, {Queloz}, {Raynard}, {Tilbrook}, {Watson}, \& {West}}]{Jackman:2021}
{Jackman}, J. A.~G., {Wheatley}, P.~J., {Acton}, J.~S., {et~al.} 2021, \bibinfo{title}{{Stellar flares detected with the Next Generation Transit Survey},} \mnras, 504, 3246, \dodoi{10.1093/mnras/stab979}

\bibitem[{C.~P. {Johnstone} {et~al.}(2021){Johnstone}, {Bartel}, \& {G{\"u}del}}]{Johnstone:2021}
{Johnstone}, C.~P., {Bartel}, M., \& {G{\"u}del}, M. 2021, \bibinfo{title}{{The active lives of stars: A complete description of the rotation and XUV evolution of F, G, K, and M dwarfs},} \aap, 649, A96, \dodoi{10.1051/0004-6361/202038407}

\bibitem[{B. {Klein} {et~al.}(2021){Klein}, {Donati}, {H{\'e}brard}, {Zaire}, {Folsom}, {Morin}, {Delfosse}, \& {Bonfils}}]{Klein:2021}
{Klein}, B., {Donati}, J.-F., {H{\'e}brard}, {\'E}.~M., {et~al.} 2021, \bibinfo{title}{{The large-scale magnetic field of Proxima Centauri near activity maximum},} \mnras, 500, 1844, \dodoi{10.1093/mnras/staa3396}

\bibitem[{T.~T. {Koskinen} {et~al.}(2022){Koskinen}, {Lavvas}, {Huang}, {Bergsten}, {Fernandes}, \& {Young}}]{Koskinen:2022}
{Koskinen}, T.~T., {Lavvas}, P., {Huang}, C., {et~al.} 2022, \bibinfo{title}{{Mass Loss by Atmospheric Escape from Extremely Close-in Planets},} \apj, 929, 52, \dodoi{10.3847/1538-4357/ac4f45}

\bibitem[{A.~F. {Kowalski} {et~al.}(2024){Kowalski}, {Allred}, \& {Carlsson}}]{Kowalski:2024}
{Kowalski}, A.~F., {Allred}, J.~C., \& {Carlsson}, M. 2024, \bibinfo{title}{{Time-dependent Stellar Flare Models of Deep Atmospheric Heating},} \apj, 969, 121, \dodoi{10.3847/1538-4357/ad4148}

\bibitem[{A.~F. {Kowalski} {et~al.}(2013){Kowalski}, {Hawley}, {Wisniewski}, {Osten}, {Hilton}, {Holtzman}, {Schmidt}, \& {Davenport}}]{Kowalski:2013}
{Kowalski}, A.~F., {Hawley}, S.~L., {Wisniewski}, J.~P., {et~al.} 2013, \bibinfo{title}{{Time-resolved Properties and Global Trends in dMe Flares from Simultaneous Photometry and Spectra},} \apjs, 207, 15, \dodoi{10.1088/0067-0049/207/1/15}

\bibitem[{A.~F. {Kowalski} {et~al.}(2017){Kowalski}, {Allred}, {Uitenbroek}, {Tremblay}, {Brown}, {Carlsson}, {Osten}, {Wisniewski}, \& {Hawley}}]{Kowalski:2017}
{Kowalski}, A.~F., {Allred}, J.~C., {Uitenbroek}, H., {et~al.} 2017, \bibinfo{title}{{Hydrogen Balmer Line Broadening in Solar and Stellar Flares},} \apj, 837, 125, \dodoi{10.3847/1538-4357/aa603e}

\bibitem[{A.~F. {Kowalski} {et~al.}(2019){Kowalski}, {Wisniewski}, {Hawley}, {Osten}, {Brown}, {Fari{\~n}a}, {Valenti}, {Brown}, {Xilouris}, {Schmidt}, \& {Johns-Krull}}]{Kowalski:2019}
{Kowalski}, A.~F., {Wisniewski}, J.~P., {Hawley}, S.~L., {et~al.} 2019, \bibinfo{title}{{The Near-ultraviolet Continuum Radiation in the Impulsive Phase of HF/GF-type dMe Flares. I. Data},} \apj, 871, 167, \dodoi{10.3847/1538-4357/aaf058}

\bibitem[{C.~H. {Lacy} {et~al.}(1976){Lacy}, {Moffett}, \& {Evans}}]{Lacy:1976}
{Lacy}, C.~H., {Moffett}, T.~J., \& {Evans}, D.~S. 1976, \bibinfo{title}{{UV Ceti stars: statistical analysis of observational data.},} \apjs, 30, 85, \dodoi{10.1086/190358}

\bibitem[{J.~M. {Laming}(2021){Laming}}]{Laming:2021}
{Laming}, J.~M. 2021, \bibinfo{title}{{The FIP and Inverse-FIP Effects in Solar Flares},} \apj, 909, 17, \dodoi{10.3847/1538-4357/abd9c3}

\bibitem[{R.~O.~P. {Loyd} \& K. {France}(2014){Loyd} \& {France}}]{Loyd:2014}
{Loyd}, R.~O.~P., \& {France}, K. 2014, \bibinfo{title}{{Fluctuations and Flares in the Ultraviolet Line Emission of Cool Stars: Implications for Exoplanet Transit Observations},} \apjs, 211, 9, \dodoi{10.1088/0067-0049/211/1/9}

\bibitem[{R.~O.~P. {Loyd} {et~al.}(2018{\natexlab{a}}){Loyd}, {Shkolnik}, {Schneider}, {Barman}, {Meadows}, {Pagano}, \& {Peacock}}]{Loyd:2018b}
{Loyd}, R.~O.~P., {Shkolnik}, E.~L., {Schneider}, A.~C., {et~al.} 2018{\natexlab{a}}, \bibinfo{title}{{HAZMAT. IV. Flares and Superflares on Young M Stars in the Far Ultraviolet},} \apj, 867, 70, \dodoi{10.3847/1538-4357/aae2ae}

\bibitem[{R.~O.~P. {Loyd} {et~al.}(2018{\natexlab{b}}){Loyd}, {France}, {Youngblood}, {Schneider}, {Brown}, {Hu}, {Segura}, {Linsky}, {Redfield}, {Tian}, {Rugheimer}, {Miguel}, \& {Froning}}]{Loyd:2018a}
{Loyd}, R.~O.~P., {France}, K., {Youngblood}, A., {et~al.} 2018{\natexlab{b}}, \bibinfo{title}{{The MUSCLES Treasury Survey. V. FUV Flares on Active and Inactive M Dwarfs},} \apj, 867, 71, \dodoi{10.3847/1538-4357/aae2bd}

\bibitem[{M.~A. {MacGregor} {et~al.}(2021){MacGregor}, {Weinberger}, {Loyd}, {Shkolnik}, {Barclay}, {Howard}, {Zic}, {Osten}, {Cranmer}, {Kowalski}, {Lenc}, {Youngblood}, {Estes}, {Wilner}, {Forbrich}, {Hughes}, {Law}, {Murphy}, {Boley}, \& {Matthews}}]{MacGregor:2021}
{MacGregor}, M.~A., {Weinberger}, A.~J., {Loyd}, R.~O.~P., {et~al.} 2021, \bibinfo{title}{{Discovery of an Extremely Short Duration Flare from Proxima Centauri Using Millimeter through Far-ultraviolet Observations},} \apjl, 911, L25, \dodoi{10.3847/2041-8213/abf14c}

\bibitem[{H. {Maehara} {et~al.}(2021){Maehara}, {Notsu}, {Namekata}, {Honda}, {Kowalski}, {Katoh}, {Ohshima}, {Iida}, {Oeda}, {Murata}, {Yamanaka}, {Takagi}, {Sasada}, {Akitaya}, {Ikuta}, {Okamoto}, {Nogami}, \& {Shibata}}]{Maehara:2021}
{Maehara}, H., {Notsu}, Y., {Namekata}, K., {et~al.} 2021, \bibinfo{title}{{Time-resolved spectroscopy and photometry of M dwarf flare star YZ Canis Minoris with OISTER and TESS: Blue asymmetry in the H{\ensuremath{\alpha}} line during the non-white light flare},} \pasj, 73, 44, \dodoi{10.1093/pasj/psaa098}

\bibitem[{A.~A. {Medina} {et~al.}(2022){Medina}, {Winters}, {Irwin}, \& {Charbonneau}}]{Medina:2022}
{Medina}, A.~A., {Winters}, J.~G., {Irwin}, J.~M., \& {Charbonneau}, D. 2022, \bibinfo{title}{{Galactic Kinematics and Observed Flare Rates of a Volume-complete Sample of Mid-to-late M Dwarfs: Constraints on the History of the Stellar Radiation Environment of Planets Orbiting Low-mass Stars},} \apj, 935, 104, \dodoi{10.3847/1538-4357/ac77f9}

\bibitem[{N. {Mondrik} {et~al.}(2019){Mondrik}, {Newton}, {Charbonneau}, \& {Irwin}}]{Mondrik:2019}
{Mondrik}, N., {Newton}, E., {Charbonneau}, D., \& {Irwin}, J. 2019, \bibinfo{title}{{An Increased Rate of Large Flares at Intermediate Rotation Periods for Mid-to-late M Dwarfs},} \apj, 870, 10, \dodoi{10.3847/1538-4357/aaee64}

\bibitem[{R.~H. {Munro} {et~al.}(1979){Munro}, {Gosling}, {Hildner}, {MacQueen}, {Poland}, \& {Ross}}]{Munro:1979}
{Munro}, R.~H., {Gosling}, J.~T., {Hildner}, E., {et~al.} 1979, \bibinfo{title}{{The association of coronal mass ejection transients with other forms of solar activity.},} \solphys, 61, 201, \dodoi{10.1007/BF00155456}

\bibitem[{K. {Namekata} {et~al.}(2022){Namekata}, {Ichimoto}, {Ishii}, \& {Shibata}}]{Namekata:2022}
{Namekata}, K., {Ichimoto}, K., {Ishii}, T.~T., \& {Shibata}, K. 2022, \bibinfo{title}{{Sun-as-a-star Analysis of H{\ensuremath{\alpha}} Spectra of a Solar Flare Observed by SMART/SDDI: Time Evolution of Red Asymmetry and Line Broadening},} \apj, 933, 209, \dodoi{10.3847/1538-4357/ac75cd}

\bibitem[{W.~M. {Neupert}(1968){Neupert}}]{Neupert:1968}
{Neupert}, W.~M. 1968, \bibinfo{title}{{Comparison of Solar X-Ray Line Emission with Microwave Emission during Flares},} \apjl, 153, L59, \dodoi{10.1086/180220}

\bibitem[{E.~R. {Newton} {et~al.}(2016){Newton}, {Irwin}, {Charbonneau}, {Berta-Thompson}, {Dittmann}, \& {West}}]{Newton:2016}
{Newton}, E.~R., {Irwin}, J., {Charbonneau}, D., {et~al.} 2016, \bibinfo{title}{{The Rotation and Galactic Kinematics of Mid M Dwarfs in the Solar Neighborhood},} \apj, 821, 93, \dodoi{10.3847/0004-637X/821/2/93}

\bibitem[{R.~A. {Osten} {et~al.}(2005){Osten}, {Hawley}, {Allred}, {Johns-Krull}, \& {Roark}}]{Osten:2005}
{Osten}, R.~A., {Hawley}, S.~L., {Allred}, J.~C., {Johns-Krull}, C.~M., \& {Roark}, C. 2005, \bibinfo{title}{{From Radio to X-Ray: Flares on the dMe Flare Star EV Lacertae},} \apj, 621, 398, \dodoi{10.1086/427275}

\bibitem[{T. pandas~development team(2020)pandas~development team}]{pandas:2020}
pandas~development team, T. 2020, pandas-dev/pandas: Pandas, latest Zenodo, \dodoi{10.5281/zenodo.3509134}

\bibitem[{E.~K. {Pass} {et~al.}(2024){Pass}, {Charbonneau}, {Latham}, {Berlind}, {Calkins}, {Esquerdo}, \& {Mink}}]{Pass:2024}
{Pass}, E.~K., {Charbonneau}, D., {Latham}, D.~W., {et~al.} 2024, \bibinfo{title}{{The Mass Dependence of H{\ensuremath{\alpha}} Emission and Stellar Spindown for Fully Convective M Dwarfs},} \apj, 966, 231, \dodoi{10.3847/1538-4357/ad3631}

\bibitem[{R.~R. {Paudel} {et~al.}(2021){Paudel}, {Barclay}, {Schlieder}, {Quintana}, {Gilbert}, {Vega}, {Youngblood}, {Silverstein}, {Osten}, {Tucker}, {Huber}, {Do}, {Hamaguchi}, {Mullan}, {Gizis}, {Monsue}, {Col{\'o}n}, {Boyd}, {Davenport}, \& {Walkowicz}}]{Paudel:2021}
{Paudel}, R.~R., {Barclay}, T., {Schlieder}, J.~E., {et~al.} 2021, \bibinfo{title}{{Simultaneous Multiwavelength Flare Observations of EV Lacertae},} \apj, 922, 31, \dodoi{10.3847/1538-4357/ac1946}

\bibitem[{K.~J.~H. {Phillips} {et~al.}(1992){Phillips}, {Bromage}, \& {Doyle}}]{Phillips:1992}
{Phillips}, K.~J.~H., {Bromage}, G.~E., \& {Doyle}, J.~G. 1992, \bibinfo{title}{{The Origin of the Far-Ultraviolet Continuum in Solar and Stellar Flares},} \apj, 385, 731, \dodoi{10.1086/170979}

\bibitem[{J.~S. {Pineda} {et~al.}(2021{\natexlab{a}}){Pineda}, {Youngblood}, \& {France}}]{Pineda:2021a}
{Pineda}, J.~S., {Youngblood}, A., \& {France}, K. 2021{\natexlab{a}}, \bibinfo{title}{{The Far Ultraviolet M-dwarf Evolution Survey. I. The Rotational Evolution of High-energy Emissions},} \apj, 911, 111, \dodoi{10.3847/1538-4357/abe8d7}

\bibitem[{J.~S. {Pineda} {et~al.}(2021{\natexlab{b}}){Pineda}, {Youngblood}, \& {France}}]{Pineda:2021b}
{Pineda}, J.~S., {Youngblood}, A., \& {France}, K. 2021{\natexlab{b}}, \bibinfo{title}{{The M-dwarf Ultraviolet Spectroscopic Sample. I. Determining Stellar Parameters for Field Stars},} \apj, 918, 40, \dodoi{10.3847/1538-4357/ac0aea}

\bibitem[{K.~E. {Rockcliffe} {et~al.}(2023){Rockcliffe}, {Newton}, {Youngblood}, {Duvvuri}, {Plavchan}, {Gao}, {Mann}, \& {Lowrance}}]{Rockcliffe:2023}
{Rockcliffe}, K.~E., {Newton}, E.~R., {Youngblood}, A., {et~al.} 2023, \bibinfo{title}{{The Variable Detection of Atmospheric Escape around the Young, Hot Neptune AU Mic b},} \aj, 166, 77, \dodoi{10.3847/1538-3881/ace536}

\bibitem[{K.~E. {Rockcliffe} {et~al.}(2025){Rockcliffe}, {Newton}, {Youngblood}, {Duvvuri}, {Gilbert}, {Plavchan}, {Gao}, {M{\"u}ller}, {Feinstein}, {Barclay}, \& {Lopez}}]{Rockcliffe:2025}
{Rockcliffe}, K.~E., {Newton}, E.~R., {Youngblood}, A., {et~al.} 2025, \bibinfo{title}{{Far-ultraviolet Flares and Variability of the Young M Dwarf AU Mic: A Nondetection of Planet C in Transit at Ly{\ensuremath{\alpha}}},} \aj, 169, 321, \dodoi{10.3847/1538-3881/adccc7}

\bibitem[{L. {Schaefer} {et~al.}(2016){Schaefer}, {Wordsworth}, {Berta-Thompson}, \& {Sasselov}}]{Schaefer:2016}
{Schaefer}, L., {Wordsworth}, R.~D., {Berta-Thompson}, Z., \& {Sasselov}, D. 2016, \bibinfo{title}{{Predictions of the Atmospheric Composition of GJ 1132b},} \apj, 829, 63, \dodoi{10.3847/0004-637X/829/2/63}

\bibitem[{T. {Shibayama} {et~al.}(2013){Shibayama}, {Maehara}, {Notsu}, {Notsu}, {Nagao}, {Honda}, {Ishii}, {Nogami}, \& {Shibata}}]{Shibayama:2013}
{Shibayama}, T., {Maehara}, H., {Notsu}, S., {et~al.} 2013, \bibinfo{title}{{Superflares on Solar-type Stars Observed with Kepler. I. Statistical Properties of Superflares},} \apjs, 209, 5, \dodoi{10.1088/0067-0049/209/1/5}

\bibitem[{E.~L. {Shkolnik} \& T.~S. {Barman}(2014){Shkolnik} \& {Barman}}]{Shkolnik:2014}
{Shkolnik}, E.~L., \& {Barman}, T.~S. 2014, \bibinfo{title}{{HAZMAT. I. The Evolution of Far-UV and Near-UV Emission from Early M Stars},} \aj, 148, 64, \dodoi{10.1088/0004-6256/148/4/64}

\bibitem[{A. {Skumanich}(1972){Skumanich}}]{Skumanich:1972}
{Skumanich}, A. 1972, \bibinfo{title}{{Time Scales for Ca II Emission Decay, Rotational Braking, and Lithium Depletion},} \apj, 171, 565, \dodoi{10.1086/151310}

\bibitem[{A. {Skumanich}(1986){Skumanich}}]{Skumanich:1986}
{Skumanich}, A. 1986, \bibinfo{title}{{Some Evidence on the Evolution of the Flare Mechanism in Dwarf Stars},} \apj, 309, 858, \dodoi{10.1086/164654}

\bibitem[{X. {Sun} {et~al.}(2022){Sun}, {T{\"o}r{\"o}k}, \& {DeRosa}}]{Sun:2022}
{Sun}, X., {T{\"o}r{\"o}k}, T., \& {DeRosa}, M.~L. 2022, \bibinfo{title}{{Torus-stable zone above starspots},} \mnras, 509, 5075, \dodoi{10.1093/mnras/stab3249}

\bibitem[{F. {Tian} {et~al.}(2014){Tian}, {France}, {Linsky}, {Mauas}, \& {Vieytes}}]{Tian:2014}
{Tian}, F., {France}, K., {Linsky}, J.~L., {Mauas}, P. J.~D., \& {Vieytes}, M.~C. 2014, \bibinfo{title}{{High stellar FUV/NUV ratio and oxygen contents in the atmospheres of potentially habitable planets},} Earth and Planetary Science Letters, 385, 22, \dodoi{10.1016/j.epsl.2013.10.024}

\bibitem[{I.~I. {Tristan} {et~al.}(2023){Tristan}, {Notsu}, {Kowalski}, {Brown}, {Wisniewski}, {Osten}, {Vrijmoet}, {White}, {Carter}, {Grady}, {Henry}, {Hinojosa}, {Lomax}, {Neff}, {Paredes}, \& {Soutter}}]{Tristan:2023}
{Tristan}, I.~I., {Notsu}, Y., {Kowalski}, A.~F., {et~al.} 2023, \bibinfo{title}{{A 7 Day Multiwavelength Flare Campaign on AU Mic. I. High-time-resolution Light Curves and the Thermal Empirical Neupert Effect},} \apj, 951, 33, \dodoi{10.3847/1538-4357/acc94f}

\bibitem[{M.~L. Waskom(2021)Waskom}]{seaborn:2021}
Waskom, M.~L. 2021, \bibinfo{title}{seaborn: statistical data visualization,} Journal of Open Source Software, 6, 3021, \dodoi{10.21105/joss.03021}

\bibitem[{ {W}es {M}c{K}inney(2010){W}es {M}c{K}inney}]{pandas:2010}
{W}es {M}c{K}inney. 2010, \bibinfo{title}{{D}ata {S}tructures for {S}tatistical {C}omputing in {P}ython,} in {P}roceedings of the 9th {P}ython in {S}cience {C}onference, ed. {S}t\'efan van~der {W}alt \& {J}arrod {M}illman, 56 -- 61, \dodoi{10.25080/Majora-92bf1922-00a}

\bibitem[{A. {Youngblood} {et~al.}(2021){Youngblood}, {Pineda}, \& {France}}]{Youngblood:2021}
{Youngblood}, A., {Pineda}, J.~S., \& {France}, K. 2021, \bibinfo{title}{{FUMES. II. Ly{\ensuremath{\alpha}} Reconstructions of Young, Active M Dwarfs},} \apj, 911, 112, \dodoi{10.3847/1538-4357/abe8d8}

\end{thebibliography}
\bibliographystyle{aasjournalv7}


\end{document}